\documentclass[fleqn,usenatbib]{mnras}

\usepackage{newtxtext,newtxmath}
\usepackage[T1]{fontenc}
\usepackage{ae,aecompl}
\usepackage{graphicx}
\usepackage{amsmath}
\usepackage{amssymb}
\usepackage[applemac]{inputenc} 
\usepackage{epstopdf}
\usepackage{hyperref}
\usepackage{color}
\usepackage{footnote}
\usepackage{lineno}
\usepackage{caption}
\usepackage[perpage]{footmisc}
\usepackage{algorithm}
\usepackage{algpseudocode}
\usepackage{booktabs,siunitx}

\usepackage{eso-pic}

\makesavenoteenv{table}
\newcommand{\mstellar}{\ensuremath{M_\textrm{stellar}}}
\newcommand{\msolar}{\ensuremath{\textrm{M}_{\sun}}}
\newcommand{\logmstellar}{\log{{M}_\textrm{stellar}/\textrm{M}_{\sun}}}

\newcommand{\mstep}{\ensuremath{M_\textrm{step}}}
\newcommand{\PEGASE}{P\'EGASE}
\newcommand{\PEGASEbursts}{P\'EGASE:bursts}

\newcommand{\lowz}{low-$z$}
\newcommand{\mubias}{\ensuremath{\mu_\textrm{bias}}}
\newcommand{\NSPEC}{251} 
\newcommand{\NDES}{206}  
\newcommand{\NLOST}{45}  
\newcommand{\NLOWZ}{122} 
\newcommand{\NTOT}{328}  
\defcitealias{Wiseman2020}{W20}
\defcitealias{Salpeter1955}{S55}
\defcitealias{Kroupa2001}{K01}
\defcitealias{Bruzual2003}{BC03}
\defcitealias{Maraston2005}{M05}
\defcitealias{Brout2019a}{B19}

\title[The Host Galaxy Mass Step]{First Cosmology Results using Type Ia Supernovae from the Dark Energy Survey: The Effect of Host Galaxy Properties on Supernova Luminosity}
\author[M. Smith et al.]{
\parbox{\textwidth}{
\Large
M.~Smith$^{1,\thanks{email: \href{mailto:mat.smith@soton.ac.uk}{mat.smith@soton.ac.uk}}}$,
M.~Sullivan,$^{1}$
P.~Wiseman,$^{1}$
R.~Kessler,$^{2,3}$
D.~Scolnic,$^{4}$
D.~Brout,$^{5,6}$
C.~B.~D'Andrea,$^{5,7}$
T.~M.~Davis,$^{8}$
R.~J.~Foley,$^{9}$
C.~Frohmaier,$^{10}$
L.~Galbany,$^{11}$
R.~R.~Gupta,$^{12}$
C.~P.~Guti\'errez,$^{1}$
S.~R.~Hinton,$^{8}$
L.~Kelsey,$^{1}$
C.~Lidman,$^{13,14}$
E.~Macaulay,$^{10,15}$
A.~M\"oller,$^{13,14,16}$
R.~C.~Nichol,$^{10}$
P.~Nugent,$^{12,17}$
A.~Palmese,$^{3,18}$
M.~Pursiainen,$^{1}$
M.~Sako,$^{5}$
R.~C.~Thomas,$^{12}$
B.~E.~Tucker,$^{13}$
D.~Carollo,$^{19}$
G.~F.~Lewis,$^{20}$
N.~E.~Sommer,$^{13}$
T.~M.~C.~Abbott,$^{21}$
M.~Aguena,$^{22,23}$
S.~Allam,$^{18}$
S.~Avila,$^{24}$
E.~Bertin,$^{25,26}$
S.~Bhargava,$^{27}$
D.~Brooks,$^{28}$
E.~Buckley-Geer,$^{18}$
D.~L.~Burke,$^{29,30}$
A.~Carnero~Rosell,$^{23,31}$
M.~Carrasco~Kind,$^{32,33}$
M.~Costanzi,$^{34,35}$
L.~N.~da Costa,$^{23,36}$
J.~De~Vicente,$^{31}$
S.~Desai,$^{37}$
H.~T.~Diehl,$^{18}$
P.~Doel,$^{28}$
T.~F.~Eifler,$^{38,39}$
S.~Everett,$^{9}$
B.~Flaugher,$^{18}$
P.~Fosalba,$^{40,41}$
J.~Frieman,$^{3,18}$
J.~Garc\'ia-Bellido,$^{24}$
E.~Gaztanaga,$^{40,41}$
K.~Glazebrook,$^{42}$
D.~Gruen,$^{29,30,43}$
R.~A.~Gruendl,$^{32,33}$
J.~Gschwend,$^{23,36}$
G.~Gutierrez,$^{18}$
W.~G.~Hartley,$^{28,44}$
D.~L.~Hollowood,$^{9}$
K.~Honscheid,$^{45,46}$
D.~J.~James,$^{47}$
E.~Krause,$^{38}$
K.~Kuehn,$^{48,49}$
N.~Kuropatkin,$^{18}$
M.~Lima,$^{22,23}$
N.~MacCrann,$^{45,46}$
M.~A.~G.~Maia,$^{23,36}$
J.~L.~Marshall,$^{50}$
P.~Martini,$^{45,51}$
P.~Melchior,$^{52}$
F.~Menanteau,$^{32,33}$
R.~Miquel,$^{53,54}$
F.~Paz-Chinch\'{o}n,$^{32,33}$
A.~A.~Plazas,$^{52}$
A.~K.~Romer,$^{27}$
A.~Roodman,$^{29,30}$
E.~S.~Rykoff,$^{29,30}$
E.~Sanchez,$^{31}$
V.~Scarpine,$^{18}$
M.~Schubnell,$^{55}$
S.~Serrano,$^{40,41}$
I.~Sevilla-Noarbe,$^{31}$
E.~Suchyta,$^{56}$
M.~E.~C.~Swanson,$^{33}$
G.~Tarle,$^{55}$
D.~Thomas,$^{10}$
D.~L.~Tucker,$^{18}$
T.~N.~Varga,$^{57,58}$
and A.~R.~Walker$^{21}$
\begin{center} (DES Collaboration) \end{center}
}
\vspace{0.0cm}
\\
\parbox{\textwidth}{
Author affiliations are shown in Appendix \ref{app:affiliations}
}
}

\pubyear{2020}
\begin{document}
\label{firstpage}
\pagerange{\pageref{firstpage}--\pageref{lastpage}}
\maketitle
\begin{abstract}
We present improved photometric measurements for the host galaxies of 206 spectroscopically confirmed type Ia supernovae discovered by the Dark Energy Survey Supernova Program (DES-SN) and used in the first DES-SN cosmological analysis. Fitting spectral energy distributions to the $griz$ photometric measurements of the DES-SN host galaxies, we derive stellar masses and star-formation rates. For the DES-SN sample, when considering a 5D ($z$, $x_1$, $c$, $\alpha$, $\beta$) bias correction, we find evidence of a Hubble residual `mass step', where SNe Ia in high mass galaxies ($>10^{10} \textrm{M}_{\odot}$) are intrinsically more luminous (after correction) than their low mass counterparts by $\gamma=0.040\pm0.019$mag. This value is consistent with other recent supernova samples that use a 5D correction, and is larger by $0.031$mag than the value found in the first DES-SN cosmological analysis. This difference is due to a combination of updated photometric measurements and improved star formation histories and is not from host-galaxy misidentification. When using a 1D (redshift-only) bias correction the inferred mass step is larger, with $\gamma=0.066\pm0.020$mag. The 1D-5D $\gamma$ difference for DES-SN is $0.026\pm0.009$mag. We show that this difference is due to a strong correlation between host galaxy stellar mass and the $x_1$ component of the 5D distance-bias correction. To better understand this effect, we include an intrinsic correlation between light-curve width and stellar mass in simulated SN Ia samples. We show that a 5D fit recovers $\gamma$ with $-9$mmag bias compared to a $+2$mmag bias for a 1D fit. This difference can explain part of the discrepancy seen in the data. Improvements in modeling correlations between galaxy properties and SN is necessary to determine the implications for $\gamma$ and ensure unbiased precision estimates of the dark energy equation-of-state as we enter the era of LSST.
\end{abstract}

\section{Introduction}
\label{sec:Intro}

As standardisable candles, type Ia supernovae (SNe Ia) are a geometric probe of the expansion history of the universe \citep{Riess1998,Perlmutter1999} and provide a mature, robust measure of its accelerated expansion \citep{Betoule2014,Riess2018,Scolnic2018,DES2019}. SNe Ia are not perfect standard candles: empirical \lq corrections\rq\ based on light-curve shape \citep{Phillips1993} and colour \citep{Riess1996,Tripp1998} are required to standardise their peak luminosity, reducing the observed scatter in their peak magnitudes from $\sim$0.35\,mag to $\sim$0.14\,mag, or $\sim7$ per cent in distance. With around 1000 spectroscopically confirmed SNe Ia currently published for cosmological analyses \citep{Scolnic2018}, and with the size of photometrically-classified samples ever-increasing \citep{Jones2018,LSST2012}, understanding the origin and optimal treatment of these empirical correlations is key to maximising their constraining power. Enhancing the standardisation of SNe Ia beyond corrections for light-curve shape and colour may improve measurements of the evolution of dark energy with redshift. 

The local environment in which SNe~Ia explode can provide insights into the physical mechanisms governing these events and their observed diversity. Global properties of SN~Ia host galaxies, such as the stellar mass, star-formation rate (SFR), metallicity and mean age of the stellar populations, have been observed to correlate with various properties of SNe~Ia. SNe~Ia are $\sim25$ times more common (per unit stellar mass) in highly star-forming galaxies than passive systems \citep{Mannucci2005,Sullivan2006,Smith2012}, and such star-forming galaxies also host intrinsically slower-declining and observationally brighter SNe Ia \citep{Hamuy1995,Hamuy2000,Sullivan2006,Johansson2013,Wolf2016,Moreno-Raya2018}. The origin of these differences is unknown, but may arise from multiple progenitor configurations capable of producing SNe Ia \citep{Scannapieco2005,Mannucci2006}.

Correlations between the luminosity of SNe Ia (after correction for light-curve width and colour) and the stellar mass of their host galaxies have motivated a third empirical correction \citep{Kelly2010,Sullivan2010,Lampeitl2010}. This is commonly parameterised as a \lq mass step\rq, with two absolute magnitudes for SNe Ia in the cosmological fits, depending on whether an event is located in a high stellar-mass ($\mstellar>10^{10}$\,\msolar) or low stellar-mass ($\mstellar<10^{10}$\,\msolar) host galaxy. This correction has been observed at $3-6$\,$\sigma$ confidence in multiple samples, spanning low- and high-redshift, and using different light curve fitters and distance estimation techniques. It is now ubiquitous in most cosmological analyses using SNe Ia \citep{Sullivan2011,Betoule2014,Scolnic2018}, but lacks a firm physical motivation. There has been speculation that the mass step may be driven by the age of the stellar population \citep{Childress2014} or metallicity \citep{Sullivan2010}, and similar luminosity effects have also been observed using variables beyond stellar mass, such as metallicity, stellar age \citep{Gupta2011,DAndrea2011,Hayden2013} and star-formation rate \citep{Sullivan2010}. As stellar populations evolve with redshift, and evolve differently for age and metallicity, uncovering and modelling the source of the mass step is a key challenge when using cosmological samples of $>1000$ SNe Ia over an extended phase of cosmic history. 

While the majority of early studies used SN Ia samples at cosmological distances, and thus focused on a galaxy's \textit{global} photometric properties, more recent studies have highlighted a link between the intrinsic brightness of SNe Ia and the characteristics of their \textit{local} environment. \citet{Rigault2013}, using (for example) $\textrm{H}\alpha$ nebular emission as a proxy for local SFR, have shown that locally passive environments preferentially host redder, low-stretch SNe, which appear to be intrinsically brighter than their locally star-forming counterparts after correction. The size of this local effect remains surprisingly controversial: using statistically significant datasets, \citet{Roman2018}, \citet{Kim2018}, \citet{Rigault2018} and \citet{Kelsey2020} find results consistent with \citet{Rigault2013}, while \citet{Jones2015} and \citet{Jones2018b} find no evidence of a correlation between SN Ia luminosity and local environment. 

The Dark Energy Survey (DES) \lq three-year\rq\ (DES3YR) cosmological analysis \citep{DES2019} combines data for \NSPEC\ spectroscopically confirmed SNe~Ia (\NDES\ after applying light-curve quality cuts) from the DES-SN program, with a low-redshift sample of \NLOWZ\ SNe~Ia to constrain the equation-of-state of dark energy ($w$). Using data on the global properties of its SNe~Ia, the DES3YR cosmology analysis \citep[][hereafter \citetalias{Brout2019a}]{Brout2019a}, using a \lq BEAMS with Bias Corrections\rq\ \citep[BBC;][]{Kessler2017} framework, found no significant correlations between SN~Ia luminosity and stellar mass for the DES-SN subsample. It was unclear whether this was due to the relatively small DES-SN Ia sample size, or whether some novel aspects of the DES analysis pipeline had (perhaps inadvertently) removed or corrected for the mass-step effect. In this paper, we present new host galaxy data for the \NSPEC\ spectroscopically confirmed SNe Ia from DES-SN. Using stacked DES imaging from all five years of DES-SN, excluding dates around the SN explosion, we measure the host galaxy fluxes and estimate their stellar masses and star-formation rates, and compare them to the light-curve properties of the SNe Ia they host, finding a strong correlation between \mstellar, SN~Ia light-curve width and the bias correction used to correct for survey selection effects. Using simulated samples of the DES-SN survey that include intrinsic correlations between SN parameters and host galaxy \mstellar\ we show that this correlation inadvertently leads to reduction in the `mass step' measured by DES. This result is consistent across a wide range of systematic tests. 

This paper is organised as follows. In \S\ref{sec:data}, we introduce the photometric measurements and derived galaxy parameters for the DES-SN sample, and examine the sensitivity of these measurements to alternative photometric measurements and assumptions on the template galaxy spectral energy distributions (SED) used to determine stellar masses. \S\ref{sec:lcparams} considers correlations between the light-curve parameters of SNe Ia and the derived parameters of their host galaxies. \S\ref{sec:massstep} introduces and measures the mass step for DES3YR, and studies how systematic uncertainties affect the inferred mass step. In \S\ref{sec:simulations} we use simulated samples to show that estimates of the mass step in a BBC framework are dependent on the underlying assumptions of the galaxy population and their correlation with the SNe that they host. We conclude in \S\ref{sec:conclusions}. Throughout this paper, we use AB magnitudes \citep{Oke83} and where relevant assume a reference cosmological model that is a spatially-flat $\Lambda$CDM model, with a matter density $\Omega_\mathrm{m}=0.3$ and a Hubble constant $H_0 = 70$\,km\,s$^{-1}$\,Mpc$^{-1}$. 

\section{SN and Host Galaxy Data}
\label{sec:data}

The DES-SN Program was a five-year rolling search using the 570 Megapixel Dark Energy Camera \citep[DECam]{Flaugher2015} on the 4-m Blanco telescope at the Cerro Tololo Inter-American Observatory (CTIO), giving a  $2.7$\,deg$^{2}$ field-of-view. DES-SN observed two \lq deep\rq\ fields and 8 \lq shallow\rq\ fields in $griz$ filters approximately every 7 days, to single-visit depths of $\sim$24.5\,mag and $\sim$23.5\,mag respectively. 

Transient events were detected using a difference-imaging pipeline \citep{Kessler2015}, with machine-learning algorithms used to remove spurious candidates \citep{Goldstein2015}. During the first three years, \NSPEC\ SNe Ia were spectroscopically classified \citep{DAndrea2018}. The SN Ia light curve fluxes were measured using a \lq Scene Model Photometry\rq\ (SMP) technique \citep{Brout2019b}, and the photometric calibration is described in \citet{Burke2018} and \citet{Lasker2018}. The light curves were fit with the SALT2 spectral energy distribution (SED) template \citep{Guy2007,Guy2010}, trained using the Joint Lightcurve Analysis \citep[JLA; ][]{Betoule2014} SN compilation, and implemented in the \textsc{snana} software package \citep{Kessler2009}. The light-curve fitting provides estimates of the rest-frame amplitude ($m_B$), stretch ($x_1$), and colour ($c$) for each SN. Quality cuts, based on the light-curve coverage, are applied to the sample \citep[see][for details]{Brout2019a}, which removes \NLOST\ SNe Ia. This leaves \NDES\ SNe Ia in the fiducial DES sample. Due to an updated estimate of the time of maximum light in the \textsc{snana} package, one event (SNID$=$1279500) is lost compared to the analysis of \citet{DES2019} and \citetalias{Brout2019a}. This does not impact our conclusions. 

In the DES analysis \citepalias{Brout2019a}, the DES-SN sample is combined with \NLOWZ\ \lq low-redshift\rq\ ($z<0.1$) SNe Ia from the literature to form the DES3YR sample. In this paper, we also consider other SN Ia samples from the literature: the JLA sample \citep{Betoule2014} (740 SNe Ia) and the \lq Pantheon\rq\ sample \citep{Scolnic2018}. The latter combines SNe Ia discovered by the Pan-STARRS1 (PS1) Medium Deep Survey with the JLA sample, as well as events from the \textit{Hubble Space Telescope} \citep[][]{Suzuki2012,Riess2018} to form a sample of 1048 SNe Ia. 

\subsection{SN Ia distance estimation}
\label{subsec:snIa-method}

The observed distance modulus for each SN, $\mu_\textrm{obs}$, is given by
\begin{equation}
\label{eqn:salt2formula}
\mu_\textrm{obs} = m_B + \alpha x_1 - \beta c +M_0 + \gamma G_\textrm{host} + \mubias,
\end{equation}
where
\begin{equation}
\label{eqn:gamma_def}
G_\textrm{host} = \left \{
  \begin{tabular}{ll}
  +1/2			& if $\logmstellar>\mstep$ \\
  --1/2 	        & otherwise.
  \end{tabular}
  \right.
\end{equation}
\mstellar\ is the SN host-galaxy stellar mass, and $\gamma$ is commonly referred to as the \lq mass step\rq. The value of \mstep\ is often fixed to some fiducial value, typically 10. $\alpha$, $\beta$, $\gamma$ and  $M_0$ are nuisance parameters that describe the global SN Ia population, and are usually determined simultaneously with the distances of with the cosmological parameters.

A correction, \mubias, determined from simulations, is also made to each SN Ia to account for various survey selection effects, such as Malmquist bias and spectroscopic targeting algorithms. In previous analyses \citep[e.g.,][]{Conley2011,Betoule2014}, \mubias\ is a function of redshift (a \lq 1D correction\rq), and is estimated from either image-level simulations \citep{Perrett2010} or catalogue-level simulations \citep{Betoule2014}. More recent analyses \citep{Scolnic2018,Brout2019a} have determined \mubias\ as a 5D function of ($z$, $x_1$, $c$, $\alpha$, $\beta$) using the BBC framework, splitting \mubias\ into 3 terms: ${m_B}_\textrm{bias}$, ${x_1}_\textrm{bias}$ and $c_\textrm{bias}$. The fiducial DES3YR analysis \citepalias{Brout2019a} uses the BBC formalism, which relies upon large, accurate simulations of the underlying SN Ia population determined using the \textsc{snana} package \citep{Kessler2019} combined with a model for intrinsic brightness variations, or `intrinsic scatter'. The DES3YR analysis \citepalias{Brout2019a} uses two intrinsic scatter models from \citet{Kessler2013}: \citep[G10;][]{Guy2010} and \citep[C11;][]{Chotard2011}. For simplicity, we restrict our analysis to the G10 model, which recovers consistent values of $\gamma$ for the DES-SN sample compared to the C11 model \citepalias{Brout2019a}. The residuals from a cosmological model (often termed \lq Hubble residuals\rq) are given by
\begin{equation}
\label{eqn:hubbleresiduals}
\Delta_\mu = \mu_\textrm{obs} - \mu_\textrm{theory}(z), 
\end{equation}
where $\mu_\textrm{theory}$ is the theoretical distance modulus, which is dependent on the cosmological parameters. 

A mass step has been detected in nearly all large SN Ia surveys at all redshifts \citep{Sullivan2010,Lampeitl2010}, with SNe Ia in galaxies with $\logmstellar>10$ brighter on average (after standardisation) than those in lower-mass galaxies.  Typical values for $\gamma$ using a 1D \mubias\ correction ($\gamma_{1\textrm{D}}$) include $\gamma_{1\textrm{D}}=0.070\pm0.023$\,mag \citep[$3.0\sigma$;][]{Betoule2014} for the sample of 740 JLA SNe~Ia and $\gamma_{1\textrm{D}}=0.070\pm0.013$\,mag \citep[$5.5\sigma$;][]{Roman2018} for the 882 SNLS5 SNe~Ia while \citep[][]{Scolnic2018} using a 5D \mubias\ correction ($\gamma_{5\textrm{D}}$) found $\gamma_{5\textrm{D}}=0.053\pm0.009$\,mag ($5.5\sigma$) for the 1048 SNe~Ia that comprise the Pantheon dataset and $\gamma_{5\textrm{D}}=0.039\pm0.016$\,mag ($2.4\sigma$) for the 365 SNe~Ia spectroscopically confirmed by PS1. Conversely, \citetalias{Brout2019a} found $\gamma_{5\textrm{D}}=0.009\pm0.018$\,mag ($0.5\sigma$) for the DES-SN sample when using a G10 scatter model and $\gamma_{5\textrm{D}}=0.004\pm0.017$\,mag ($0.2\sigma$) when using a C11 model for intrinsic scatter. 

\subsection{SN Ia Host Galaxy Data} 
\subsubsection{Host Galaxy Photometry}
\label{subsubsec:gal_phot}

Photometric data for the host galaxies of the DES3YR cosmology analysis \citep{Brout2019a,DES2019} were determined from the DES SVA1-GOLD catalogue. This catalogue, has 10$\sigma$ limiting magnitudes of $(g,r,i,z)=(24.0,23.8,23.0,22.3)$, as described in \citet{Rykoff2016} and \citet{Bonnett2016}. It was constructed from DES Science Verification (SV) data collected prior to the DES-SN data used in the DES3YR sample. In this paper, we upgrade from the DES SVA1-GOLD catalogue and instead determine photometric properties of the DES SNe~Ia host galaxies from DES deep stack photometry \citep[][hereafter \citetalias{Wiseman2020}]{Wiseman2020} utilising images from all 5 years of DES-SN. 

In summary, for each transient, the images used to create the deep stack photometry are selected from the five years of the DES-SN survey, excluding the season where the transient was first detected. Defining $\tau_\textrm{obs}$ as the ratio between the effective exposure time of an individual observation given the atmospheric conditions, and the true exposure time \citep{Morganson2018}, we select images with $\tau_\textrm{obs}>X$, with $0.2<X<0.5$, optimised for each field/band combination to produce final images with the greatest possible depth \citepalias{Wiseman2020}. We combine these images using \textsc{scamp} \citep{Bertin2006} and \textsc{swarp} \citep{Bertin2002}, and create catalogues using Source Extractor \citep[\textsc{SExtractor}]{Bertin1996,Bertin2011}. These coadded images have limiting magnitudes of $(griz)=(25.6,25.8,26.0,26.0)$ in the 8 shallow fields and $(griz)=(26.1,26.3,26.5,26.4)$ in the 2 deep fields. We use \textsc{SExtractor} $griz$ \lq\texttt{FLUX\_AUTO}\rq\ measurements, and correct for foreground extinction using the Milky Way (MW) dust maps of \citet{Schlegel1998}. 

The photometric catalogue of \citetalias{Wiseman2020} considers each DECam CCD individually when constructing deep stacked images. To ensure that host galaxies are not lost due to CCD gaps, which comprise 10\% of the DECam field-of-view \citep{Flaugher2015}, we supplement this catalogue with data from the DES SVA1-GOLD catalogue. Only 1 of our \NDES\ SNe Ia has host galaxy measurements determined from the SVA1-GOLD catalogue, which has consistent \lq\texttt{FLUX\_AUTO}\rq\ values with those of \citetalias{Wiseman2020} for galaxies common to both catalogues. 

The host galaxies of the DES SNe Ia were identified using the \lq Directional Light Radius\rq\ (DLR) methodology described in \citet{Sullivan2006,Smith2012,Gupta2016,Sako2018} and below in Appendix~\ref{app:dlr}. Following \citet{Gupta2016} and \citet{Sako2018}, we only consider galaxies with $d_\text{DLR}<7$ to be candidates for the true host, and also require that the potential host be classified as a galaxy based on the \texttt{CLASS\_STAR} \textsc{SExtractor} output \citep{Soumagnac2015}. SNe with no galaxy matching this criteria are denoted hostless. 201 of \NDES\ (98 per cent) of the DES-SN sample have an associated host galaxy. This fraction of hostless SN, two per cent, is less than that found for the Supernova Legacy Survey \citep[6 per cent;][]{Sullivan2006} and SDSS-SN \citep[4 per cent;][]{Sako2018} highlighting the depth of the deep-stacks relative to the redshift range probed by DES-SN. When using the shallower SVA1-GOLD catalogue, as used in \citetalias{Brout2019a}, 18 events are denoted hostless, while 5 events are associated to different galaxies, either due the detection of new sources located in close proximity to the SN or due to changes in the measured light-profile of the nearby hosts. AB magnitudes, corrected for MW extinction, for each identified host in DES-SN are given in \autoref{tab:hostmags_masses}. 

\subsubsection{Host galaxy physical parameters}
\label{subsubsec:mass_estimates}

To estimate the stellar mass (\mstellar) and star-formation rate (SFR) for each host galaxy in our sample, we use a methodology similar to that used in \citet{Sullivan2010} and \citet{Kim2018}. We use the P\'EGASE.2 spectral synthesis code \citep{Fioc1997,LeBorgne2002} to calculate the SED of a galaxy as a function of time, using 9 smooth, exponentially declining star formation histories (SFHs), with $\mathrm{SFR}(t)=\exp^{-t/\tau}/\tau$, where $t$ is the age of the galaxy and $\tau$\footnotemark\ is the e-folding time; each SFH is therefore normalised to produce 1\,\msolar. The SED of each SFH is calculated at 102 timesteps from 0 to 14\,Gyr, and we include the standard P\'EGASE.2 prescription for nebular emission. Each SFH has an initial metallicity ($Z$) of 0.004 that evolves consistently, with new stars formed with the metallicity of the ISM. We use a \citet[][hereafter \citetalias{Kroupa2001}]{Kroupa2001} initial mass function (IMF). (In \S\ref{subsubsec:mass_systematics} and~\ref{subsec:massstep_systematics} we investigate potential systematic uncertainties associated with this IMF choice.) At each timestep, P\'EGASE.2 provides the total mass in stars, and following \citet{Sullivan2006}, we calculate the average SFR over the previous $250$\,Myr of the SFH. For each SED we also use 7 foreground dust screens with a colour excess, $E(B-V)$, ranging from 0.0 to 0.30\,mag in steps of 0.05\,mag. This grid effectively creates 63 unique host-galaxy models, each with 102 timesteps (i.e., 6426 unique SEDs). We note that the rest-frame wavelength range probed by the DES filters, limits our ability to accurately constrain the dust content of galaxies, which can impact the estimates of \mstellar\ and SFR by 0.1dex \citep{Mitchell2013,Laigle2019}, although \citet{Palmese2019} show that this effect is negligible for early type galaxies. 
\footnotetext{Where $\tau=100, 200, 300, 400, 500, 750, 1000, 1500, 2000$\,Myr}

For each host galaxy, the fluxes of each model SED at the redshift of the SN in the DES $g,r,i,z$ filters are calculated (giving 6426 sets of model fluxes, $F_\mathrm{model}$), and for each $F_\mathrm{model}$ we minimise the $\chi^2$ as 
\begin{equation}
\chi^2=\sum_{x\in \it{griz}} \left(\frac{AF_\mathrm{model;x}-F_\mathrm{obs;x}}{\sigma_\mathrm{obs;x}}\right)^2
\end{equation}
where $A$ is a scale factor determined from a global $\chi^2$ minimization. To ensure consistency with our assumed cosmological model, we enforce that the age of the best-fit template must be less than the age of the Universe at the redshift of the SN. \mstellar\ and SFR are calculated from $A$ and the best-fit SED. From these, we calculate the specific SFR (sSFR) as $\mathrm{sSFR}=\mathrm{SFR}/\mstellar$.

We use a Monte Carlo approach to estimate the statistical uncertainties in our derived parameters. For each galaxy, we perform 1000 random realisations of $F_\mathrm{obs}$, drawing a new $F^\prime_\mathrm{obs}$ randomly from a Normal distribution with a mean $F_\mathrm{obs}$ and $\sigma=\sigma_\mathrm{obs}$, and repeating the minimisation procedure described above. The quoted uncertainties on the best-fit parameters are the standard deviation of the best-fit parameters over all realisations. Derived values for \mstellar\ and sSFR for each identified host in DES-SN is given in \autoref{tab:hostmags_masses}. 

For comparison, the DES3YR analysis in \citetalias{Brout2019a} used a P\'EGASE.2 template library comprised of 9 spectral types, described in \citet{Smith2012}, evaluated at 200 age steps and a \citetalias{Kroupa2001} IMF. The best-fit SED, stellar mass and star-formation rate were determined with the code ZPEG \citep{LeBorgne2002} using $\chi^2$-minimization. In \S\ref{subsubsec:massstep_sys_masses} we investigate how the mass estimates for this study compare to those determined in our fiducial analysis. Further, while the DES-SN estimates of \mstellar\ and sSFR are based only on 4 band photometry, with no information on the rest-frame infrared contribution, \citet{Palmese2016}, for cluster galaxies with a known redshift, found no evidence of an offset in $\logmstellar$ estimated from 5 band DES-SV photometry compared to that estimated from 17 band photometry. This suggests that while the inclusion of near infrared data would improve constraints on the underlying galaxy SED, our best fit models are likely unlikely unbiased. 

\begin{figure}
\centering
\includegraphics[width=0.45\textwidth]{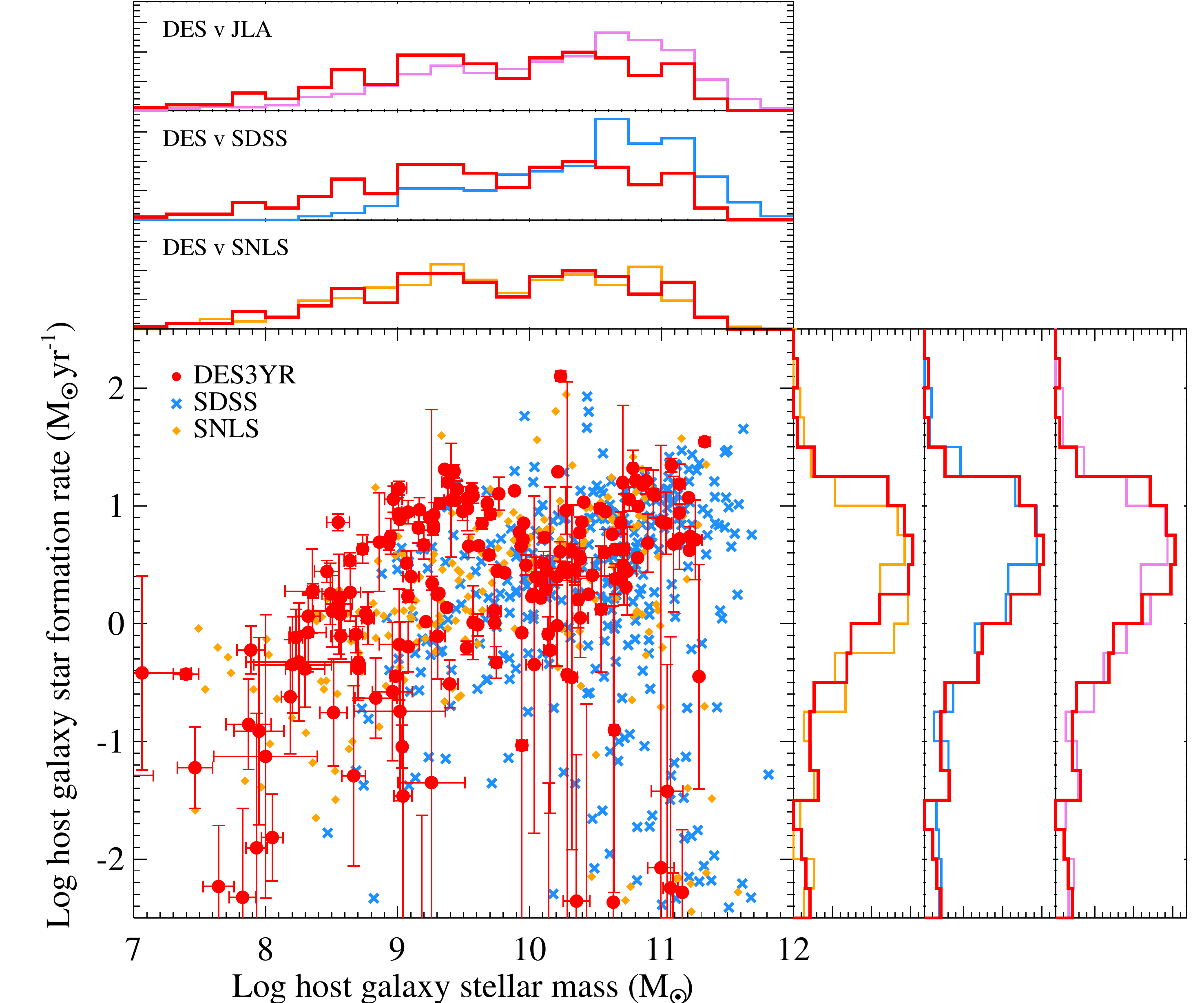}\\
\vspace{1cm}
\includegraphics[width=0.45\textwidth]{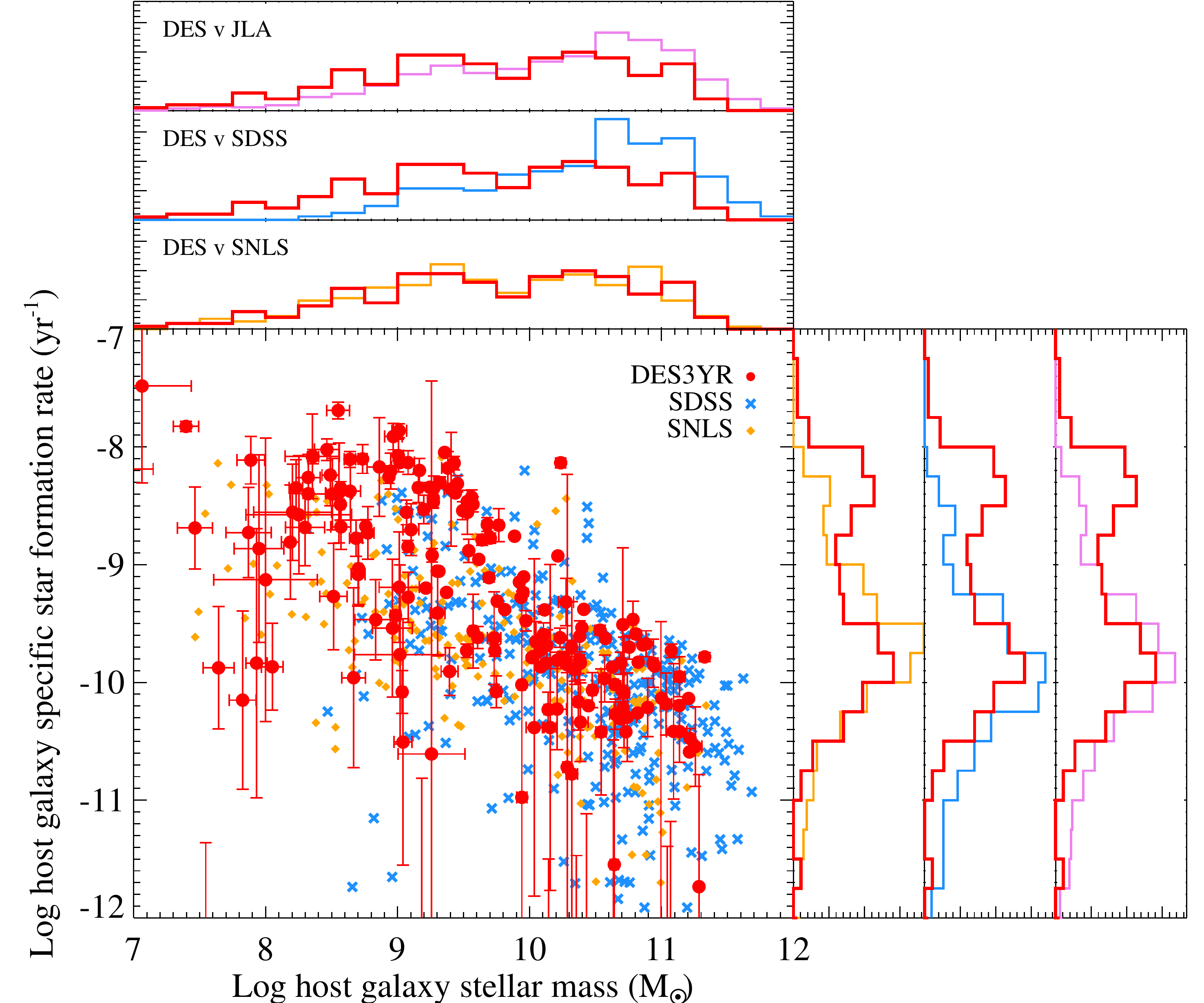}
\caption{\textit{Top}: The relationship between \mstellar\ and SFR for the DES-SN sample (red circles). Overplotted are the values for the SDSS (blue crosses) and SNLS (orange diamonds) samples, combined as the JLA sample (violet) and analysed in a consistent manner. \textit{Bottom}: As left, showing the relationship between \mstellar\ and sSFR. The parameter distributions are normalised to contain an equal area. 
\label{fig:mass_v_sfr}}
\end{figure}

For our DES host galaxies, the relationships between \mstellar\ and SFR, and \mstellar\ and sSFR, are shown in \autoref{fig:mass_v_sfr}, together  with the distributions of \mstellar, SFR and sSFR. For comparison, we also show the values for SN Ia hosts discovered by the SDSS \citep{Sako2018} and SNLS \citep{Conley2011} surveys; for consistency, we have re-fitted the host galaxy data from \citet{Sullivan2010} and \citet{Sako2018} using the above framework. As anticipated, there is a strong correlation between host galaxy \mstellar, SFR and sSFR (defined in part by our underlying SFHs). The most massive galaxies typically have a lower sSFR, while lower mass galaxies consistently have a higher sSFR. The correlation between \mstellar\, SFR and sSFR for the DES hosts are consistent with those found for the SDSS and SNLS samples.

The \mstellar\ distribution for the DES sample is consistent with the SNLS sample, which also probes a wide redshift range. The SDSS sample tends to have more massive host galaxies. The SDSS sample probes lower redshifts (with a mean of 0.20) compared DES-SN (a with mean redshift of 0.39) and SNLS (0.64). The increased contribution from high mass galaxies for the SDSS sample may be a consequence of this, as galaxies at lower redshifts tend to be more massive, or a selection effect reflecting the fact that SNe Ia in bright host galaxies are harder to spectroscopically confirm at higher redshift. The SFR distributions for the DES sample are consistent with the SDSS and SNLS samples, while for sSFR, there is an over-density of high sSFR ($\log\textrm{sSFR}>-9$) hosts in the DES sample compared to the SDSS and SNLS samples. The hosts of these events are preferentially low mass, with mean $\logmstellar=8.86\pm0.09$, and moderately star-forming, with mean  $\log{\textrm{SFR}/\msolar{\textrm{yr}^{-1}}}=0.46\pm0.08$. 

\subsubsection{Systematic uncertainties of the stellar mass estimates}
\label{subsubsec:mass_systematics}

\begin{figure*}
\centering
\includegraphics[width=0.32\textwidth]{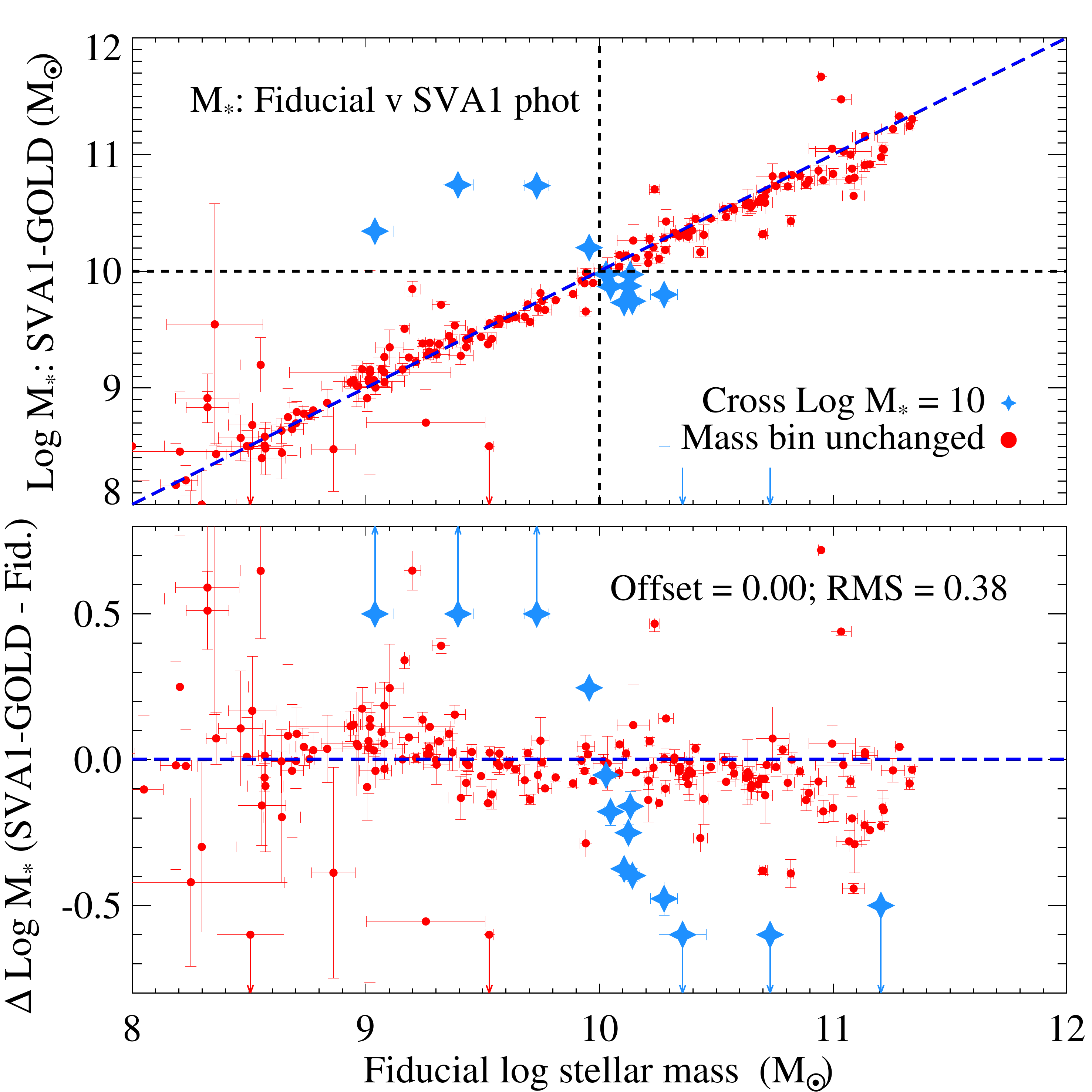}
\includegraphics[width=0.32\textwidth]{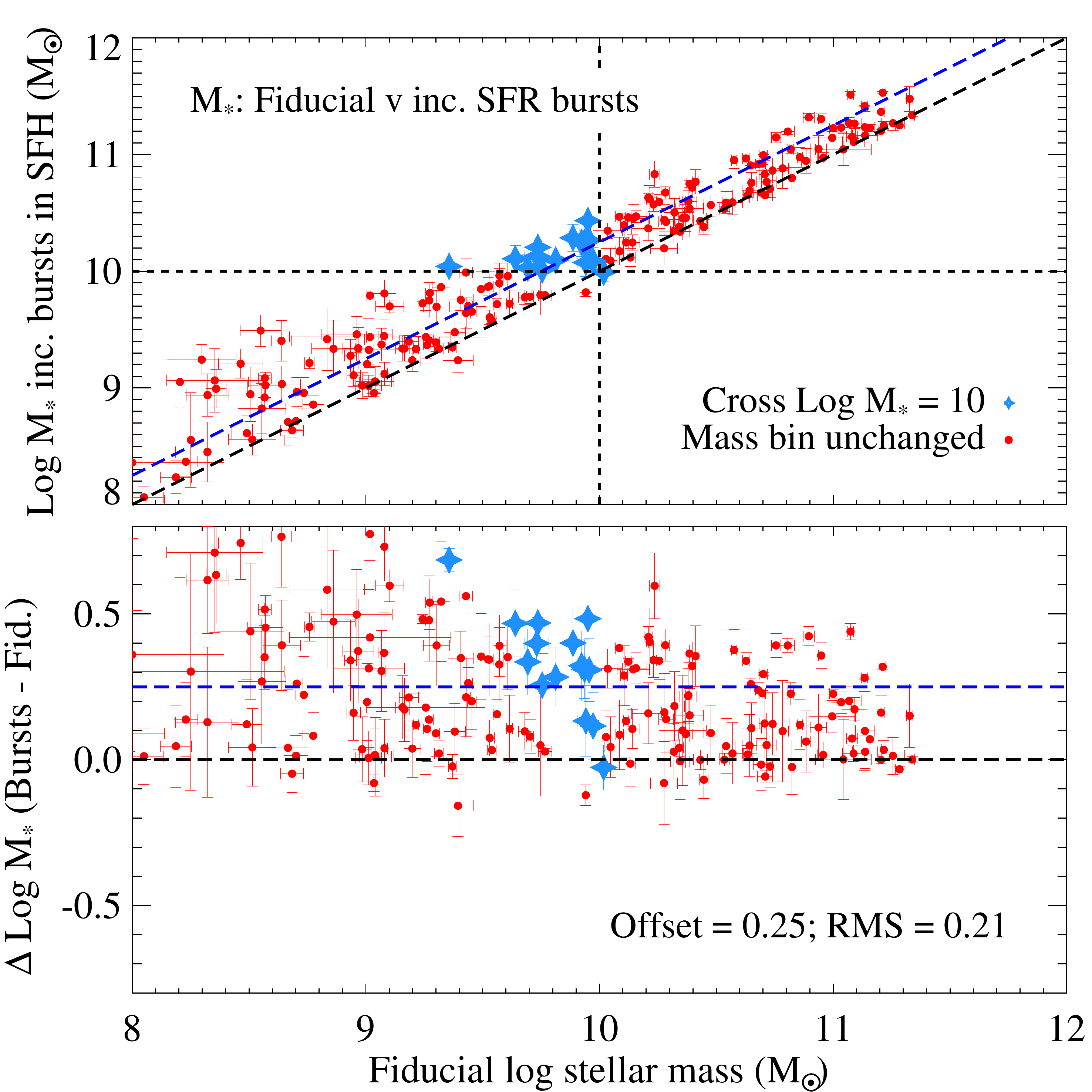}
\includegraphics[width=0.32\textwidth]{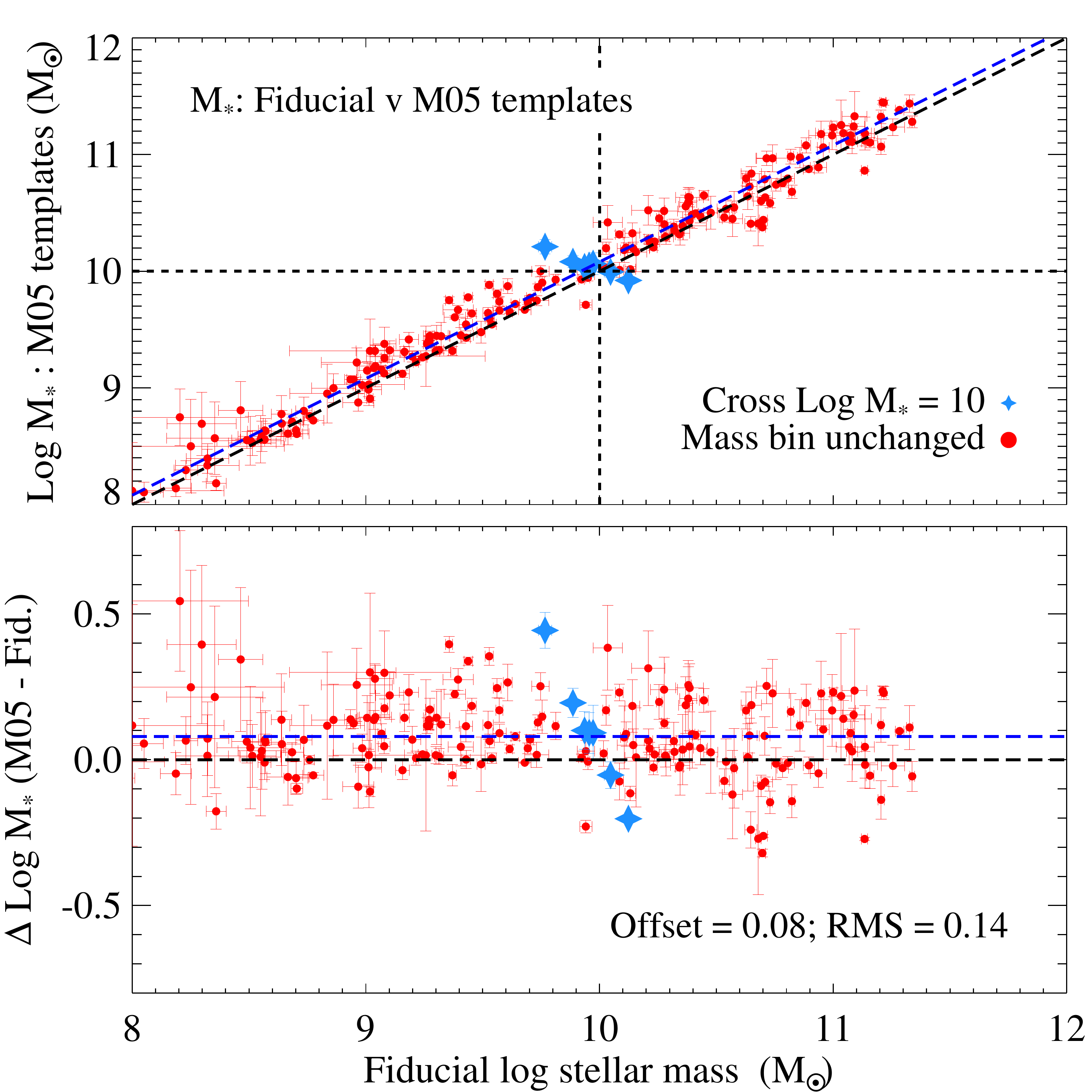}
\caption{Testing the robustness of \mstellar\ estimates. Left panel: Fiducial \mstellar\ estimates compared to those estimated using $griz$ galaxy magnitudes taken from the DES SVA1-GOLD catalogue \citep{Rykoff2016}. The lower panel shows the difference in \mstellar\ as a function of stellar mass. No linear trend as a function of stellar mass is observed. Centre panel: As left panel, but considering the effect of extra bursts of star-formation in the template SEDs used to determine derived galaxy parameters. Including additional bursts of star-formation increases the inferred $\logmstellar$ by $0.25\pm0.02$. Right panel: As left panel, but showing the inferred stellar masses when alternative templates \citep{Maraston2005} are used in the fit. These templates decrease the inferred $\logmstellar$ by $0.11\pm0.01$, but no trend is observed. In all panels, DES-SN objects are plotted in red, with galaxies that have inferred $\logmstellar>10$ in one axis but $\logmstellar<10$ in another (i.e., those that would cross the $M_\mathrm{step}$ in \autoref{eqn:gamma_def}) plotted as blue diamonds. The mean offset between the two values is highlighted by a blue dashed line. 
\label{fig:mass_systematics}}
\end{figure*}

\begin{table*}
{\centering
\caption{Comparison between \mstellar\ derived for the host galaxies of the \NDES\ spectroscopically confirmed SNe~Ia that comprise the DES-SN sample and those derived with different assumptions.}
\label{tab:mass_systematics}
\begin{tabular}{lllcccc}
  \hline
  Row \# & Photometric Catalogue & Templates$^{1}$ & IMF$^{1}$ & $\left< \Delta \logmstellar \right>$$^{2}$ (r.m.s.) & \multicolumn{2}{c}{\# hosts moving class} \\
  & & & & & high mass$^{3}$ & low mass$^{4}$\\
  \hline
  1; Fiducial result & \citetalias{Wiseman2020} & \PEGASE & \citetalias{Kroupa2001} & - & - & - \\
  2; \citetalias{Brout2019a} & SVA1-GOLD: \texttt{mag\_detmodel} & ZPEG & \citetalias{Kroupa2001} & $0.12\pm0.02$ (0.38) & 8 ($4.3\%$) & 11 ($5.9\%$) \\
  3 & \citetalias{Wiseman2020} & \PEGASEbursts & \citetalias{Kroupa2001} & $0.25\pm0.02$ (0.21) & 14 ($7.0\%$) & 1 ($0.5\%$) \\
  4 & \citetalias{Wiseman2020} & \PEGASE & \citetalias{Salpeter1955} & $0.17\pm0.01$ (0.09) & 11 ($5.5\%$) & 0 ($0.0\%$) \\
  5 & \citetalias{Wiseman2020} & \PEGASEbursts & \citetalias{Salpeter1955} & $0.43\pm0.02$ (0.24) & 25 ($12.4\%$) & 0 ($0.0\%$) \\
  6 & \citetalias{Wiseman2020} & \citetalias{Maraston2005} & \citetalias{Kroupa2001} & $-0.11\pm0.01$ (0.15) & 0 ($0.0\%$) & 8 ($4.0\%$)\\
  7 & \citetalias{Wiseman2020} & \citetalias{Maraston2005} & \citetalias{Salpeter1955} & $0.08\pm0.01$ (0.14) & 5 ($2.5\%$) & 2 ($1.0\%$) \\
  8 & \citetalias{Wiseman2020} & \citetalias{Bruzual2003} & \citetalias{Salpeter1955} & $0.18\pm0.01$ (0.09) & 10 ($5.0\%$) & 0 ($0.0\%$) \\
  9 & \citetalias{Wiseman2020} & ZPEG & \citetalias{Kroupa2001} & $0.08\pm0.02$ (0.20) & 8 ($4.0\%$) & 3 ($1.5\%$) \\
  10 & SVA1-GOLD: \texttt{mag\_auto} & \PEGASE & \citetalias{Kroupa2001} & $0.00\pm0.02$ (0.38) & 10 ($5.3\%$) & 4 ($2.1\%$)\\
  11 & SVA1-GOLD: \texttt{mag\_detmodel} & \PEGASE & \citetalias{Kroupa2001} &$0.03\pm0.02$ (0.37) & 9 ($4.8\%$) & 4 ($2.1\%$)\\
\hline
\end{tabular}\\
}
\raggedright
\footnote{a}{Galaxy templates, assumptions of the SFH and Initial Mass Function used. See \S\ref{subsubsec:mass_systematics} for details.}\\
\footnote{b}{$\left< \Delta ({\log{{M}_\textrm{stellar}}} - {\log{{M}_\textrm{stellar;fid}}}) \right>$, where ${\log{{M}_\textrm{stellar;fid}}}$ is derived from the PEGASE templates with a \citetalias{Kroupa2001} IMF.}\\
\footnote{c}{Number of hosts with ${\log{{M}_\textrm{stellar;estimate}/\textrm{M}_{\sun}}}>10$ and ${\log{{M}_\textrm{stellar;fid}/\textrm{M}_{\sun}}}<10$.}\\
\footnote{d}{Number of hosts with ${\log{{M}_\textrm{stellar;estimate}/\textrm{M}_{\sun}}}<10$ and ${\log{{M}_\textrm{stellar;fid}/\textrm{M}_{\sun}}}>10$.}\\
\end{table*}

Our \mstellar\ estimates depend on the photometric catalogue considered and assumptions on the SFH, IMF and SED templates used to describe the galaxy population, all of which are of debate in the literature. We here test the sensitivity of our \mstellar\ estimates to these assumptions. The results are shown in \autoref{fig:mass_systematics} and \autoref{tab:mass_systematics}.

The left-hand panel of \autoref{fig:mass_systematics} and row 10 of \autoref{tab:mass_systematics}, show the correlation between our fiducial \mstellar, derived using photometry determined from deep stacks, compared to those obtained from the SVA1-GOLD catalogue as described in \S\ref{subsubsec:gal_phot}. There is no evidence of a systematic offset between the two measurements, and the best-fit linear fit has a slope of $0.98\pm0.03$. There is a  mean difference in $\log\mstellar$ of $0.002\pm0.016$\,dex between the two measurements, and an r.m.s. scatter of $0.38$\,dex. An increased scatter is observed for galaxies with $\logmstellar<9.5$ due to the increased scatter in the fluxes for the faintest objects in our sample, but no systematic trend as a function of stellar mass is observed. The blue crosses in \autoref{fig:mass_systematics} correspond to galaxies that cross the threshold of $\logmstellar=10$ between the two analysis; i.e., those that have $\logmstellar>10$ in one mass estimate, but have $\logmstellar<10$ in the other. These objects have implications for the inferred mass step (see \S\ref{sec:massstep} for details), where $\logmstellar=10$ is used to differentiate between two classes of SNe Ia with differing absolute magnitudes. 4 of 188 SN hosts (two per cent) are classified as high mass when considering the SVA1-GOLD catalogue, but are considered low-mass hosts in our fiducial analysis using deep coadds. Ten objects (five per cent) satisfy the reverse criteria. 

Our fiducial analysis uses \lq\texttt{FLUX\_AUTO}\rq\ measurements derived from deep stack images. These flux estimates are determined from model fits where each passband is treated independently. An alternative approach is to use a fixed apertures across all filters. These, \lq\texttt{FLUX\_DETMODEL}\rq\ measurements will better represent the colour of each galaxy, but as a consequence, can underestimate the total flux. Row 11 of \autoref{tab:mass_systematics} shows the consequence of using \lq\texttt{FLUX\_DETMODEL}\rq\ measurements instead of \lq\texttt{FLUX\_AUTO}\rq\ from the SVA1-GOLD catalogue. Consistent with the estimates using \lq\texttt{FLUX\_AUTO}\rq\ measurements, no residual offset with stellar mass is observed. 

The central panel of \autoref{fig:mass_systematics} and row 3 of \autoref{tab:mass_systematics} show the correlation between our fiducial \mstellar\ estimates and those derived when using SFHs that contain bursts of star formation. In this analysis, we use the same 9 exponentially declining SFHs, but superimpose a burst of star-formation on each underlying SFH. These bursts occur randomly between 1 and 10\,Gyr into the smooth, exponentially declining, SFH, and can form between 0.05 and 25 per cent of the total stellar mass in the SFH. Each burst also has an exponentially declining SFH, with $\tau=10$, 50 or 100\,Myr \citep[selected with equal probability;][]{Childress2013b}. We generate 4000 such SFHs, with an increased time resolution around the time of the bursts, calculate a new set of $F_\textrm{model}$ with the same foreground dust screens as before, and repeat the $\chi^2$ minimisation, retaining the original 9 SFHs for consideration. With differing age profiles, these burst models break the degeneracy between age and metallicity in the SFHs. 

From \autoref{fig:mass_systematics}, the inclusion of additional bursts of star-formation typically increases the inferred \mstellar\ estimate, with a mean offset of $0.25\pm0.02$\,dex and an r.m.s.$=0.21$\,dex. 189 (94 per cent) of the host galaxies in our sample \lq prefer\rq\ (i.e., have a smaller minimum $\chi^2$ for) SFHs with a recent burst of star-formation in the last 10\,Gyr. We find strong evidence (at 4.4$\sigma$) that our fiducial stellar mass estimates are not one-to-one correlated with those determined when recent bursts of star-formation are allowed in the galaxy SED. The increase in stellar mass for lower mass galaxies is proportionally higher than that observed in high mass systems. 14 of 201 (seven per cent) of the SN Ia hosts move from the low-mass to high-mass class when recent bursts of star-formation are allowed, with one galaxy (one per cent) moving into the low mass class. To further test the effect of our choice of SED modelling parameters, in \autoref{tab:mass_systematics}, row 4, we show how assuming a \citetalias{Kroupa2001} IMF affects the estimated values of \mstellar. Repeating our fiducial analysis (with no additional bursts of star formation) with a Salpeter IMF \citep[][hereafter \citetalias{Salpeter1955}]{Salpeter1955} results in a systematic offset of $0.17\pm0.01$\,dex (with the masses derived from a \citetalias{Salpeter1955} IMF being more massive), and r.m.s. of 0.09\,dex. There are 11 additional high-mass hosts (six per cent) when a \citetalias{Salpeter1955} IMF is used, while no hosts move from the high-mass to low-mass class.
 
Our final test of the robustness of our \mstellar\ estimates concerns the population model considered. The \citet{Maraston2005} population synthesis models include contributions from the thermally pulsing asymptotic giant branch (TP-AGB) phase of stellar evolution. We use 19 exponentially declining SFHs based on these models, each evaluated at 61 time steps. Generating SFHs using a \citetalias{Kroupa2001} IMF, the right-hand panel of \autoref{fig:mass_systematics} and row 6 of \autoref{tab:mass_systematics}, shows the correlation between \mstellar\ derived by our fiducial technique compared to those derived using the templates of  \citet{Maraston2005}. A strong correlation is observed between the two mass estimates, with a systematic offset of $0.11\pm0.01$\,dex (with our fiducial \mstellar\ values being more massive) and an r.m.s. of 0.15\,dex. No evidence of a residual correlation between the two mass estimates and our fiducial stellar masses is observed, with a best fitting linear relationship having a slope of 0.99$\pm$0.01. There are 8 additional low mass hosts (4 per cent) when using the \mstellar\ estimates from \citetalias{Maraston2005}, with no objects moving into the high mass bin. \autoref{tab:mass_systematics}, row 7, also shows the effect of using a \citetalias{Salpeter1955} IMF in this analysis, with a mean offset of $0.08\pm0.01$\,dex (with the \citetalias{Salpeter1955} IMF masses being more massive) and an r.m.s. of 0.14. In this case, only 7 galaxies move across the $\logmstellar=10$ division: 5 (3 per cent) listed as high mass when a Salpeter IMF is considered compared to 2 (1 per cent) which are better fit as being low mass. 

To further test the effect of our choice of template SFH, in \autoref{tab:mass_systematics}, row 8, we show the results when using the \citet{Bruzual2003} single stellar populations (SSPs) with a Salpeter IMF. A mean offset of $0.18\pm0.01$\,dex, with an r.m.s. of 0.20\,dex is seen with the \mstellar\ values being more massive for the Bruzual \& Charlot models. As a result, 10 galaxies (5 per cent) move into the high mass class, while no extra events are identified as low mass hosts. This result is consistent with the result when using the \PEGASE\ templates with a \citetalias{Salpeter1955} IMF (\autoref{tab:mass_systematics}, row 7), suggesting that this difference is driven solely by the choice of IMF. 

These tests show that of our estimates of \mstellar\ are robust to the choice of photometric catalogue and the SED models used in our fiducial analysis. Considering all systematic tests a mean of 13.3 (6.8 per cent) galaxies move across the $\logmstellar=10$ boundary, with a maximum of 25 (12.4 per cent). 

\section{SN Ia properties as a function of Host Galaxy Properties} 
\label{sec:lcparams}

Here we examine the demographics of the SN Ia host galaxies, and correlations between the SN Ia host galaxy properties and the SNe Ia they host. Of particular importance is identifying and understanding differences between the host galaxies of the DES SN Ia sample and other SN Ia samples at a similar redshifts, as these differences can result in discrepancies between measured mass steps. 

\begin{figure}
\centering
\includegraphics[width=0.48\textwidth]{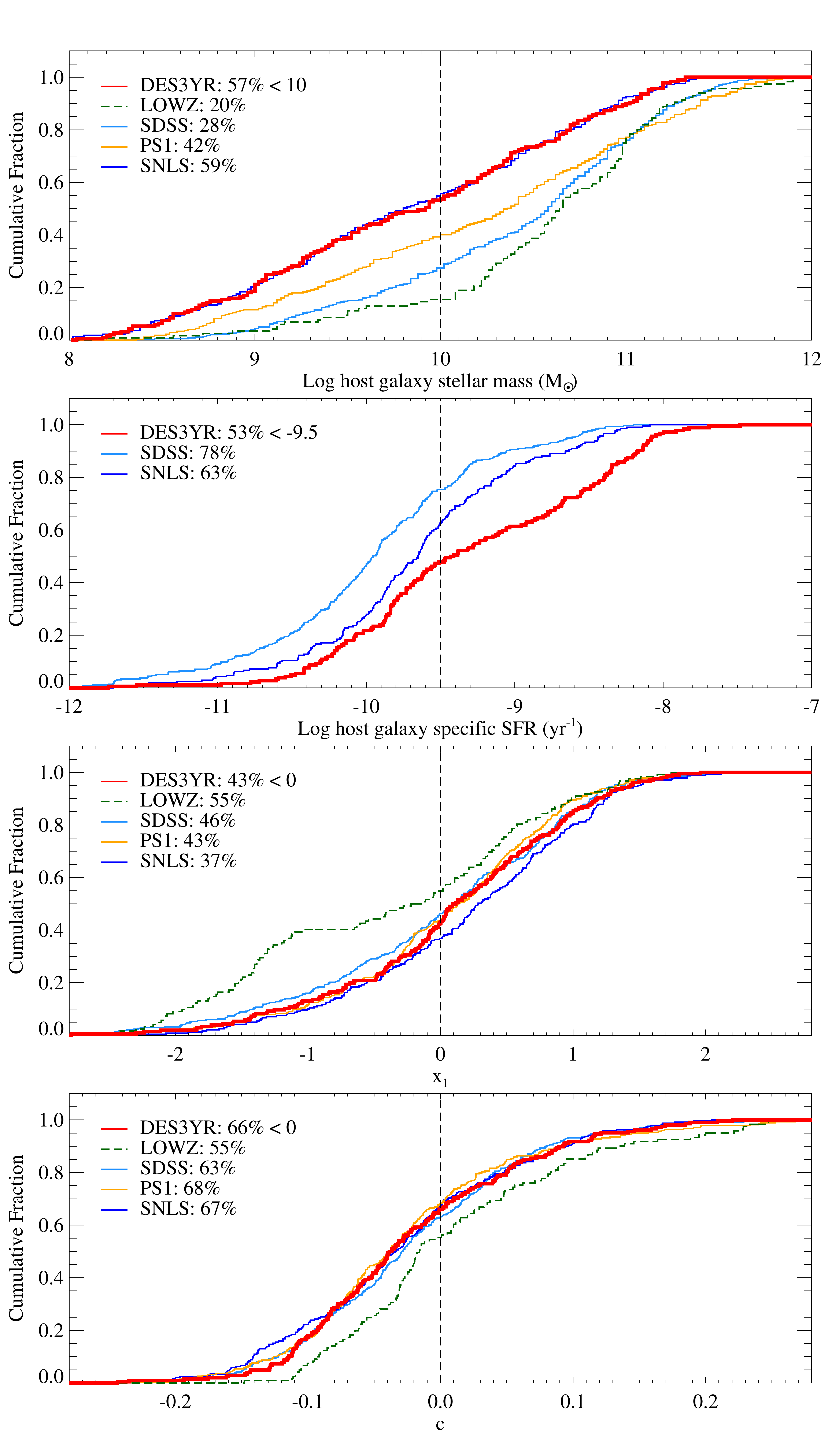}
\caption{Cumulative distributions of \mstellar, sSFR, $x_1$ and $c$ for the DES-SN sample (shown in red) compared to literature datasets (SDSS, light-blue; SNLS, dark blue; PS1, yellow; \lowz, green). The fraction of SNe Ia with $\logmstellar<10$, sSFR$<-9.5$, $x_1<0$ and $c<0$ is also shown. 
\label{fig:sample_dists}}
\end{figure}

\autoref{fig:sample_dists} shows the cumulative distribution of \mstellar, sSFR, $x_1$ and $c$ for the DES-SN sample compared to literature datasets, with the mean sample properties given in \autoref{tab:sample_info}.  The distribution of \mstellar\ for the DES-SN sample is consistent with that of the SNLS sample, with a Kolmogorov-Smirnov (KS) probability 0.78. These two high-redshift samples are both untargeted searches probing a wide redshift range, able to locate SNe Ia in all but the most extreme host galaxy environments. By contrast, the DES \mstellar\ distribution is different to that found for the low-redshift sample (with KS test probability $1.2\times{10}^{-8}$), where the SNe Ia are predominately found in high-mass ($\logmstellar>10$) host galaxies. This is expected, and is due in part to selection effects in low-redshift galaxy-targeted transient surveys, and in part to evolution in the galaxy population \citep[see discussion in][]{Pan2014}.

At intermediate redshift, the distribution of \mstellar\ for the PS1 and SDSS samples are consistent, with KS probability 0.11. We find a KS probability of 0.037 (0.0001) between the DES and PS1 (SDSS) samples, with 57$\%$ of SNe Ia found in low mass ($\mstellar<10$) hosts for the DES-SN sample, compared to 42$\%$ and 28$\%$ for PS1 and SDSS, respectively. This is likely a selection effect of the DES-SN sample. SNe in faint (and thus lower mass) hosts are preferentially targeted for real-time spectroscopic follow-up in DES \citep{DAndrea2018} as these host galaxies are more challenging to measure redshifts for once the SN light has faded, potentially biasing the DES-SN sample to lower-mass hosts compared to those determined by other surveys. 

sSFR measurements are available for the DES-SN, SDSS and SNLS samples \citep{Kim2018}. Galaxies with lower sSFR have smaller amounts of star-formation relative to their stellar mass, and are thus dominated by an older stellar population. As shown in \autoref{fig:mass_v_sfr}, there is an excess of high-sSFR ($\textrm{sSFR}>-9.5$) hosts in the DES-SN sample compared to the SDSS and SNLS samples, with KS probabilities of 0.00002 (0.008) between the DES and SDSS (SNLS) samples, indicating that the DES-SN sample is dominated by a younger stellar population. This again can be attributed to the spectroscopic targeting algorithm utilised by DES-SN \citep{DAndrea2018}, which focused on SNe in faint, low mass hosts. These, younger stellar environments, typically exhibit higher star-formation rates potentially biasing the DES-SN sample to galaxies with higher sSFR compared to literature samples.

The SN Ia properties ($x_1$, $c$) of the cosmological samples (DES-SN, SNLS, SDSS, PS1) are consistent, indicating little evolution in the population parameters, and little evidence of SN specific selection techniques. The only inconsistency is with the low-redshift sample, which is over-represented with redder ($c>0.1$), faster-declining ($x_1<0.0$) SNe Ia. These differences have been seen previously \citep{Scolnic2016,Scolnic2018}, but again are expected as the low-redshift sample is primarily SNe Ia obtained from targeted surveys, and hence in high-mass galaxies. These galaxies preferentially host fainter (lower $x_1$), redder SNe Ia \citep{Sullivan2010,Smith2012}. 

\begin{table*}
{\centering
\caption{The mean properties of samples used in this analysis}
\label{tab:sample_info}
\begin{tabular}{ccccccc}
  \hline
  Survey & $N_\textrm{SN}$ & $\overline{z}$ & $\overline{x_1}$ & $\overline{c}$ & $\overline{\logmstellar}$ & \% Low mass hosts$^{2}$\\
  \hline
DES-SN$^{1}$ & $\NDES$ & $0.364$ & $0.115$ & $-0.0367$ & $9.70$ & $57.3$ \\
SDSS \citep{Betoule2014} & $374$ & $0.198$ & $0.152$ & $-0.0307$ & $10.23$ & $40.9$ \\
SDSS \citep{Scolnic2018} & $335$ & $0.202$ & $0.170$ & $-0.0277$ & $10.40$ & $37.6$ \\
SNLS \citep{Betoule2014} & $239$ & $0.640$ & $0.285$ & $-0.0339$ & $9.64$ & $59.0$ \\
SNLS \citep{Scolnic2018} & $236$ & $0.642$ & $0.306$ & $-0.0318$ & $9.64$ & $59.3$ \\
PS1 & $279$ & $0.292$ & $0.138$ & $-0.0377$ & $10.32$ & $41.6$ \\
\lowz\ & $124$ & $0.0288$ & $-0.132$ & $-0.0172$ & $10.64$ & $19.4$ \\
\hline
\end{tabular}\\
}
\raggedright
\footnote{a}{Passing selection criteria in \citetalias{Brout2019a}.}\\
\footnote{b}{Percentage of hosts with $\logmstellar < 10$.}\\
\end{table*}

\subsection{Correlating SN and host galaxy properties}
\label{subsec:x1c_v_mass}

\begin{figure*}
\centering
\includegraphics[width=0.45\textwidth]{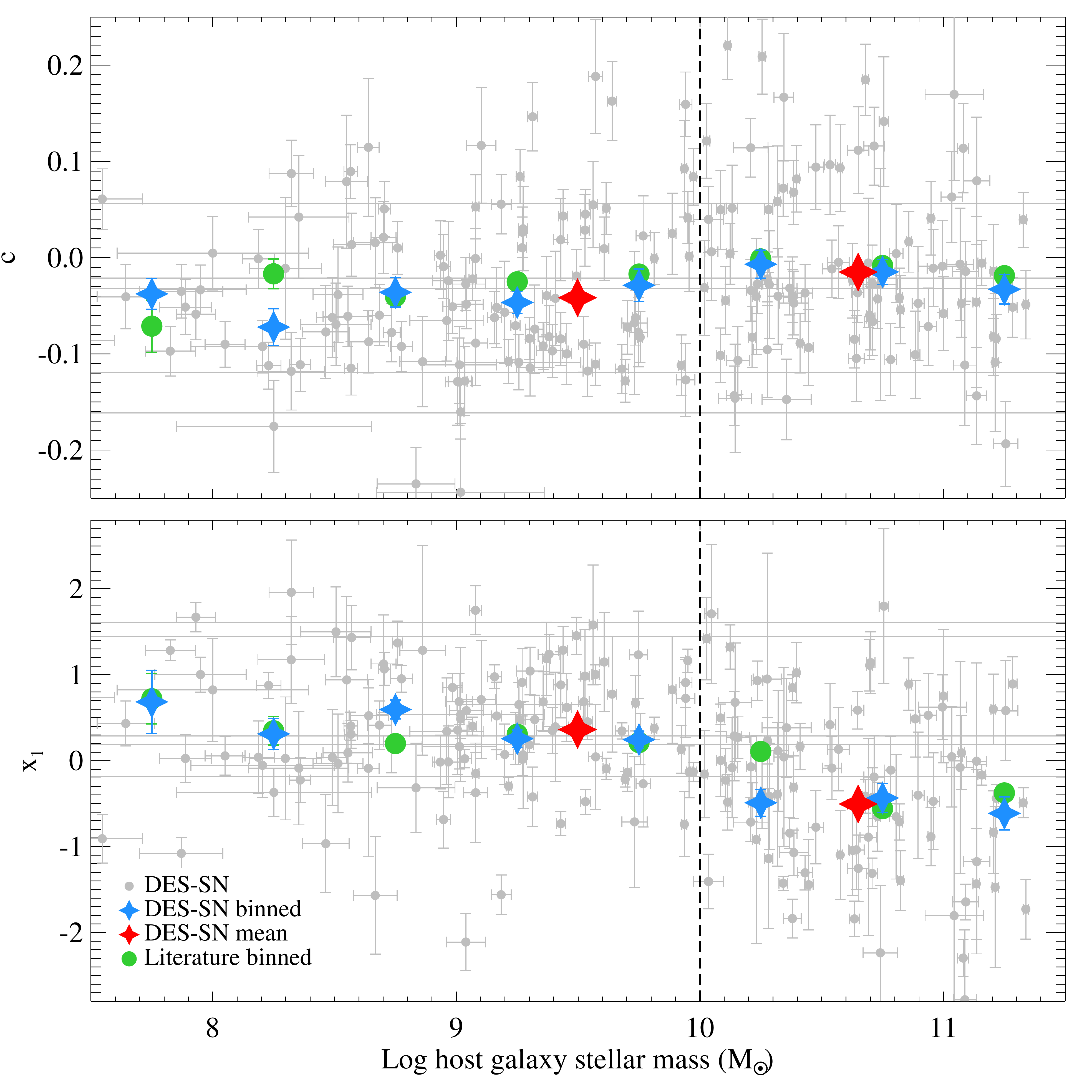}
\includegraphics[width=0.45\textwidth]{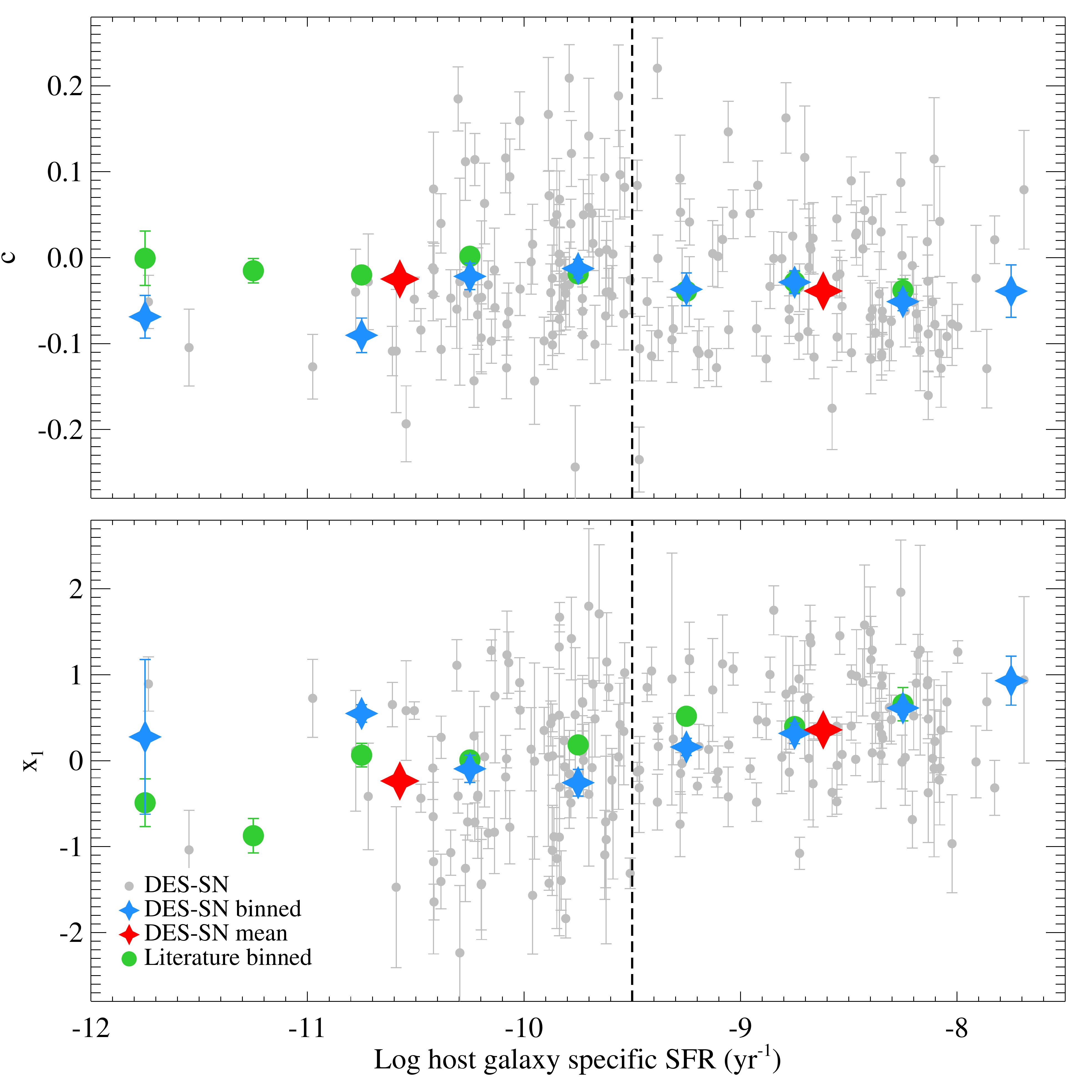}
\caption{Left: The relationship between \mstellar\ and $c$ (top panel) and $x_1$ (lower panel) for the DES-SN sample. Data points are shown in grey, with the mean value in bins of stellar mass are shown as blue diamonds. The overall mean values for high and low mass galaxies are shown as red diamonds. Means for the literature sample are plotted as closed green circles. Right: Same as left panels, only as a function of host galaxy specific star-formation rate. 
\label{fig:galprop_v_snprop}}
\end{figure*}

Correlations between the light-curve shape ($x_1$) and host galaxy properties have been observed in many previous studies \citep[e.g.][]{Sullivan2010,Lampeitl2010,Gupta2011,Childress2013,Wolf2016}: low mass, high star-forming, low metallicity, young stellar populations preferentially host broader (high $x_1$), brighter SNe Ia.  

The DES-SN dataset recovers these trends (\autoref{fig:galprop_v_snprop}). At 2.5$\sigma$ significance, we find evidence that higher stellar-mass ($\logmstellar>10$) galaxies host redder SNe Ia than those found in in lower-mass galaxies, with a mean difference of $\Delta c = 0.027\pm0.011$. This is consistent with a difference of $0.022\pm0.005$ measured by \citetalias{Brout2019a} and $0.012\pm0.004$ found by \citet{Scolnic2018}. For the DES-SN sample, there is no evidence of a difference in dispersion in $c$ as a function of \mstellar. The SNe~Ia colour distribution in high mass galaxies has an r.m.s. of 0.086 compared to 0.081 for those in low mass hosts. 

As expected, there is a strong correlation between light-curve width ($x_1$) and galaxy properties, with high $x_1$ SNe Ia preferentially found in low \mstellar\ ($\logmstellar<10$), high sSFR (sSFR$>-9.5$) galaxies: the mean $x_1$ differs between high and low \mstellar\ galaxies at 7.6$\sigma$, and at $5.3\sigma$ between low and high sSFR galaxies. The $x_1$ distribution is also narrower for SNe Ia found in low stellar mass galaxies compared to high stellar mass galaxies, with an r.m.s. of 0.73 compared to 0.95; consistent results are found as a function of sSFR. 

\section{The mass step in DES3YR}
\label{sec:massstep}

\begin{table*}
{\centering
\caption{Best-fit $\gamma$ determined from various samples as a function of the parameters varied. For a 5D \mubias\ correction, all sub-samples recover a positive $\gamma$ at a consistent value, with the exception of \citetalias{Brout2019a}, as discussed in \S\ref{subsec:brout_comparison}. For a 1D \mubias\ correction, a higher value of $\gamma$ is found, as discussed in \S\ref{subsec:massstep_systematic_mubias}}
\label{tab:gamma_fitvals}
\begin{tabular}{llllccl}
 \hline
Sample & Biascor & Fixed & Fitted & Best-fit $\gamma$ & Significance & Reference \\
& & parameters & parameters & (mag) & &  \\
\hline
DES-SN & 5D & $\alpha$, $\beta$, $\mstep$ & $\gamma$ & 0.030 $\pm$ 0.017 & 1.8$\sigma$ & This work \\
DES-SN & 5D & $\mstep$ & $\alpha$, $\beta$, $\gamma$ & 0.040 $\pm$ 0.019 & 2.1$\sigma$ & This work \\
DES-SN \citepalias{Brout2019a} & 5D & $\mstep$ & $\alpha$, $\beta$, $\gamma$ & 0.009 $\pm$ 0.018 & 0.5$\sigma$ & \citet{Brout2019a} \\
DES3YR & 5D & $\mstep$ & $\alpha$, $\beta$, $\gamma$ & 0.043 $\pm$ 0.018 & 2.4$\sigma$ & This work \\
low-$z$ & 5D & $\mstep$ & $\alpha$, $\beta$, $\gamma$ & 0.068 $\pm$ 0.038 & 1.8$\sigma$ & This work \\
Pantheon & 5D & $\mstep$ & $\alpha$, $\beta$, $\gamma$ & 0.053 $\pm$ 0.009 & 5.5$\sigma$ & \citet{Scolnic2018} \\
PS1 & 5D & --- & $\alpha$, $\beta$, $\gamma$, $\mstep$ & 0.039 $\pm$ 0.016 & 2.4$\sigma$ & \citet{Scolnic2018} \\
\hline
DES-SN & 1D & $\mstep$ & $\alpha$, $\beta$, $\gamma$ & 0.066 $\pm$ 0.020 & 3.3$\sigma$ & This work \\
DES3YR & 1D & $\mstep$ & $\alpha$, $\beta$, $\gamma$ & 0.064 $\pm$ 0.019 & 3.4$\sigma$ & This work \\
SNLS5YR & 1D & $\mstep$ & $\alpha$, $\beta$, $\gamma$ & 0.070 $\pm$ 0.013 & 5.5$\sigma$ & \citet{Roman2018} \\
JLA & 1D & $\mstep$ & $\alpha$, $\beta$, $\gamma$ & 0.070 $\pm$ 0.023 & 3.0$\sigma$ & \citet{Betoule2014} \\
Pantheon & 1D & --- & $\alpha$, $\beta$, $\gamma$, $\mstep$ & 0.072 $\pm$ 0.010 & 7.2$\sigma$ & \citet{Scolnic2018} \\
PS1 & 1D & --- & $\alpha$, $\beta$, $\gamma$, $\mstep$ &  0.064 $\pm$ 0.018 & 3.6$\sigma$ & \citet{Scolnic2018} \\
\hline
\end{tabular}\\
}
\flushleft
\end{table*}

\begin{figure}
\centering
\hspace{-1.2cm}\includegraphics[width=0.54\textwidth]{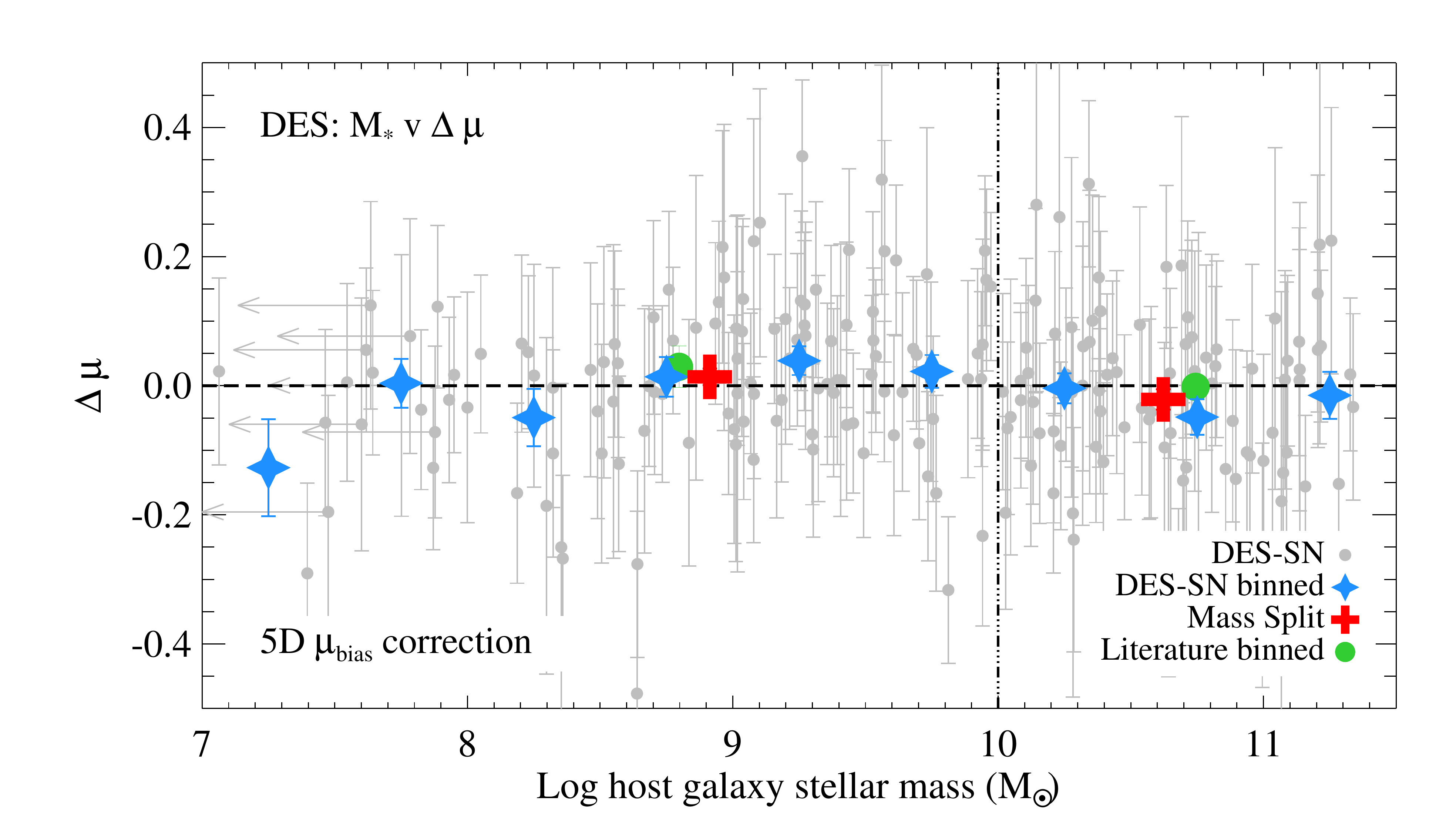}\\
\hspace{-1.2cm}\includegraphics[width=0.54\textwidth]{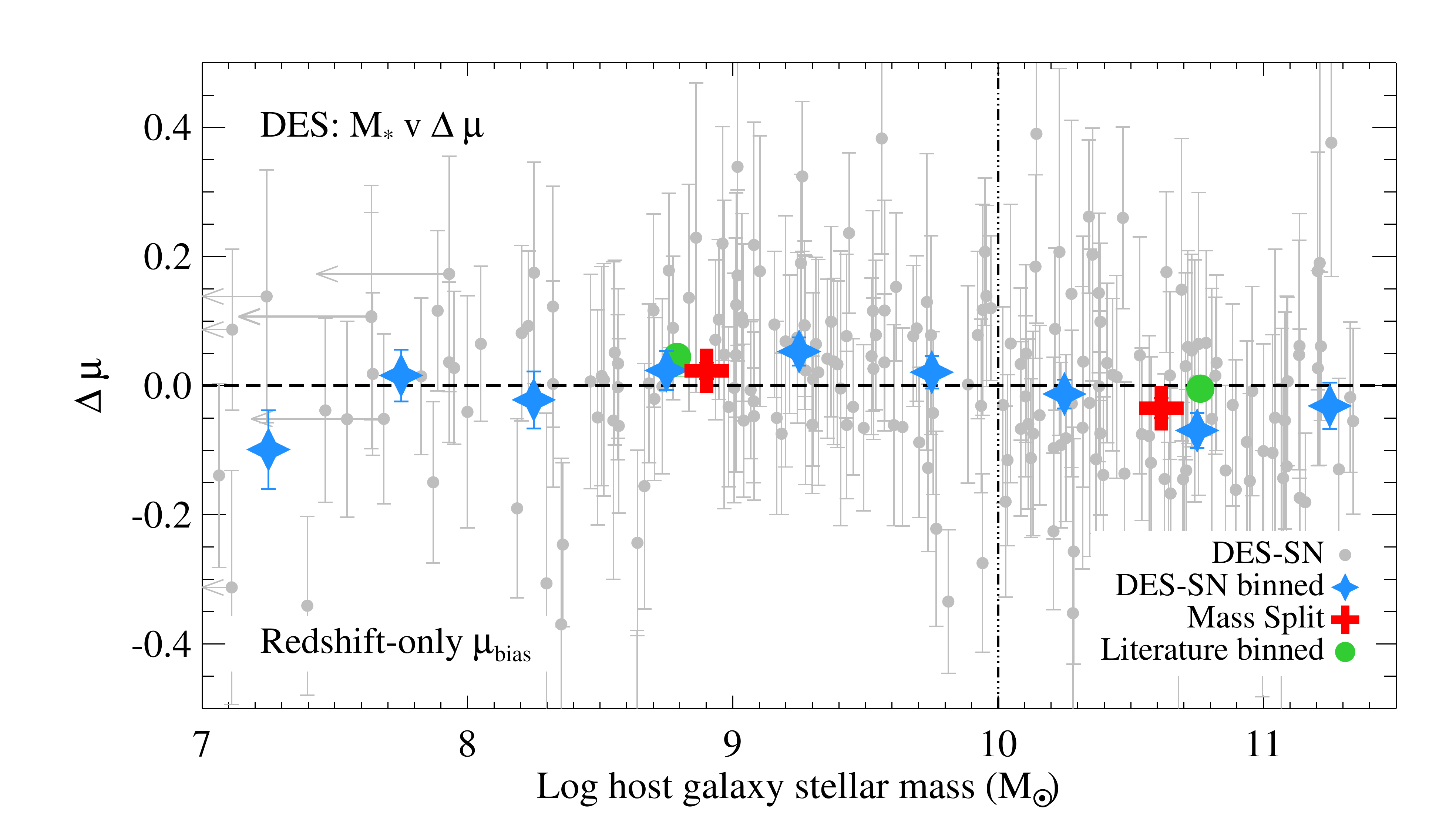}
\caption{\textbf{The DES3YR mass step:} Hubble residuals as a function of host galaxy stellar mass (\mstellar) for the DES-SN sample. Residuals are calculated using the best-fit nuisance parameters determined from the combined DES3YR and low-redshift sample. DES-SN data points are shown in grey. Mean values in bins of stellar mass are plotted as blue diamonds, with the overall values for high mass ($\logmstellar>10$) and low-mass galaxies plotted as large red crosses for the DES-SN sample and green diamonds for the low-redshift data. The \emph{top} panel shows the results when a 5D \mubias\ correction is used as described in \citepalias{Brout2019a}, while the \emph{lower} panel shows the results for a 1D \mubias\ correction as discussed in \S\ref{subsec:massstep_systematic_mubias}.
\label{fig:mass_step}}
\end{figure}

Correlations between \mstellar\ and SN Ia Hubble residuals have been reported in the literature. For example, the JLA analysis \citep{Betoule2014} found $\gamma=0.070\pm0.023$\,mag, a detection at 3.04$\sigma$, while the \citet{Roman2018} analysis measured $\gamma=0.070\pm0.013$\,mag, a detection at 5.4$\sigma$. The DES3YR cosmology analysis \citepalias{Brout2019a}, using galaxy photometry from DES-SVA1, found no significant correlation, with $\gamma=0.021\pm0.018$\,mag for the DES3YR (DES-SN and \lowz\ combined) sample and $\gamma=0.009\pm0.018$\,mag for the DES-SN subsample alone.

\autoref{fig:mass_step} shows the correlation between \mstellar\ and SN Ia Hubble residuals ($\Delta_\mu = \mu_\textrm{obs} - \mu_\textrm{theory}$) for the DES-SN sample. In this analysis, to calculate $\mu_\textrm{theory}$, we fix the cosmological parameters ($\Omega_\mathrm{m}$, $\Omega_\Lambda$)=(0.30,0.70). To calculate $\mu_\textrm{obs}$ we set the SN Ia nuisance parameters  to the best-fit values determined from fitting the DES3YR sample assuming no correction for stellar mass such that ($\alpha$,$\beta$,$\gamma$)=(0.142,3.03,0.0) in \autoref{eqn:gamma_def}. The top panel of \autoref{fig:mass_step} shows the results with a 5D \mubias\ correction (see \S\ref{subsec:snIa-method}), as used in \citetalias{Brout2019a}, with the bottom panel showing the results when a 1D \mubias\ correction is applied. The implications of this choice are discussed in detail in \S\ref{subsec:massstep_systematic_mubias}.  

\autoref{tab:gamma_fitvals} shows the best-fit value of $\gamma$ from this analysis compared to values determined in the literature. For the DES-SN sample, no significant correlation with \mstellar\ is observed: fitting only for $\gamma$ and keeping the location of the mass step at $\mstep=10$, we find $\gamma=0.030\pm0.017$\,mag (inconsistent with zero at 1.8$\sigma$). When $\alpha$, $\beta$ and $\gamma$ are all floated in the fit, we recover $\gamma=0.040\pm0.019$\,mag (2.1$\sigma$) for the the DES-SN sample, $\gamma=0.043\pm0.018$\,mag (2.4$\sigma$) for the DES3YR sample and $\gamma=0.068\pm0.038$\,mag (1.8$\sigma$) for the \lowz\ sub-sample alone. The value for the DES-SN sample is higher, at 1.3$\sigma$, than the value found in the previous DES3YR analysis \citepalias{Brout2019a}. The value found here for the DES-SN sample is consistent with $\gamma$ derived from the JLA analysis at $<1\sigma$ and with $\gamma=0$ at 2.1$\sigma$. 

\subsection{Comparison to \citet{Brout2019a}} 
\label{subsec:brout_comparison}

For this analysis of the DES-SN sample, we find $\gamma=0.040\pm0.019$\,mag, compared to $\gamma=0.009\pm0.018$\,mag as determined by \citetalias{Brout2019a}, a difference of $\Delta \gamma_\textrm{sys} = \gamma_\textrm{sys}-\gamma_\textrm{fid}= -0.031$\,mag. While statistically consistent at 1.3$\sigma$, these two measurements use the same sample of 206 SNe~Ia, each with identical SMP light-curves, analysed consistently with the BBC framework (using a G10 intrinsic scatter model), suggesting a larger tension. These two analyses differ in two distinct ways: here we use deep stack photometry \citep{Wiseman2020} and improved SFHs in the determination of \mstellar.  

\begin{figure}
\centering
\includegraphics[width=0.45\textwidth]{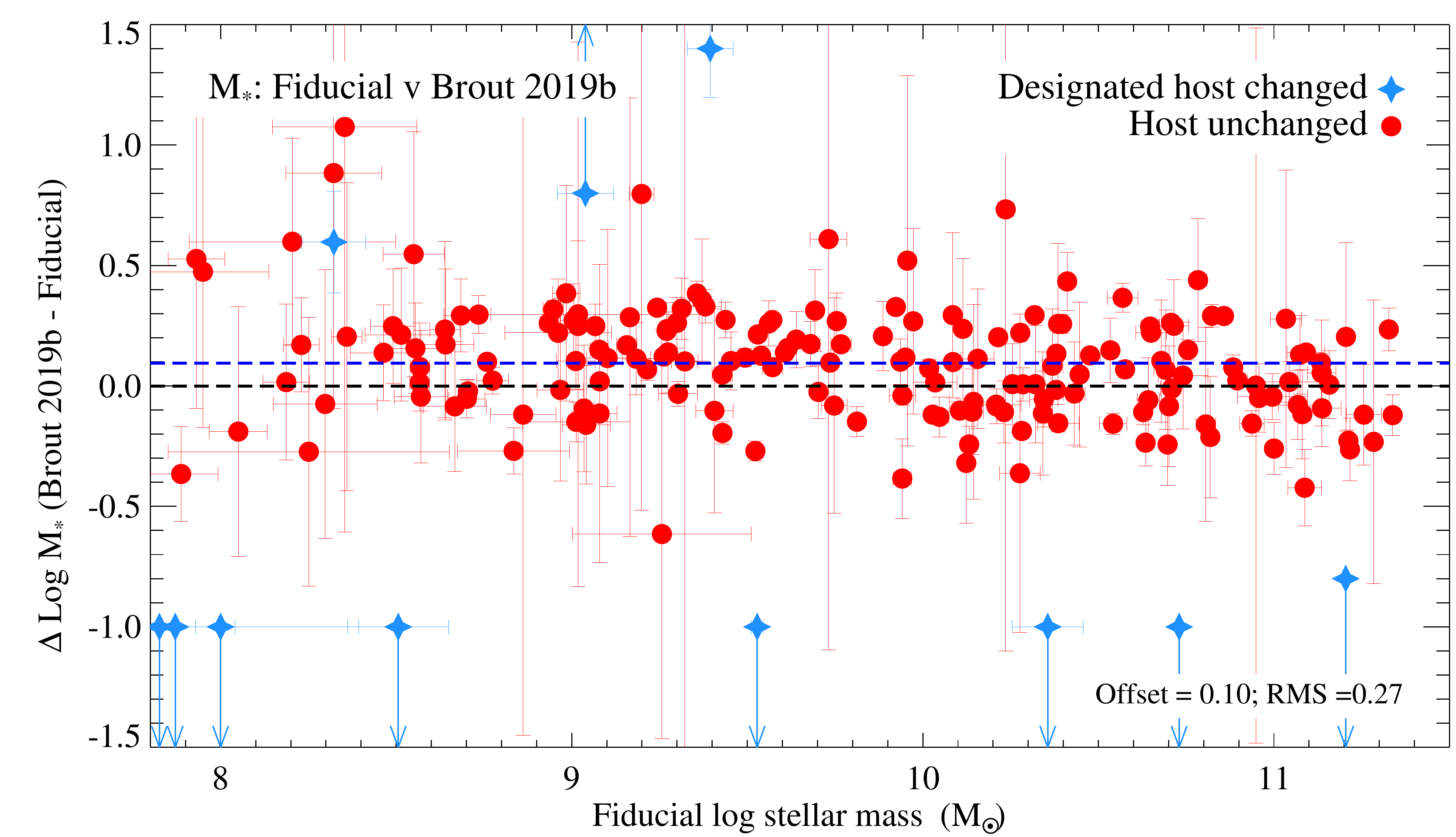}
\caption{Fiducial \mstellar\ estimates compared to those determined by \citetalias{Brout2019a}, using $griz$ galaxy magnitudes taken from the DES SVA1-GOLD catalogue and estimated using the P\'EGASE.2 template library combined with the ZPEG code. No trend as a function of stellar mass is observed, with a mean offset of $0.10\pm0.02$ and an r.m.s of 0.24 for galaxies present in both catalogues. The mean offset between the two values is highlighted by a blue dashed line. Galaxies associated as host galaxies in the deep stacks that differ from those of SVA1-GOLD catalogue, either due to the detection of galaxies below the SVA1-GOLD detection limit or due to differing DLR ratios, are plotted as blue diamonds. 
\label{fig:fid_brout_mass_comp}}
\end{figure}

To probe the sensitivity of our results to these effects, \autoref{fig:fid_brout_mass_comp} shows the difference between our fiducial \mstellar\ estimates and those used in the analysis of \citetalias{Brout2019a}. No evidence of a correlation with stellar mass is observed, with a mean offset of $\Delta\mstellar = 0.10\pm0.02$\,dex and an r.m.s. of 0.24 for galaxies present in both catalogues, with the estimates from \citetalias{Brout2019a} being marginally higher. This difference is consistent with our analysis of the sensitivity of our mass estimates, as discussed in \S\ref{subsubsec:mass_systematics}. Compared to the \citetalias{Brout2019a} sample, 11 previously high-mass hosts ($\logmstellar>10$) are reclassified as low-mass ($\logmstellar<10$) in our analysis, with 8 galaxies moving in the reverse direction. The 11 reclassified low-mass hosts have smaller uncertainties on distance, with a mean uncertainty on $\mu$ of 0.11 compared to 0.15 for the 8, now high-mass hosts. Of the 18 SNe~Ia that were designated as hostless in \citetalias{Brout2019a}, 13 are matched with a galaxy in the \citetalias{Wiseman2020} catalogue, of which only 2 have $\logmstellar>10$, potentially impacting the value of $\gamma$, as all hostless objects were designated `low-mass' in the \citetalias{Brout2019a} analysis. Due to the increased depth and updated galaxy profile information provided by the deep stacked images, 5 SNe~Ia are associated to different galaxies in the \citetalias{Wiseman2020} catalogue compared to the SVA1-GOLD catalogue. Of these, 3 cross the $\logmstellar=10$ boundary, with 2 designated as high-mass based on the photometry of \citetalias{Wiseman2020}. Galaxies associated as host galaxies in the deep stacks that differ from those of SVA1-GOLD catalogue are highlighted as blue diamonds on \autoref{fig:fid_brout_mass_comp}. 

To test how host galaxy misidentification affects our results we remove the 5 events with differing associated host galaxies that cross the $\logmstellar=10$ boundary and reanalyse the DES-SN sample. For the 201 events that pass this criteria we measure $\gamma=0.044\pm0.019$\,mag, while removing these events from the \citetalias{Brout2019a} sample we recover $\gamma=0.009\pm0.019$\,mag. These values are consistent with results for the full sample, suggesting that host galaxy association is not the cause of $\Delta\gamma_\textrm{sys}=-0.031$\,mag between this analysis and that of \citetalias{Brout2019a}. 

\autoref{tab:massstep_systematics}, row 6, shows the effect of varying our host galaxy template library. Using the deep-stack photometry of \citetalias{Wiseman2020} combined with the methodology used in \citetalias{Brout2019a} to estimate $\logmstellar$, we find $\gamma=0.036\pm0.018$\,mag, consistent with our fiducial result. Conversely, \autoref{tab:massstep_systematics}, row 11 shows the results using photometric measurements from the SVA1-GOLD catalogue, as used by \citetalias{Brout2019a}, but the methodology used here, and described in \S\ref{subsubsec:mass_estimates} to estimate \mstellar. In this case, we recover $\gamma=0.031\pm0.020$\,mag. This value is also consistent, if marginally smaller than our fiducial result. These tests suggest that no single cause explains the $\Delta\gamma_\textrm{sys}=-0.031$\,mag observed between this analysis and that of \citetalias{Brout2019a}, and therefore the reduced value of $\gamma$ found by \citetalias{Brout2019a} is likely caused by a combination of the photometric catalogue and template library.

\subsubsection{Cosmological Implications}

To study how our \mstellar\ estimates affect the cosmological parameters, we replicate the analysis of \citetalias{Brout2019a}, and combine the DES3YR sample with a CMB prior from \citet{Planck2016}. Considering a statistical-only covariance matrix, we find a shift in the dark energy equation-of-state of $\Delta w = 0.011$ when using a G10 intrinsic model ($\Delta w = 0.015$ for the C11 model) when using our \mstellar\ estimates compared to those used in \citetalias{Brout2019a}. This shift, while non-negligible, is sub-dominant to the astrophysical systematic uncertainty of $\sigma_w = 0.026$ determined for the DES3YR cosmological analysis \citepalias[][Table 8]{Brout2019a}.  

\subsection{Systematic tests of the mass step}
\label{subsec:massstep_systematics}

We next study the sensitivity of our $\gamma$ estimate to various assumptions in our analysis. Determining $\gamma$ depends on two measurements: the host galaxy mass estimates and the estimated distance to each event. We discuss each in turn.

\subsubsection{Sensitivity of the mass step to stellar mass estimates}
\label{subsubsec:massstep_sys_masses}

In \S\ref{subsubsec:mass_systematics} we showed that our stellar mass estimates have a small sensitivity to choices in our analysis (e.g., galaxy photometry, stellar libraries used, SFHs) with at most 15 per cent of SNe moving between the high and low stellar mass bins as we vary these choices (\autoref{tab:mass_systematics}). In \autoref{tab:massstep_systematics} we show the implications these choices have on the best-fit value of $\gamma$. In all cases, we vary $\alpha$, $\beta$ and $\gamma$ simultaneously in the BBC fit, and find no statistically significant variation in $\alpha$ or $\beta$. We consider two samples: the DES-SN sample alone, and then combined with the low-redshift SN Ia data: the DES3YR sample. 

For the DES-SN sample, $\gamma$ is maximally inconsistent from $\gamma=0$ at 2.3$\sigma$ (\autoref{tab:massstep_systematics}, row 4). There is no significant difference from our fiducial result for any of the systematic tests considered. Averaged over all systematic tests considered in \autoref{tab:massstep_systematics}, we recover $\left<\gamma\right>= 0.030$\,mag with a mean uncertainty of $\left<\sigma\right>=0.019$\,mag and r.m.s. of 0.009\,mag. These results confirm that our assumptions on the underlying SFHs and photometric catalogue used to estimate the DES-SN host stellar masses do not significantly impact the best-fit value of $\gamma$. 

When the low-redshift sample is included in this analysis, $\gamma$ is maximally inconsistent from zero at 2.6$\sigma$ (\autoref{tab:massstep_systematics}, row 14). Combining all estimates of $\gamma$, we recover $\left<\gamma\right> = 0.037$\,mag with $\left<\sigma\right>=0.018$\,mag and an r.m.s. of 0.008\,mag, again consistent with our fiducial value.

\begin{table*}
{\centering
\caption{Comparison between $\gamma$ determined using various photometric catalogues and SFHs  to estimate \mstellar.}
\label{tab:massstep_systematics}
\begin{tabular}{llllccc}
  \hline
  Row \# & SN Sample & Photometric Catalogue & Templates & IMF & $\gamma$ (mag) & $\Delta \gamma$ (mag)$^{1}$ \\ 
  \hline
  1 Fiducial result & DES-SN & \citetalias{Wiseman2020} & \PEGASE & \citetalias{Kroupa2001} & $0.040\pm0.019$ & 0.0 \\
  2 \citetalias{Brout2019a}$^{2,3,4}$ & DES-SN & SVA1-GOLD:\texttt{mag\_detmodel} & ZPEG & \citetalias{Kroupa2001} & $0.009\pm0.019$ & -0.031 \\
  3 & DES-SN & \citetalias{Wiseman2020} & \PEGASEbursts & \citetalias{Kroupa2001} & $0.030\pm0.018$ & -0.010 \\
  4 & DES-SN & \citetalias{Wiseman2020} & \PEGASE & \citetalias{Salpeter1955} & $0.042\pm0.019$ & +0.002\\
  5 & DES-SN & \citetalias{Wiseman2020} & \PEGASEbursts & \citetalias{Salpeter1955} & $0.019\pm0.018$ & -0.021 \\
  6 $^{2}$ & DES-SN & \citetalias{Wiseman2020} & ZPEG & \citetalias{Kroupa2001} & $0.036\pm0.018$ & -0.004 \\
  7 & DES-SN & \citetalias{Wiseman2020} & \citetalias{Maraston2005} & \citetalias{Kroupa2001} & $0.032\pm0.019$ & -0.008 \\
  8 & DES-SN & \citetalias{Wiseman2020} & \citetalias{Maraston2005} & \citetalias{Salpeter1955} & $0.030\pm0.019$ & -0.010 \\
  9 & DES-SN & \citetalias{Wiseman2020} & \citetalias{Bruzual2003} & \citetalias{Salpeter1955} & $0.030\pm0.019$ & -0.010 \\
  10 & DES-SN & SVA1-GOLD: \texttt{mag\_auto} & \PEGASE & \citetalias{Kroupa2001} & $0.032\pm0.020$ & -0.008 \\
  11 $^{3}$ & DES-SN & SVA1-GOLD: \texttt{mag\_detmodel} & \PEGASE & \citetalias{Kroupa2001} & $0.031\pm0.020$ & -0.009 \\
  \hline\hline
  12 Fiducial result & DES3YR & \citetalias{Wiseman2020} & \PEGASE & \citetalias{Kroupa2001} & $0.043\pm0.018$ & 0.0 \\
  13 \citetalias{Brout2019a} $^{2,3,4}$ & DES3YR & SVA1-GOLD:\texttt{mag\_detmodel} & ZPEG & \citetalias{Kroupa2001} & $0.024\pm0.018$ & -0.020 \\
  14 & DES3YR & \citetalias{Wiseman2020} & \PEGASEbursts & \citetalias{Kroupa2001} & $0.037\pm0.018$ & -0.006 \\
  15 & DES3YR & \citetalias{Wiseman2020} & \PEGASE & \citetalias{Salpeter1955} & $0.045\pm0.018$ & +0.002 \\
  16 & DES3YR & \citetalias{Wiseman2020} & \PEGASEbursts & \citetalias{Salpeter1955} & $0.029\pm0.017$ & -0.015 \\
  17 $^{2}$ & DES3YR & \citetalias{Wiseman2020} & ZPEG & \citetalias{Kroupa2001} & $0.042\pm0.018$ & -0.001 \\
  18 & DES3YR & \citetalias{Wiseman2020} & \citetalias{Maraston2005} & \citetalias{Kroupa2001} & $0.038\pm0.018$ & -0.005 \\
  19 & DES3YR & \citetalias{Wiseman2020} & \citetalias{Maraston2005} & \citetalias{Salpeter1955} & $0.037\pm0.018$ & -0.006 \\
  20 & DES3YR & \citetalias{Wiseman2020} & \citetalias{Bruzual2003} & \citetalias{Salpeter1955} & $0.038\pm0.018$ & -0.006 \\
  21 & DES3YR & SVA1-GOLD: \texttt{mag\_auto} & \PEGASE & \citetalias{Kroupa2001} & $0.038\pm0.018$ & -0.006 \\
  22 $^{3}$ & DES3YR & SVA1-GOLD: \texttt{mag\_detmodel} & \PEGASE & \citetalias{Kroupa2001} & $0.038\pm0.018$ & -0.006 \\
\hline
\end{tabular}\\
}
\flushleft
\footnote{a}{$\gamma - \gamma_\textrm{fid}$ where $\gamma_\textrm{fid}$ is given in row 1 or 12 depending upon sample.}\\
\footnote{b}{Matches the methodology used in \citet{Betoule2014} and \citet{Scolnic2018}.}\\
\footnote{c}{Matches the photometry used in the analysis of \citetalias{Brout2019a}.}\\
\footnote{d}{The value of $\gamma$ matches that in Table 5 of \citetalias{Brout2019a} (considering the G10 intrinsic scatter model) for the DES3YR analysis, but differs by 0.001 for the DES-SN sample due to the loss of CID$=$1279500. See text for details.} \\
\end{table*}

\subsubsection{Sensitivity of the mass step to light-curve systematics}

There are four major sources of uncertainty from the light curves that could impact the value of $\gamma$: (1) the photometric technique used to estimate light-curve fluxes, (2) the light-curve cuts used to generate the DES3YR sample, (3) the parameterisation of the mass step, and (4) the methodology used to estimate distances and nuisance parameters. \autoref{tab:massstep_lc_systematics} shows the best-fit value of $\gamma$ for each systematic test considered.

\paragraph{Photometry}
The DES SN Ia analysis uses a \lq Scene Model Photometry\rq\ (SMP) technique \citep{Brout2019b} to measure light-curve fluxes and uncertainties. This technique forward models a time dependent flux from the transient with an underlying constant host galaxy component, and compares to the DES images. This method differs from traditional \lq difference imaging\rq, where a deep reference image is subtracted from each SN observation. As a crosscheck of $\gamma$ to SMP photometry, we consider flux estimates using the DES real-time difference-imaging pipeline \citep[\texttt{DIFFIMG}; ][]{Kessler2015}. Propagating these light curves through the DES3YR cosmology pipeline, we find $\gamma=0.019\pm0.021$\,mag for the DES-SN SNe, and $0.030\pm0.019$\,mag when combined with the low-redshift sample (\autoref{tab:massstep_lc_systematics}, rows 3 and 16). These values differ from our fiducial values of $\gamma$ by -0.021\,mag and -0.013\,mag, respectively. Analysing the DES-SN sample with the DES real-time difference-imaging pipeline reduces the number of SN that pass the light-curve coverage criteria defined in \citetalias{Brout2019a} by 6 and increases the r.m.s. dispersion of our sample from 0.126\,mag to 0.134\,mag. 

Considering only the 193 DES-SN common to both datasets we measure $\gamma=0.028\pm0.020$\,mag when using \texttt{DIFFIMG} photometry compared to $0.030\pm0.019$\,mag for the SMP photometry. These values are consistent, suggesting that the value of $\gamma$ determined using \texttt{DIFFIMG} photometry, smaller than our fiducial analysis, is driven by the complement of the two datasets, not the photometric measurements themselves.  The 7 SNe~Ia in the \texttt{DIFFIMG} sample that do not pass the SMP criteria have mean $\mstellar=9.94\pm0.20$, consistent with the DES-SN sample (\autoref{tab:sample_info}), and mean $\Delta_\mu=0.142\pm0.070$, indicating that these events are responsible for the additional scatter in this sample. The 3 events with $\mstellar>10.0$ have mean $\Delta\mu=0.285\pm0.111$, compared to $0.036\pm0.045$ for SNe~Ia in low mass hosts, suggesting that these outlying events, excluded from the SMP analysis, are responsible for the reduced value of $\gamma$ when analysing the DES-SN sample with \texttt{DIFFIMG} photometry. 

\paragraph{SN selection cuts}
Our analysis requires all SNe~Ia to have well-observed light-curves to reliably constrain the light-curve fit parameters, and we require $-3<x_1<3$ and $-0.3<c<0.3$ matching the range over which the SALT2 model has been trained \citep{Guy2010}. 

To test the effect that our selection criteria has on $\gamma$, in rows 4-7 of \autoref{tab:massstep_lc_systematics}, we split the DES-SN sample into subsamples of $x_1$ and $c$. For SNe~Ia with $x_1<0$ we recover $\gamma=0.000\pm0.029$\,mag for the DES sample alone, compared to $\gamma=0.026\pm0.028$\,mag for those with $x_1>0$, different at 1.2$\sigma$. From \autoref{fig:galprop_v_snprop}, SNe Ia with $x_1<0$ are preferentially found in high mass galaxies, while those with $x_1>0$ are dominated by low mass galaxies. For the analogous test with $c$ we find $\gamma=-0.001\pm0.021$\,mag for events with $c<0$ and $\gamma=0.106\pm0.039$\,mag for those with $c>0$, different at 2.4$\sigma$. We find consistent results when combining the DES-SN sample with the low redshift sample (\autoref{tab:massstep_lc_systematics} rows 17-20). From \autoref{fig:galprop_v_snprop} there is some evidence that high mass hosts preferentially host redder ($c>0$) SNe~Ia. Averaging over all mass estimates derived from deep stack photometry we find a mean difference of 1.2 and 1.7$\sigma$ between the value of $\gamma$ determined for high and low $x_1$ and $c$, respectively. 

\paragraph{Parameterising the Mass Step}
Our fiducial analysis considers the mass step to be parameterised by \autoref{eqn:gamma_def} with $\mstep=10$. To test how this assumption affects the value of $\gamma$, in row 8 of \autoref{tab:massstep_lc_systematics}, we simultaneously fit for $\gamma$ and $\mstep$, finding $\mstep=9.68\pm0.06$ and $\gamma=0.046\pm0.018$\,mag ($\Delta \gamma_\textrm{sys} = +0.006$\,mag) for the DES sample alone. These values are consistent with those found when combining with the low redshift sample and with our fiducial result. 

In \autoref{eqn:gamma_def}, the mass step is parameterised as a step function at $\mstellar=\mstep$. To test the sensitivity of our results on this assumption, we re-parameterise $G_\textrm{host}$ in \autoref{eqn:salt2formula} as a smooth function around a transition mass \citep{Childress2014} such that

\begin{equation}
\label{eqn:gamma_tau}
G_\textrm{host} = \left[\frac{1}{1+\exp\left(\frac{-(\mstellar - \mstep)}{\gamma_\tau}\right)} -0.5\right],
\end{equation}
where $\gamma_\tau$ parameterises the \mstellar\ scale of the mass step. Fitting for $\gamma_\tau$ and $\gamma$ simultaneously (while holding \mstep\ fixed at $\mstep=10$), we recover $\gamma_\tau=0.003\pm0.016$ and $\gamma=0.040\pm0.019$\,mag, while fitting for $\gamma_\tau$, $\mstep$ and $\gamma$ simultaneously, we recover $\gamma=0.047\pm0.018$\,mag, $\gamma_\tau=0.001\pm0.019$ and $\mstep=9.70\pm0.00$ (\autoref{tab:massstep_lc_systematics}, rows 10-12). The fits including the low-redshift sample are consistent with these values. For these systematic tests we recover $\Delta \gamma_\textrm{sys} = +0.000, +0.006, +0.007$ and $+0.007$\,mag, indicating that there is no evidence that a different mass step parametrisation affects $\gamma$.

\paragraph{Distance estimates}
The DES3YR cosmological analysis uses the BBC framework \citep{Kessler2017,Brout2019a} which differs from earlier analyses (such as JLA) by implementing 5D bias-corrections determined from large simulations of the DES survey \citep{Kessler2019}. In the BBC framework \mubias\ \lq corrects\rq\ the observed values of $m_B$, $x_1$ and $c$ for each SN Ia and includes a correction for the distance uncertainty.  

When we use a 1D \mubias\ correction dependent only on $z$ \citep[e.g.,][]{Betoule2014}; we recover $\gamma=0.066\pm0.020$\,mag ($\Delta \gamma_\textrm{sys} = +0.026$\,mag) for the DES-SN sample, and $\gamma=0.064\pm0.019$\,mag ($\Delta \gamma_\textrm{sys} = +0.021$\,mag) when including the low-redshift SNe. These are the highest values of $\gamma$ measured for the DES-SN sample, and consistent with the values found by \citet{Betoule2014,Roman2018}. To test this further, \autoref{tab:massstep_lc_systematics}, rows 9, 11 and 13, show the results when a 1D bias correction is used and various combinations of $\mstep$ and $\gamma_\tau$ are varied. In all cases, the best-fit value of $\gamma$ is larger than that found in the fiducial analysis and the corresponding systematic test using a 5D \mubias\ correction. 

\begin{figure}
\centering
\includegraphics[width=0.45\textwidth]{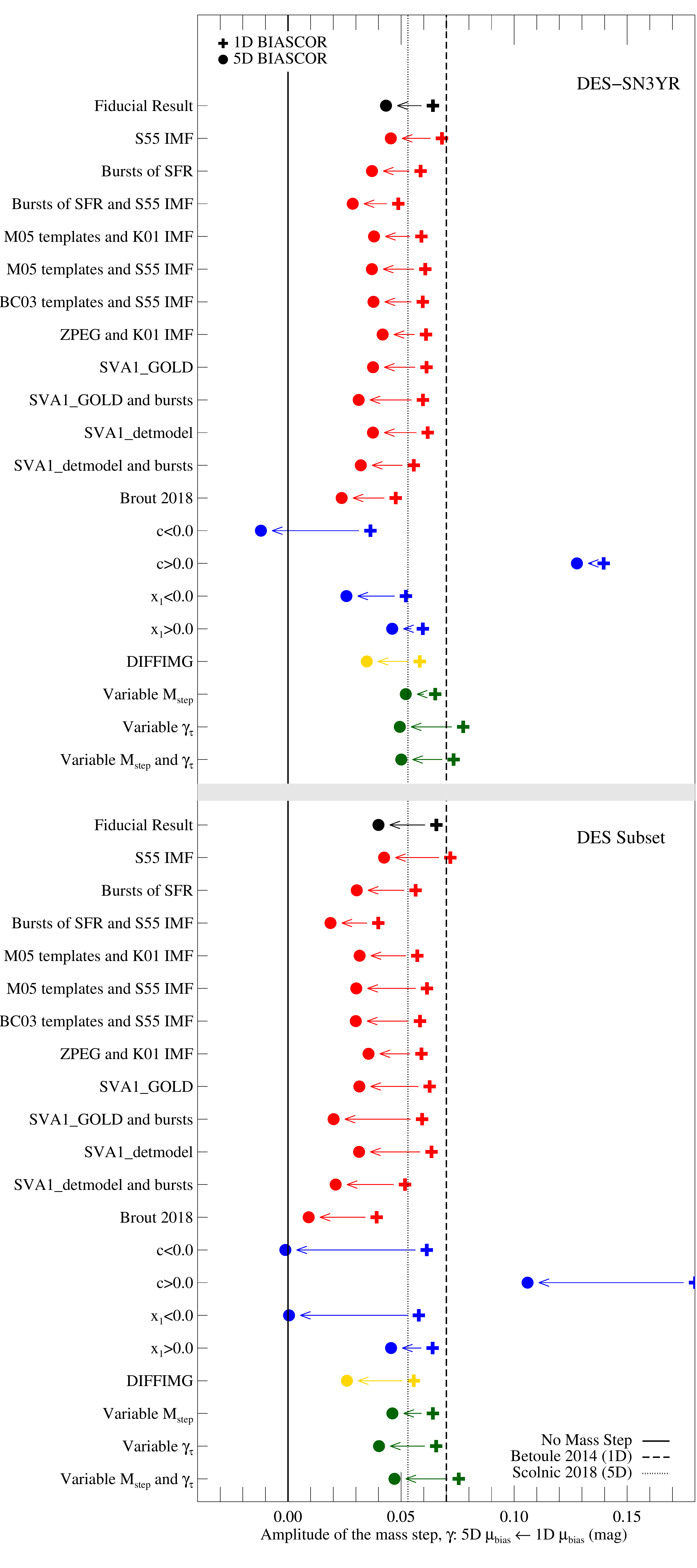}
\caption{The best-fit values of $\gamma$ considering different systematic uncertainties. For each entry, the right-hand value (plotted as a plus symbol) indicates the value when a 1D bias correction is used, while the left-hand entry (plotted as a filled circle) is for the 5D correction. Red entries denote alternative mass estimates (see \S\ref{subsubsec:mass_systematics}), blue denote systematics in the sample selection, yellow  the results when \texttt{DIFFIMG} photometry is used in the light-curve fitting, and green when various assumptions about the mass step parametrisation are considered. The dashed and dotted lines show the values of $\gamma$ determined by \citet{Betoule2014} (assuming a 1D bias correction) and \citet{Scolnic2018} assuming a 5D bias correction, while the solid line indicates the case where $\gamma=0$. The top panel shows the results for the DES3YR sample, while the bottom panel presents results for the DES-SN subset. 
\label{fig:gamma_dist}}
\end{figure}

\autoref{fig:gamma_dist} shows the effect that the 5D bias correction has over all systematic tests considered. The top panel shows the results for the DES3YR sample, while the bottom panel highlights the results for the DES-SN subset. This figure shows the best-fit value of $\gamma$ for both 5D and 1D bias corrections, when alternative estimates of \mstellar\ are used along with different photometric estimates and light-curve cuts. In all cases, the 1D bias correction produces a higher value of $\gamma$. Over all 42 systematic tests, a 1D bias correction recovers a larger value of $\gamma$ compared to a 5D bias correction with offsets between 0.012 and 0.082\,mag, with a mean of 0.028\,mag and standard deviation 0.013\,mag. 

To estimate an uncertainty on this measurement, we simulate 100 realisations of the DES-SN sample (using the prescription described in \S\ref{sec:simulations}). For each simulated sample, we determine the best-fit values of $\alpha$, $\beta$, $\gamma$ using both a 5D and 1D \mubias\ correction. Averaging over all samples, we find a mean value of $\Delta \gamma = 0.014$\,mag (see \S\ref{subsec:sim_mubias_diff} for details) with a standard deviation of $0.009$\,mag. Our results are unaffected if we further require that the 5D and 1D samples comprise exactly the same SNe after cuts.  

Overall, for the DES-SN sample, we find an offset of 

\begin{align}\label{eq:data_result}
    \Delta_\gamma = [\gamma_\textrm{1D}-\gamma_\textrm{5D}]_\textrm{data} = 0.026\pm0.009\,\textrm{mag}. 
\end{align}

This value consistent with a difference of $\Delta\gamma=0.025$\,mag observed for the PS1 sample \citep[][Section~3.7]{Scolnic2018}. The cause of this offset is explored in \S\ref{subsec:massstep_systematic_mubias}. 

\begin{table*}
{\centering
\caption{The best-fit value of $\gamma$ considering systematic uncertainties in the light-curve fitting procedure. The fiducial results from this study are highlighted in bold}
\label{tab:massstep_lc_systematics}
\begin{tabular}{llllccccccc}
  \hline
  Row \# & SN Sample & Phot. & Cuts$^{1}$ & \mstep$^{2}$ & $\gamma_\tau^{2}$ & BiasCor & $\textrm{N}_\textrm{SN}$ & $\gamma$ (mag) & $\Delta \gamma$ (mag)$^{3}$ & r.m.s.$^{4}$ \\
  \hline
  \textbf{1} & \textbf{DES-SN} & \textbf{SMP} & \textbf{None} & \textbf{Fixed} & \textbf{Fixed} & \textbf{5D} & \textbf{\NDES} & \textbf{0.040$\pm$0.019} & \textbf{0.0} & \textbf{0.126}\\%
  \textbf{2} & \textbf{DES-SN} & \textbf{SMP} & \textbf{None} & \textbf{Fixed} & \textbf{Fixed} & \textbf{1D} & \textbf{208} & \textbf{0.066$\pm$0.020} & \textbf{+0.026} & \textbf{0.153} \\
  \cline{1-11}
  3 & DES-SN & \texttt{DIFFIMG} & None & Fixed & Fixed & 5D & 200 & $0.019\pm0.021$ & -0.021 & 0.134\\%
  4 & DES-SN & SMP & C$<$0.0 & Fixed & Fixed & 5D & 136 & $-0.001\pm0.021$ & -0.041 & 0.108 \\%
  5 & DES-SN & SMP & C$>$0.0 & Fixed & Fixed & 5D & 70 & $0.106\pm0.039$ & +0.066  & 0.154 \\%
  6 & DES-SN & SMP & $x_1<$0.0 & Fixed & Fixed & 5D & 88 & $0.000\pm0.029$ & -0.040 & 0.136 \\%
  7 & DES-SN & SMP & $x_1>$0.0 & Fixed & Fixed & 5D & 118 & $0.046\pm0.026$ & +0.006 & 0.117 \\
  8 & DES-SN & SMP & None & $9.68\pm0.06$ & Fixed & 5D & \NDES\ & $0.046\pm0.018$ & +0.006 & 0.126 \\
  9 & DES-SN & SMP & None & $10.17\pm0.13$ & Fixed & 1D & 208 & $0.064\pm0.022$ & +0.024 & 0.153 \\
  10 & DES-SN & SMP & None & Fixed & $0.003\pm0.016$ & 5D & \NDES\ & $0.040\pm0.019$ & +0.000 & 0.126 \\
  11 & DES-SN & SMP & None & Fixed & $0.003\pm0.143$ & 1D & 208 &  $0.066\pm0.020$ & +0.026 & 0.153 \\
  12 & DES-SN & SMP & None & $9.70\pm0.01$ & $0.001\pm0.019$ & 5D & \NDES\ & $0.047\pm0.018$ & +0.007 & 0.127 \\
  13 & DES-SN & SMP & None & $9.70\pm0.01$ & $0.001\pm0.006$ & 1D & 208 & $0.076\pm0.020$ & +0.035 & 0.154 \\
  \hline\hline
  14 & DES3YR & SMP & None & Fixed & Fixed & 5D & \NTOT\ & $0.043\pm0.018$ & 0.0 & 0.144\\
  15 & DES3YR & SMP & None & Fixed & Fixed & 1D & 336 & $0.064\pm0.019$ & +0.021 & 0.157 \\
  16 & DES3YR & \texttt{DIFFIMG} & None & Fixed & Fixed & 5D & 322 &  $0.030\pm0.019$ & -0.013 & 0.151 \\
  17 & DES3YR & SMP & C$<$0.0 & Fixed & Fixed & 5D & 203 & $-0.012\pm0.021$ & -0.055 & 0.126 \\
  18 & DES3YR & SMP & C$>$0.0 & Fixed & Fixed & 5D & 125 & $0.128\pm0.034$ & +0.084 & 0.170 \\
  19 & DES3YR & SMP & $x_1<$0.0 & Fixed & Fixed & 5D & 155 & $0.026\pm0.028$ & -0.017 & 0.140 \\
  20 & DES3YR & SMP & $x_1>$0.0 & Fixed & Fixed & 5D & 173 & $0.046\pm0.024$ & +0.003 & 0.141 \\
  21 & DES3YR & SMP & None & $10.89\pm0.04$ & Fixed & 5D & \NTOT\ & $0.052\pm0.021$ & +0.009 & 0.145 \\
  22 & DES3YR & SMP & None & $10.89\pm0.03$ & Fixed & 1D & 336 &  $0.065\pm0.022$ & +0.022 & 0.157 \\
  23 & DES3YR & SMP & None & Fixed & $0.151\pm0.083$ & 5D & \NTOT\ & $0.049\pm0.021$ & +0.006 & 0.145 \\
  24 & DES3YR & SMP & None & Fixed & $0.164\pm0.122$ & 1D & 336 & $0.077\pm0.023$ & +0.034 & 0.158 \\
  25 & DES3YR & SMP & None & $10.15\pm0.02$ & $0.001\pm0.021$ & 5D & \NTOT\ & $0.050\pm0.018$ & +0.007 & 0.145 \\
  26 & DES3YR & SMP & None & $10.15\pm0.00$ & $0.001\pm0.000$ & 1D & 336 & $0.073\pm0.019$ & +0.030 & 0.158 \\
\hline
\end{tabular}\\
}
\flushleft
\footnote{a}{The fiducial analysis includes cuts of $-3.0<x_1<3.0$ and $-0.3<c<0.3$.}\\
\footnote{b}{Fixed to $\mstep=10.0$ and $\gamma_\tau=0.01$ in the fiducial analysis.}\\
\footnote{c}{$\gamma - \gamma_\textrm{fid}$ where $\gamma_\textrm{fid}$ is given in row 1 or 2 depending upon sample.}\\
\footnote{d}{r.m.s. of Hubble diagram residuals from LCDM model after correction ($\Delta \mu$ in \autoref{eqn:hubbleresiduals}).}\\
\end{table*}

\subsection{The dependence of the mass step on the bias correction} 
\label{subsec:massstep_systematic_mubias}

\begin{figure}
\centering
\includegraphics[width=0.45\textwidth]{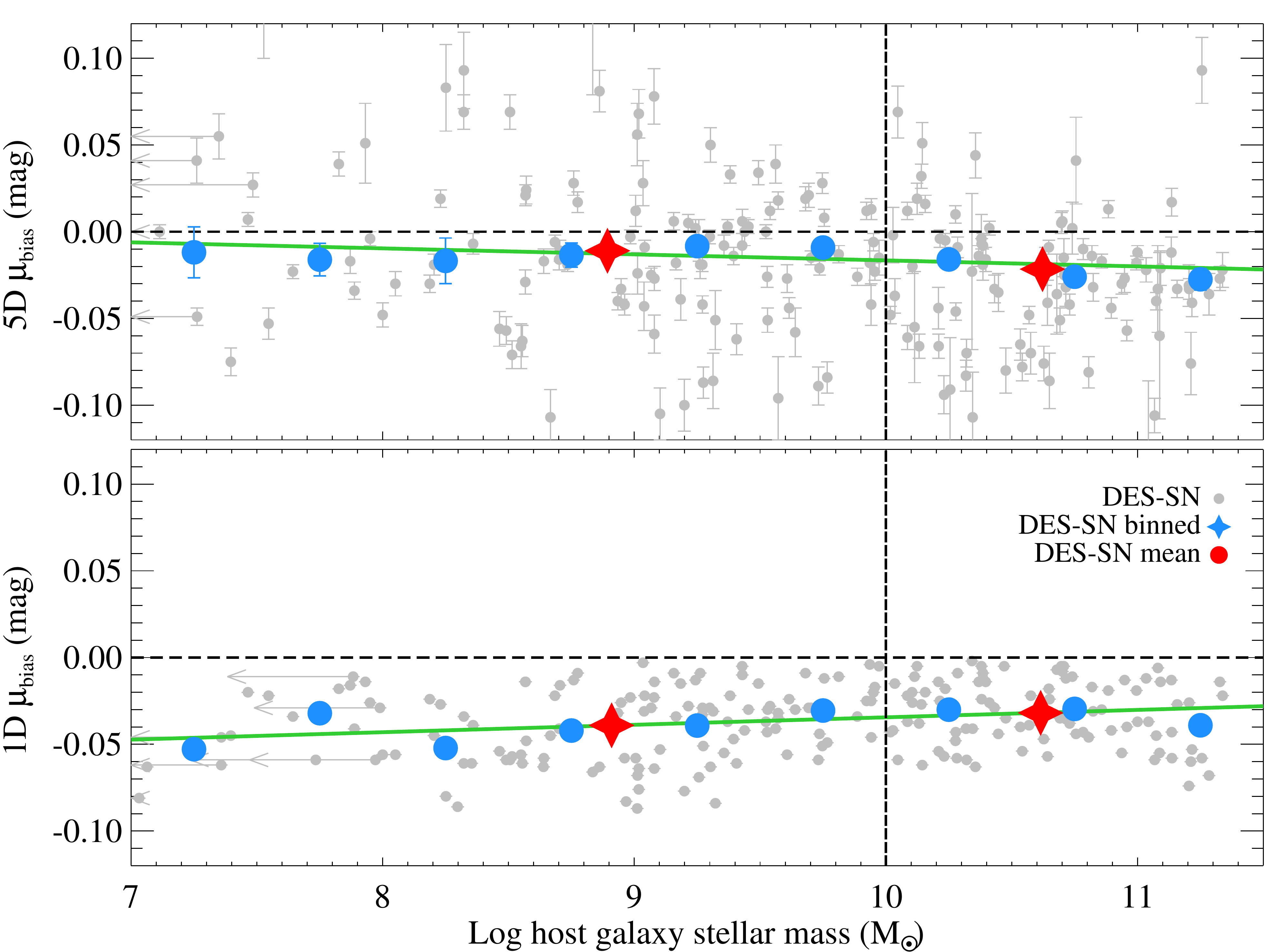}
\caption{The relationship between \mstellar\ and \mubias\ for the DES-SN sample. The top panel shows the results for a 5D \mubias\ correction, with the lower panel showing the results for a 1D \mubias\ correction. Data points are shown in grey. The mean value in bins of stellar mass are shown as blue filled circles, with the value for high and low mass samples shown as red diamonds. The best fitting linear relationship is shown in green.  
\label{fig:mass_v_biascor}}
\end{figure}

Systematic offsets between the value of $\gamma$ when using 1D and 5D bias corrections implies a difference in \mubias\ between SNe~Ia found in high mass galaxies compared to their low mass counterparts. \autoref{fig:mass_v_biascor} shows the correlation between the SN host stellar mass and the bias correction applied to that SN distance, \mubias, for both the 1D and 5D bias corrections. For the 5D bias correction, there is a correlation between \mstellar\ and \mubias\ with a slope $-0.004\pm0.001$. There is a difference in the mean value of \mubias\ between high- and low-mass galaxies of $\Delta \mubias = 0.011\pm0.004$\,mag. The 1D bias correction shows the opposite correlation, with a mean difference of $\Delta \mubias = -0.007\pm0.003$\,mag. 

\begin{figure}
\centering
\includegraphics[width=0.45\textwidth]{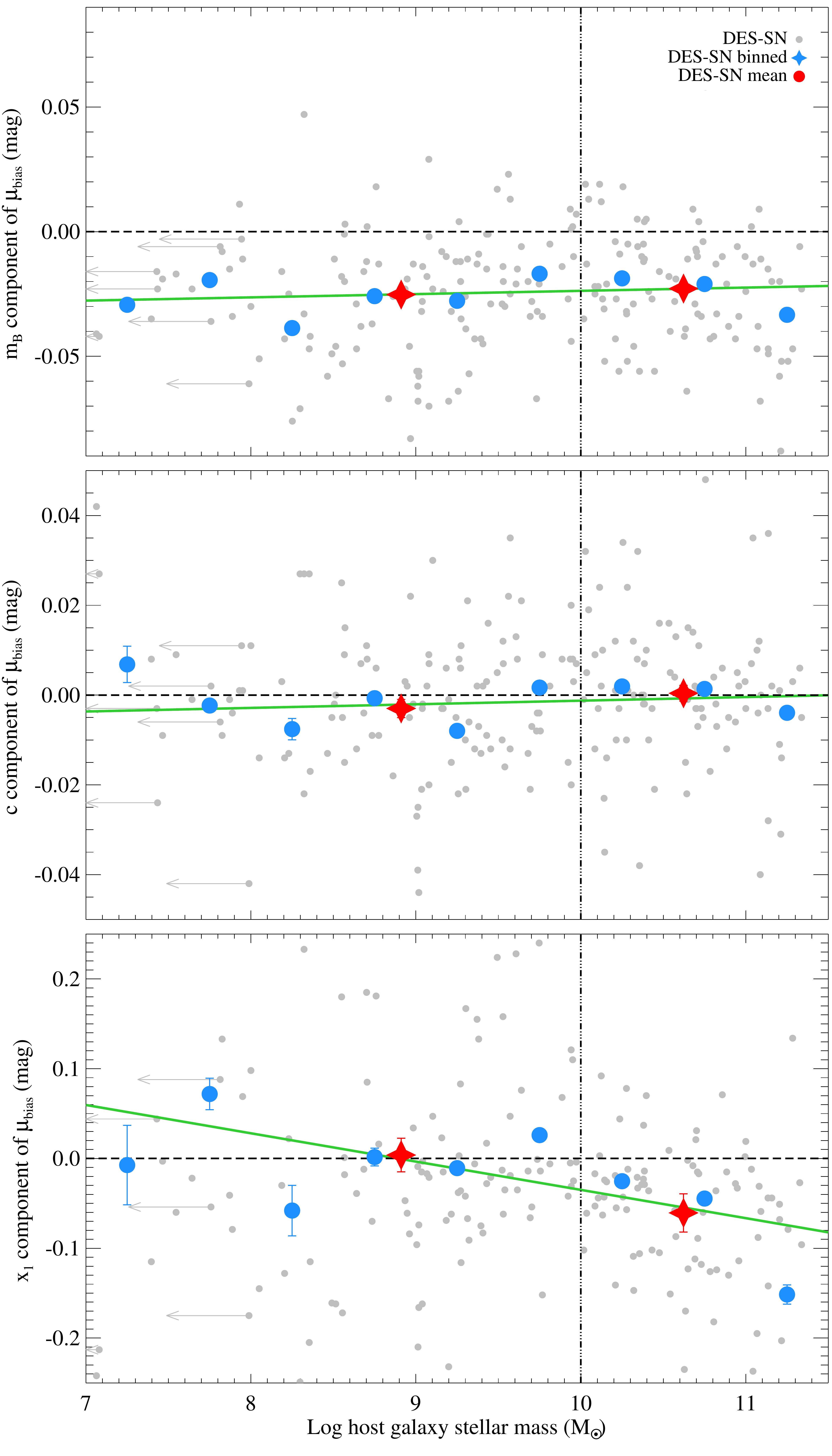}
\caption{The relationship between \mstellar\ and the individual components of the 5D \mubias\ ($m_B$, $c$ and $x_1$) correction for the DES-SN sample. Data points are shown in grey. The mean value in bins of stellar mass are shown as blue filled circles, with the value for high and low mass samples shown as red diamonds. The best fitting linear relationship is shown in green. A correlation coefficient of 0.152, inconsistent from 0 at 2.2$\sigma$ is seen between \mstellar\ and ${x_1}_{\textrm{bias}}$.
\label{fig:mass_v_biascor_breakdown}
}
\end{figure}

\autoref{fig:mass_v_biascor_breakdown} shows the origin of the 5D \mubias\ correlation: the correction to the observed values of $m_B$, $x_1$ and $c$ for each event, denoted ${m_B}_\textrm{bias}$, ${x_1}_\textrm{bias}$ and $c_\textrm{bias}$. No evidence of a relationship between \mstellar\ and ${m_B}_\textrm{bias}$ or $c_\textrm{bias}$ is observed, but we find a correlation between \mstellar\ and ${x_1}_\textrm{bias}$ with a difference of $\Delta {x_1}_\textrm{bias}$ of $0.064\pm0.028$\,mag (2.3$\sigma$) between SNe in high- and low-mass galaxies for the DES-SN sample. Fixing $\alpha=0.150$ (the value derived for the DES3YR sample), this corresponds to $\Delta \mubias = \alpha \times \Delta {x_1}_\textrm{bias} = 0.010\pm0.004$\,mag, consistent with the offset of $\Delta \mubias = 0.011$\,mag determined above. 

This result is consistent with \autoref{fig:galprop_v_snprop}, where high-mass galaxies predominately host low-$x_1$ SNe Ia. These events require a different bias correction compared to the higher-$x_1$ SNe Ia in low-mass hosts (\autoref{fig:mass_v_biascor_breakdown}). This comparison suggests that, in the BBC framework, a fraction of \mstep\ as measured by a 1D \mubias\ correction, is not an independent offset in SN~Ia luminosity related to \mstellar, but an uncorrected contribution related to $x_1$, as deduced by a 5D \mubias\ correction. In \S\ref{sec:simulations} we test this inference by imprinting realistic correlations between SN and \mstellar\ into simulations of DES-SN, independent on \mstep, and test for potential biases in the recovered value of $\gamma$ for both 5D and 1D \mubias\ corrections. 

\section{Simulating the Mass Step}
\label{sec:simulations}

The low $\gamma$-value observed for the DES3YR and DES-SN samples when using a 5D \mubias\ correction compared to a 1D \mubias\ correction is a result of a correlation between ${x_1}_\textrm{bias}$ and \mstellar. This correlation is likely inferred from the correlation between ${x_1}$ and \mstellar\ (\autoref{fig:galprop_v_snprop}). The simulated \mubias\ corrections used in existing cosmological analyses, e.g. \citetalias{Brout2019a}, do not include correlations between SN and host, so we now turn to simulating DES-SN with correlations between \mstellar\ and ($x_1$,$c$) to see if we can predict a correlation between ${x_1}_\textrm{bias}$ and \mstellar\ and to evaluate the impact this has on the measured value of $\gamma$. 

In \S\ref{subsec:sim_general_input} we outline the \textsc{snana} methodology used to simulate DES-SN while in \S\ref{subsec:sim_mstellar_input} we describe the procedure used to generate galaxy libraries, that match the characteristics of the DES-SN sample. In \S\ref{subsec:sim_x1c_input} we use a near complete sample of cosmological SNe~Ia drawn from the SDSS and SNLS samples to produce simulated SNe with intrinsic correlations between SN and \mstellar. Having simulated large realistic representations of the DES-SN sample we show the consistency in light-curve properties between our simulated samples and DES-SN in \S\ref{subsec:simoutput_properties}. Finally, in \S\ref{subsec:simoutput_massstep} we discuss the effect that correlations between \mstellar\ and ($x_1$,$c$) have on the inferred mass step for simulated samples and compare these results to those observed in the DES3YR dataset. From analysing our simulated samples, we find $\Delta \gamma = 0.011$\,mag and a systematic offset of $-0.009$\,mag for a 5D \mubias\ correction relative to the simulated value. 

\subsection{Simulations of DES-SN}
\label{subsec:sim_general_input}

Simulations of the DES-SN sample are performed using the \lq SuperNova ANAlysis\rq\ (\textsc{snana}) software package \citep{Kessler2009}.  The simulation inputs include a rest-frame SALT-II SED model \citep{Guy2010,Betoule2014}, an intrinsic scatter model \citep{Guy2010,Kessler2013}, SNIa population parameters \citep{Scolnic2016}, the volumetric rate of SNIa and its evolution with redshift \citep{Perrett2010}, a library of survey observations with measured observing parameters (sky noise, PSF, zero point), DECam filter transmission curves and a model of the DES detection and spectroscopic follow-up efficiency \citep[Figure 4]{Kessler2019}. The application of the \textsc{snana} methodology to simulating the DES3YR sample is described in detail in \citet{Kessler2019}, while a detailed analysis of the effect that systematic uncertainties have on the resulting cosmological constraints is given in \citetalias{Brout2019a}. To achieve statistical uncertainties of $<0.001$\,mag on $\gamma$, we simulated samples of $\sim$250,000 events after light-curve cuts.

\subsection{Producing realistic simulations}

\subsubsection{Estimates of stellar mass}
\label{subsec:sim_mstellar_input}

Host galaxy information is imprinted in \textsc{snana} simulations using a host galaxy library (HOSTLIB) where each simulated SN is associated with a random galaxy with consistent redshift. The HOSTLIB for DES-SN subset is generated from a catalogue of 380,000 galaxies derived from the DES-SV data as described in \citep{Gupta2016}. Each HOSTLIB galaxy contains information on the coordinates, heliocentric redshift, observer-frame magnitudes and S\'ersic profile components. To include the effect of a mass step and host galaxy correlations in our DES-SN simulations, we estimate the mass for each HOSTLIB galaxy using the methodology described in \S\ref{subsubsec:mass_estimates}. To test the effect that our galaxy sample has on our conclusions we also use a HOSTLIB generated from the DES SVA1-GOLD catalogue \citep{Rykoff2016}. This catalogue only includes objects with spectroscopic redshifts, and thus is significantly smaller (14,000 entries compared to 380,000).

The HOSTLIBs described above represent a complete sample of galaxies as determined from DES data. As SNe~Ia preferentially occur in low \mstellar\ galaxies compared to the overall galaxy population \citep{Smith2012}, we weight our HOSTLIB galaxies to match the distribution of \mstellar\ observed for SN~Ia hosts. To generate this mass function we require an unbiased, near complete sample of SN~Ia hosts.

As part of the real-time survey operations, DES preferentially targeted SN-like events in low luminosity environments \citep{DAndrea2018}, potentially biasing the DES-SN sample with respect to host galaxy properties. In contrast, the SDSS and SNLS surveys spectroscopically confirmed SNe Ia using targeting programs principally agnostic to local environment. Therefore, to compile a near complete sample of SNe Ia hosts we combine the SDSS and SNLS samples \citep{Betoule2014}, with redshift limits of $z=0.25$ for the SDSS sample and $z=0.70$ for the SNLS sample to ensure that each subsample of SNe~Ia is spectroscopically complete \citep{Perrett2010,Sullivan2010,Sako2018}. As anticipated, this sample of 417 SN~Ia hosts (denoted `SDSS+SNLS') closely resembles the DES-SN sample for high mass events, but shows fewer events in low \mstellar\ environments, with a mean stellar mass of 9.74 compared to 9.70 for DES-SN. To generate a galaxy mass function representative of SN~Ia hosts, we determine the cumulative distribution function (CDF) of the `SDSS+SNLS' host masses in bins of $\logmstellar$ with width 0.25, and draw galaxies from our HOSTLIB to match this. 

\subsubsection{Including intrinsic correlations between SN and host} 
\label{subsec:sim_x1c_input}

As shown in \S\ref{subsec:x1c_v_mass} and \autoref{fig:galprop_v_snprop}, the light-curve width of an SN~Ia is correlated with the \mstellar\ of its host galaxy, and from \S\ref{subsec:massstep_systematics}, this correlation affects the inferred \mubias\ correction which drives the low best-fit value of $\gamma$ for the DES-SN sample. Here we attempt to predict this effect in simulations by introducing a correlation between \mstellar\ and ($x_1$, $c$) in our host galaxy library. 

\begin{figure}
\centering
\includegraphics[width=0.45\textwidth]{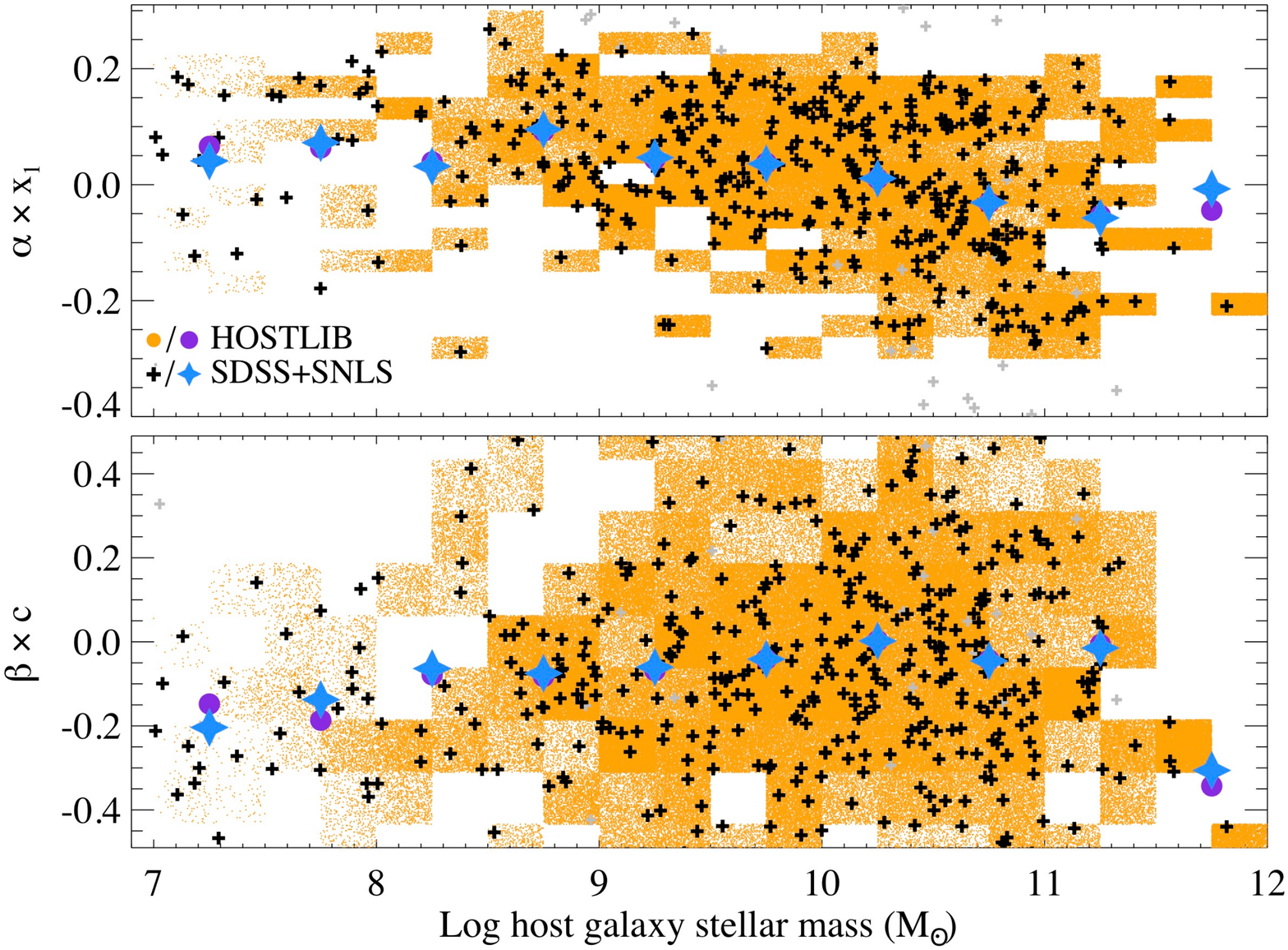}
\caption{\textbf{Simulation Input}: \textit{Top}: $\mstellar$ vs. $\alpha \times x_1$ (light-curve width) for our HOSTLIB (shown in orange) compared to a combined SDSS+SNLS sample (shown in black). For a given \mstellar, the corresponding $x_1$ is determined from the distribution of the SDSS+SNLS sample in that mass bin. The mean value of $x_1$ in bins of \mstellar\ for the HOSTLIB and SDSS+SNLS samples are plotted as violet circles and blue diamonds, respectively. \textit{Bottom}: Same as above, only for as a function of light-curve colour ($\beta \times c$).
\label{fig:sim_mass_sn_correlations}}
\end{figure}

With the DES-SN sample likely biased with respect to \mstellar\ \citep{DAndrea2018}, to do this, we use instead the near-complete SDSS+SNLS sample of SN~Ia hosts (as described in \S\ref{subsec:sim_mstellar_input}). For each galaxy in the HOSTLIB, with given stellar mass, we draw a random value of $x_1$ and $c$ from the corresponding CDF in bins of $\logmstellar$ with width 0.25. To account for our use of measured values of $x_1, c$, which probe the underlying distribution of $x_1, c$ only after the inclusion of intrinsic scatter and measurement uncertainty \citep{March2011}, we exclude events that lie in regions that contribute less than $10\%$ of the total probability. This cut, predominantly removes SNe~Ia with $x_1<-2.0$ and $x_1>2.0$. The resulting correlation between \mstellar\ and $x_1$, $c$ for our HOSTLIB is shown in \autoref{fig:sim_mass_sn_correlations} (plotted as \mstellar\ vs. the standardized contribution to $\mu$: $\alpha\times x_1$ and $\beta\times c$) where each shaded region is scaled based on the number of events contained within it. The SDSS+SNLS sample itself is overplotted for comparison. As anticipated, a correlation between \mstellar\ and $x_1$ is observed, with SNe~Ia with $x_1<0$ preferentially found in high mass hosts. There is some evidence of a reduced scatter in $c$ for low mass galaxies ($\mstellar<9$), which preferentially host SN with $c<0$. 

Each galaxy in our HOSTLIB now has an estimate of \mstellar, $x_1$ and $c$, with \mstellar\ correlated with ($x_1$, $c$) based on the SDSS+SNLS SN~Ia sample. To simulate DES-SN, we use the HOSTLIB $x_1$ and $c$ values instead of generating values for each simulated event from a parent population \citep[see][for details]{Scolnic2016}.

To determine how the value of $\gamma$ is affected when correlations between \mstellar\ and ($x_1$, $c$) are introduced, for comparison we also simulate DES-SN with no underlying correlations. Here, we draw a value of $x_1$ and $c$ from the distributions described in the high-$z$ row from Table 1 of \citet{Scolnic2016}, matching the analysis of \citetalias{Brout2019a} and \mstellar\ from our fiducial HOSTLIB.  

To ensure consistency between the underlying distributions of $x_1$ and $c$ between our simulated samples and the simulations used to calculate \mubias, denoted `\texttt{BIASCOR}' samples, we generate our own \texttt{BIASCOR} simulations of 300,000 SNe~Ia self consistently from each HOSTLIB to ensure that \mubias\ is determined correctly. Finally, we include a mass step in our analysis, by enforcing an absolute magnitude shift of $\gamma_\textrm{sim}=0.05$\,mag between SN in high ($\logmstellar>10$) and low ($\logmstellar<10$) mass galaxies in both our simulations with intrinsic correlations and correlation-free simulations. To test the consistency of our results to the value of $\gamma_\textrm{sim}$, we also produce both correlated and uncorrelated simulated samples with no mass step, i.e. $\gamma_\textrm{sim}=0$\,mag.

In summary, we have simulated two samples, with a mass step of $\gamma=0.05$\,mag. One `correlated' sample includes a correlation between \mstellar\ and ($x_1$, $c$), while for our other `uncorrelated' sample \mstellar\ and ($x_1$, $c$) are independent. Two more simulations, with and without correlations but with $\gamma_\textrm{sim}=0$\,mag, completes our simulation set. In all cases, the mass step is independent of of the underlying correlation between \mstellar\ and ($x_1$, $c$), and thus an unbiased estimator of distance should recover the simulated value of $\gamma$ for all simulated samples.

\subsection{Comparison to data}
\label{subsec:sim_data_comp}

\subsubsection{Population parameters}
\label{subsec:simoutput_properties}

\begin{figure}
\centering
\includegraphics[width=0.45\textwidth]{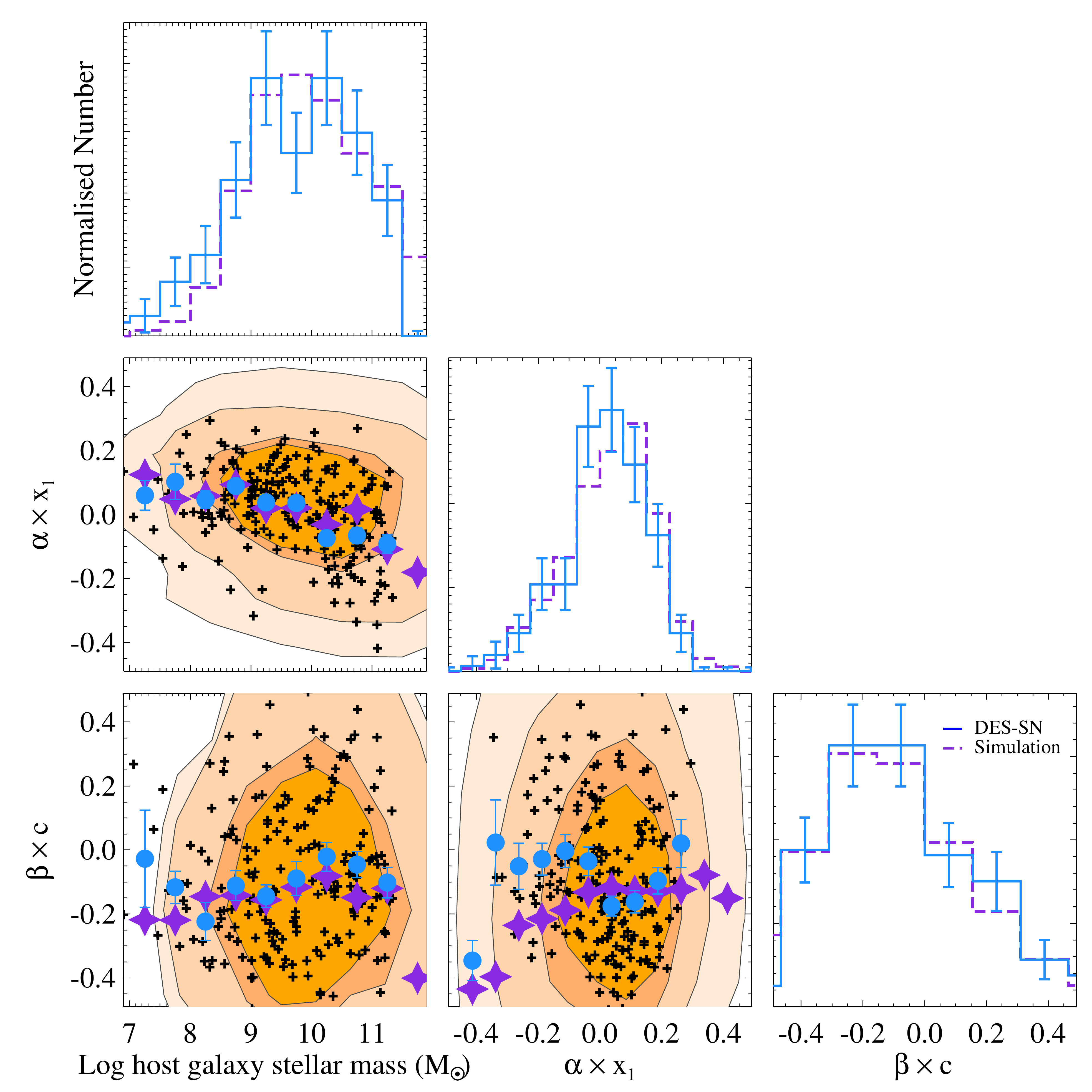}
\caption{\textbf{Simulation Output}: Corner plot showing the distributions of \mstellar, $x_1$ and $c$ for the DES-SN sample (in blue) compared to our simulated sample that includes correlations between \mstellar\ and ($x_1$, $c$) (shown as violet dashed histograms). The mean value of $x_1$ and $c$ as a function of \mstellar\ for the simulated sample is shown as violet crosses with the DES-SN sample shown as blue closed circles. Contours highlighting the area enclosed by 99.7, 95.5, 68.2 and 50\% of the simulated sample are shown in orange. 
\label{fig:sim_cornerplot}}
\end{figure}

After applying selection requirements and light curve fitting to the simulated DES-SN sample, \autoref{fig:sim_cornerplot} shows the distributions of \mstellar, $x_1$ and $c$ for our DES-SN simulation, with intrinsic correlations, of 250,000 SNIa compared to the DES-SN dataset. As anticipated, we observe a strong dependence between \mstellar\ and $x_1$ matching that observed from the data (\autoref{fig:sim_mass_sn_correlations}) and that from the SDSS+SNLS sample input into the simulation (\autoref{fig:galprop_v_snprop}). The dispersion in $\beta\times c$ is larger than that observed for $\alpha\times x_1$, but with limited evidence of a correlation between \mstellar\ and $c$, consistent with that observed for DES-SN (\autoref{fig:galprop_v_snprop}). With the simulated values of $x_1$ and $c$ being independently drawn from the SDSS+SNLS sample (see \S\ref{subsec:sim_x1c_input}), no significant correlation is observed between these two parameters. The resulting distributions of \mstellar, $x_1$ and $c$ are closely matched to the DES-SN sample, with the simulated sample marginally favouring SNe~Ia in lower mass galaxies compared to the DES-SN sample. This is driven by the lack of high mass galaxies in our HOSTLIB as described in \S\ref{subsec:sim_mstellar_input}. As shown in \autoref{fig:sim_mass_sn_correlations}, SN found in these environments preferentially exhibit low values of $x_1$ and marginally higher values of $c$. 

\subsubsection{The inferred distances: \mubias}
\label{subsec:sim_mubias_diff}

\begin{figure}
\centering
\includegraphics[width=0.48\textwidth]{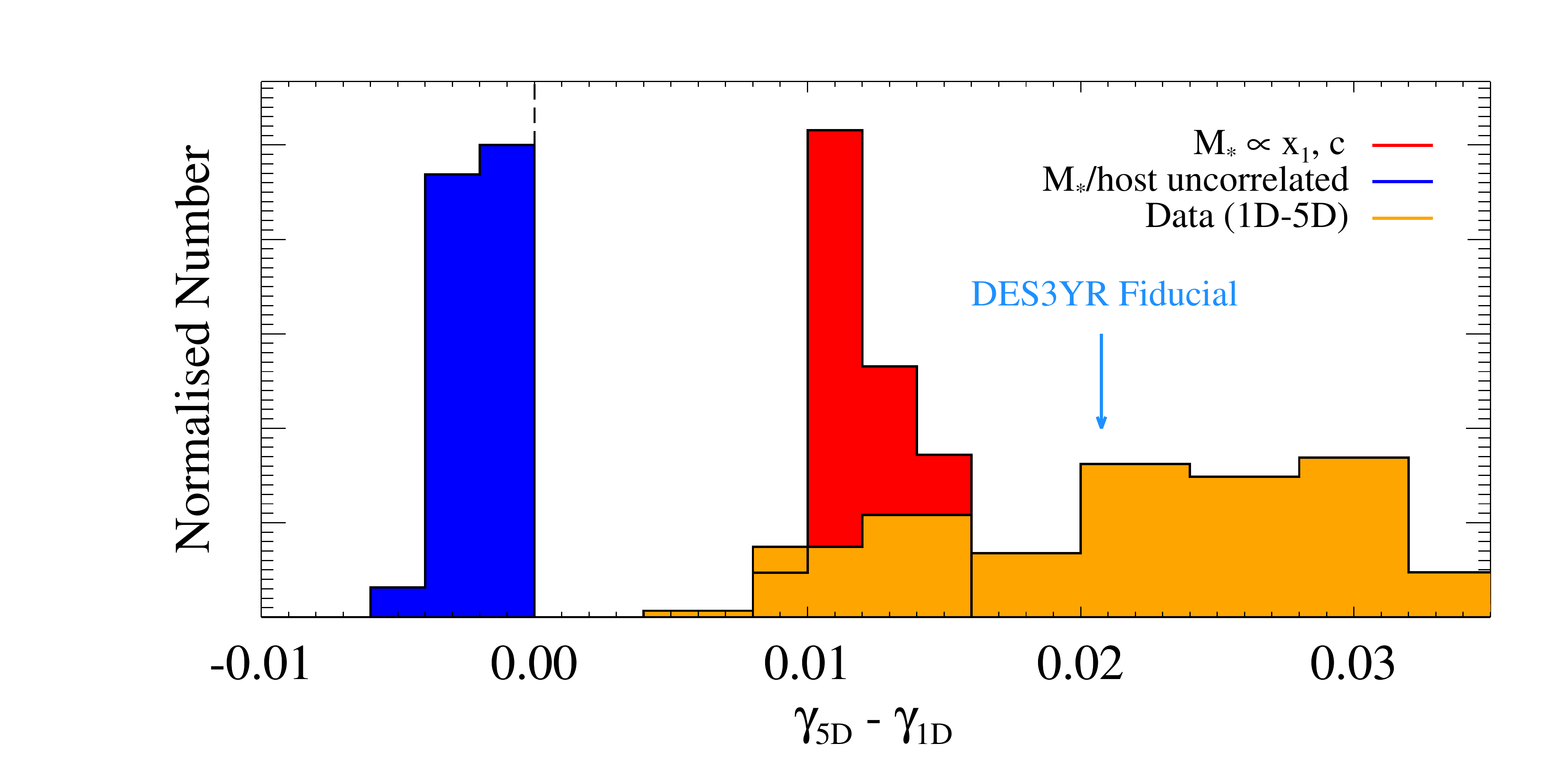}
\caption{Histograms showing the difference in the best-fit value of $\gamma$ for simulated samples when a 5D \mubias\ correction is used compared to the results assuming a 1D \mubias\ correction for differing assumptions on the relationship between \mstellar\ and ($x_1$,$c$). The results assuming no correlation between \mstellar\ and ($x_1$,$c$) are shown in blue, while those where a correlation is enforced using the prescription described in \S\ref{subsec:sim_x1c_input} are shown in red. Each histogram shows the results of all systematic tests (input HOSTLIB, \texttt{BIASCOR} simulation used to estimate \mubias, sample weighting and mass step parametrisation) as outlined in \autoref{tab:massstep_sim_systematics}. An input correlation between \mstellar\ and ($x_1$,$c$) suppresses the value of $\gamma$ for the 5D \mubias\ by 0.012\,mag compared to a 1D \mubias\ correction, while considering no correlation increases the value by 0.002\,mag. Also shown in yellow is the difference between $\gamma$ when assuming a 5D \mubias\ correction compared to a 1D \mubias\ correction for the DES-SN sample when various systematic tests (as highlighted in \S\ref{subsec:massstep_systematics}) are considered. This distribution has a mean of 0.028\,mag. The fiducial result for the DES3YR analysis is highlighted with a blue arrow.
\label{fig:sim_gamma_diff}}
\end{figure}

\begin{table}
{\centering
\caption{The difference between $\gamma_\textrm{1D}$ and $\gamma_\textrm{5D}$ for our simulated samples when intrinsic correlations between \mstellar\ and ($x_1$,$c$) are and are not considered in our simulated samples. Also shown is the result for DES-SN}
\label{tab:massstep_sim_gammadiffs}
\begin{tabular}{lcc}
  \hline
  Sample & SN / host & $\gamma_\textrm{1D}-\gamma_\textrm{5D}$\\
& correlations & (mag) \\

   \hline
\textbf{DES-SN} & --- & $\mathbf{0.026\pm0.009}$ \\
\textbf{Correlated simulation} & {$\mathbf{x_1, c}$} & $\mathbf{0.011\pm0.001}$ \\
Uncorrelated simulation & None & $-0.002\pm0.001$\\
\hline
\end{tabular}\\
}
\flushleft
\end{table}

\autoref{tab:massstep_sim_gammadiffs} shows the difference between $\gamma_\textrm{1D}$ and $\gamma_\textrm{5D}$ for our simulated samples of DES-SN. When a correlation between \mstellar\ and ($x_1$, $c$) is included in our simulated samples (as described in \S\ref{subsec:sim_x1c_input}) we find $\gamma_\textrm{1D}-\gamma_\textrm{5D} = 0.012\pm 0.001$\,\textrm{mag}, with $\gamma_\textrm{1D}-\gamma_\textrm{5D} = -0.001\pm0.001$\,\textrm{mag} for the case of no intrinsic correlations. 

\autoref{fig:sim_gamma_diff} and \autoref{tab:massstep_sim_gammadiffs} compares these results to DES-SN. To test the robustness of our results, \autoref{fig:sim_gamma_diff} shows the distribution of $\gamma_\textrm{1D}-\gamma_\textrm{5D}$ from varying our assumptions on the underlying \mstellar\ distribution, including outlying values of ($x_1$,$c$), and using different parameterisations of the mass step. For a simulated sample without correlations between SN and host, averaging over all systematic tests, we find $\gamma_\textrm{1D}-\gamma_\textrm{5D} = -0.002\pm0.001$\,\textrm{mag}. When a correlation between \mstellar, $x_1$ and $c$ is included in our simulated sample, averaging over all systematic tests, we find a mean offset of $0.011\pm0.001$\,\textrm{mag}. Given uncertainties on the true relationship between SN and host, this is well matched with the offset found in \S\ref{subsec:massstep_systematics} for DES-SN of $0.026\pm0.009$\,\textrm{mag} (\autoref{eq:data_result}), suggesting that the correlation between \mstellar\ and ($x_1$,$c$) is a significant source of the low $\gamma$-value measured for DES-SN using a 5D \mubias\ correction. 

\subsection{Biases in the recovered value of $\gamma$} 
\label{subsec:simoutput_massstep}

In \S\ref{subsec:sim_mubias_diff} we found a difference between $\gamma_\textrm{5D}$ and $\gamma_\textrm{1D}$ of 0.012\,mag for our simulated samples when intrinsic correlations between \mstellar\ and ($x_1$, $c$) are included in our simulations. Given that our simulated samples include a mass step that is independent of this correlation, this points to a bias in the recovered value of $\gamma$ for either, or both, analyses. 

\autoref{tab:massstep_sim_results} shows how the fitted value of $\gamma$ for our DES-SN simulations compares to the simulated value, for both 1D and 5D \mubias\ corrections. When an intrinsic correlation between \mstellar\ and ($x_1$,$c$) is included in our simulated samples, the value of $\gamma$ assuming a 5D \mubias\ correction is reduced relative to the simulated value of $\gamma$, with an offset of $\Delta \gamma_\textrm{5D} = \gamma_\textrm{5D;fit} - \gamma_\textrm{sim} = -0.012\pm0.001$\,mag, compared to $\Delta \gamma_\textrm{1D} = \gamma_\textrm{1D;fit} - \gamma_\textrm{sim} = 0.000\pm0.001$\,mag. 

\begin{figure}
\centering
\includegraphics[width=0.45\textwidth]{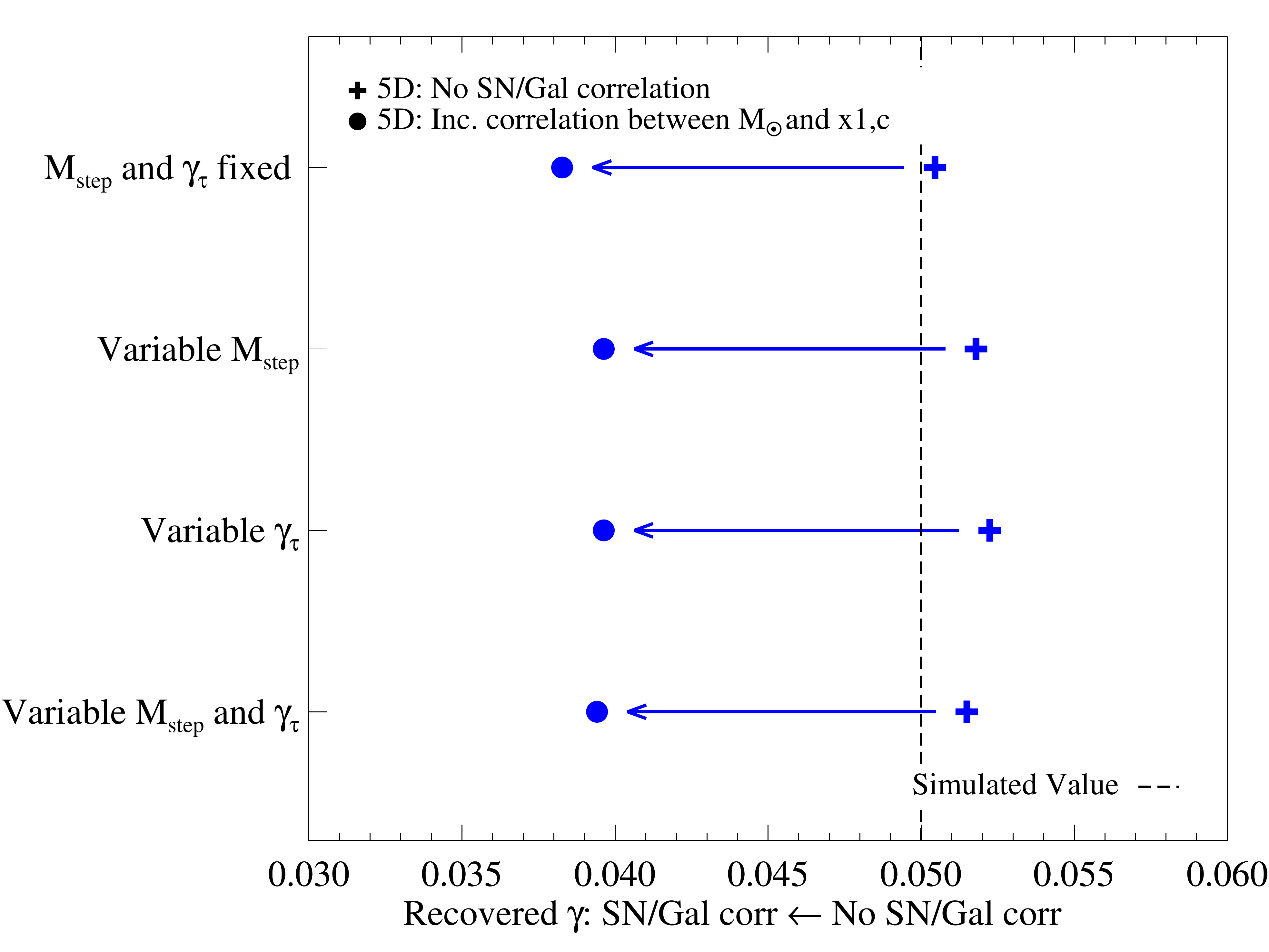}
\caption{\textbf{Simulation Output}: The best-fit value of $\gamma_\textrm{fit}$ for our simulated DES-SN samples considering different systematic uncertainties when assuming a 5D \mubias\ correction. For each entry, the right-hand value (plotted as a plus symbol) indicates the value when our simulated sample does not contain a correlation between \mstellar\ and ($x_1$, $c$), while the left-hand entry (plotted as a filled circle) is the result when a correlation is enforced using the prescription described in \S\ref{subsec:sim_x1c_input}. The simulated value of $\gamma_\textrm{sim}=0.05$ is shown as dashed vertical line. Entries are for our fiducial HOSTLIB, derived from DES-SV data, matched to the mass distribution. 
Four results with differing assumptions about the mass step parameterisation are shown. In all cases, $\gamma_\textrm{fit}$ is reduced by $\sim$0.10\,mag when correlations between \mstellar\ and $x_1,c$ are included. As shown in \autoref{tab:massstep_sim_results} and \autoref{tab:massstep_sim_systematics}, using alternative galaxy catalogues and weighting schemes results in consistent results. 
\label{fig:sim_gamma_dist}}
\end{figure}

\autoref{tab:massstep_sim_systematics} and \autoref{fig:sim_gamma_dist} show the robustness of this result by varying our assumptions on the source and underlying \mstellar\ distribution, varying the input value of $\gamma_\textrm{sim}$, including outlying values of ($x_1$,$c$), and using different parameterisations of the mass step. We find a average offset of $\Delta \gamma_\textrm{5D} = -0.0093\pm0.0013$\,mag (where the uncertainty is derived from the scatter of the results) inconsistent with zero at 6.9$\sigma$, compared to $\Delta \gamma_\textrm{1D} = 0.0019\pm0.0011$\,mag (1.8$\sigma$). When we include no mass step in our simulations (i.e. $\gamma_\textrm{sim}=0.0$\,mag), but leave $\gamma$ as a free parameter in the fit, we find a best-fit value of $\gamma_\textrm{5D}=-0.008\pm0.001$\,mag, and $\gamma_\textrm{1D}=0.003\pm0.001$\,mag, indicating that the offset in $\gamma_\textrm{5D}$ is independent of the value of $\gamma_\textrm{sim}$.

When we consider the case without intrinsic correlations between \mstellar\ and ($x_1$,$c$), the measured value of $\gamma$ is consistent with the simulated value for both 5D and 1D \mubias\ corrections. For our fidicual analysis we find best-fit values of $\Delta \gamma_\textrm{5D} = 0.000\pm0.001$\,mag and $\Delta \gamma_\textrm{1D} = -0.001\pm0.001$\,mag (\autoref{tab:massstep_sim_results}). Averaged over all systematic tests, we recover $\Delta \gamma_\textrm{5D} = 0.0024\pm0.0012$\,mag and $\Delta \gamma_\textrm{1D} = 0.0004\pm0.0007$\,mag. The 5D \mubias\ correction is inconsistent with the simulated value of $\gamma$ at 2.0$\sigma$, compared to 0.6$\sigma$ for the 1D \mubias\ correction. For our simulations with no mass step (i.e. $\gamma_\textrm{sim}=0.0$\,mag), we find a best-fit value of $\gamma_\textrm{5D}=0.003\pm0.001$\,mag and $\gamma_\textrm{1D}=0.001\pm0.001$\,mag, showing that our results are consistent independent of the input value of $\gamma_\textrm{sim}$.

From \autoref{tab:massstep_sim_results} and \autoref{tab:massstep_sim_systematics}, there is some evidence that the reduced value of $\gamma_\textrm{5D}$ is offset by an increase in the value of $\beta$, but averaging over all possible combinations, we find no evidence of an offset in the value of $\beta$, with $\Delta\beta_\textrm{5D} = 0.010 \pm 0.004$ (2.6$\sigma$) and $\Delta\beta_\textrm{1D} = 0.034 \pm 0.018$ (1.9$\sigma$) when intrinsic correlations between \mstellar, $x_1$ and $c$ are included in our simulated samples. 

\subsubsection{Implications for 5D \mubias\ corrections} 
\label{subsec:gamma_bias_consequences}

From our simulated samples, when intrinsic correlations between \mstellar\ and ($x_1$,$c$) are included, a 5D \mubias\ correction recovers a reduced value of $\gamma$ relative to the simulated value, with an offet of 0.009\,mag. To test for the source of this bias, we search for correlations between \mstellar\ and (${m_B}_\textrm{bias}$, $\Delta {x_1}_\textrm{bias}$, $c_\textrm{bias}$). We find strong evidence of correlation between \mstellar\ and ${x_1}_\textrm{bias}$, with a difference of $\Delta {x_1}_\textrm{bias}$ of $0.052\pm0.001$\,mag between SNe in low and high-mass galaxies. This is consistent with $0.062\pm0.028$\,mag measured for DES-SN in \S\ref{subsec:massstep_systematic_mubias}. 
We find offsets of 0.0005$\pm$0.0001\,mag and and -0.0009$\pm$0.0001\,mag between SNe in low and high-mass galaxies for ${m_B}_\textrm{bias}$ and $c_\textrm{bias}$, respectively. These values are consistent with those observed for DES-SN. When correlations between \mstellar\ and ($x_1$, $c$) are not included in our simulations, we find no evidence of a correlation between \mstellar\ and ${x_1}_\textrm{bias}$, ${m_B}_\textrm{bias}$ or $c_\textrm{bias}$, as expected. 

For our simulated samples, $\gamma$ is independent of ($x_1$, $c$). However, when intrinsic correlations between \mstellar\ and ($x_1$,$c$) are included in our simulations, a 5D \mubias\ correction misinterprets $\gamma$ as being caused by these correlations, subsuming 0.009\,mag of $\gamma$ into ${x_1}_\textrm{bias}$. This result suggests that a fraction of the decrease in $\gamma$ seen for the DES-SN sample, when using a 5D \mubias\ correction compared to a 1D \mubias\ correction has been incorrectly attributed to be an uncorrected contribution to $x_1$. This is further confirmed by the DES-SN sample, where there is no evidence of differing nuisance parameters for high stretch SNe~Ia compared to their low stretch counterparts. Fixing $\gamma=0$ and splitting the DES-SN sample in to high and low bins of $x_1$, we measure

\begin{align*}
    x_1>0 : \alpha,\beta,M_0=& \,0.140\pm0.028,\,3.11\pm0.18,\,-19.348\pm0.014
\end{align*}
and
\begin{align*}
     x_1<0 : \alpha,\beta,M_0=& \,0.155\pm0.023,\,2.88\pm0.19,\,-19.369\pm0.016.
\end{align*}

These values are consistent at $<1\sigma$, suggesting that high stretch SNe~Ia follow the same correction as low stretch SNe~Ia, when no correction for \mstellar\ is allowed. We find some evidence of a difference in the distribution of $c$ for SNe~Ia with $x_1<0$ compared to $x_1>0$, with mean $c=-0.010\pm0.009$\,mag for SNe Ia with $x_1<0$ compared to $c=-0.046\pm0.006$\,mag for those with $x_1>0$. For our simulated sample we find $c=-0.0425\pm 0.0003$\,mag for SNe~Ia with $x_1<0.$ and $c=-0.0358\pm 0.0002$\,mag $x_1>0.$, consistent for low-stretch SNe~Ia, but inconsistent for high-stretch events at 2.97$\sigma$, suggesting that this is likely an uncorrected for selection effect. From this test, there is little evidence from the DES-SN sample that high and low-stretch SNe~Ia follow different standardisation relationships. As a consequence, there is no evidence that an additional $x_1$ dependent correction, beyond $\alpha$, is required for the DES-SN sample, as inferred by the 5D \mubias, which considers a fraction of the mass step to be an uncorrected contribution to $x_1$. Overall, we find that the value of $\gamma$ found using a 5D \mubias\ correction is reduced relative to the true, underlying value by $\sim$0.01\,mag. 

Offsets in $\Delta\gamma$ have been found for all cosmological analyses that use a 5D \mubias\ correction, with $\Delta\gamma=0.026\pm0.009$\,mag for DES-SN and $\Delta\gamma=0.025$\,mag for the PS1 sample \citep{Scolnic2018}. With this offset likely caused by the correlation between SN and host galaxy parameters, this suggests the need for a 7D \mubias\ correction, with additional terms linked to $\gamma$ and \mstellar. The cosmological implications of this offset, while subdominant to the current statistical and systematic error budget from SNe~Ia, will likely be important for future experiments, such as the Large Synoptic Survey Telescope (LSST). The ramifications of this offset on the equation-of-state of dark energy will be explored in an upcoming analysis of the DES5YR sample of $\sim$2,000 SNe~Ia. 

\begin{table*}
{\centering
\caption{Measured nuisance parameters from simulations for the DES-SN sample when intrinsic correlations between $x_1$, $c$ and \mstellar\ are and are not included. The fiducial results from this study are highlighted in bold}
\label{tab:massstep_sim_results}
\begin{tabular}{cccccc}
  \hline
   SN / host \footnotemark[1] & BiasCor & $\sigma_\textrm{int}$ & $\Delta\alpha$ \footnotemark[2] & $\Delta\beta$ \footnotemark[2] & $\Delta\gamma$ \footnotemark[2] \\
 correlations & &  & & & (mag) \\
   \hline
 $\mathbf{x_1, c}$ & \textbf{5D} & \textbf{0.100} & \textbf{-0.001} & \textbf{0.009} & \textbf{-0.012}\\
 $\mathbf{x_1, c}$ & \textbf{1D} & \textbf{0.098} & \textbf{0.001} & \textbf{0.032} & \textbf{0.000}\\
 None & 5D & 0.099 & 0.001 & 0.006 & 0.000\\
 None & 1D & 0.098 & 0.002 & 0.040 & -0.001\\
\hline
\end{tabular}\\
}
\flushleft
\footnotetext{1}{$^{1}$ For each simulated event, the SN parameters are either linked to \mstellar\ through the SDSS+SNLS sample or drawn from a parent population following the methodology of \citet{Scolnic2016}. See \S\ref{subsec:sim_x1c_input} for details.}\\
\footnotetext{2}{$^{2}$ $\Delta x = x_\textrm{fit} - x_\textrm{sim}$ where ($\alpha_\textrm{sim},\beta_\textrm{sim}\gamma_\textrm{sim}$)=(0.15, 3.1, 0.05).}\\
\end{table*}

\begin{table*}
{\centering
\caption{Systematic tests on the best-fit value of $\gamma$ from simulated samples}
\label{tab:massstep_sim_systematics}
\begin{tabular}{lcccllcccccc}
  \hline
HOSTLIB & SN / host & BiasCor & $\gamma_\textrm{sim}$ & Mass & Outlier & $\sigma_\textrm{int}$ & $M_\textrm{step}$ \footnotemark[4] & $\gamma_\tau$ \footnotemark[4] & $\Delta\alpha$ \footnotemark[5] & $\Delta\beta$ \footnotemark[5] & $\Delta\gamma$ \footnotemark[5] \\
 & correlations \footnotemark[1] & & & distribution \footnotemark[2] & Removal \footnotemark[3] & & & & & & (mag) \\
\hline
DES-SV & $x_1, c$ & 5D & 0.05 & SDSS+SNLS & Trimmed & 0.100 & Fixed & Fixed & -0.001 & 0.009 & -0.012 \\
DES-SV & $x_1, c$ & 5D & 0.05 & Unweighted & Trimmed & 0.100 & Fixed & Fixed & -0.001 & 0.011 & -0.009 \\
DES-SV & $x_1, c$ & 5D & 0.05 & SDSS+SNLS & Full & 0.100 & Fixed & Fixed & -0.001 & 0.016 & -0.009 \\
DES-SV & $x_1, c$ & 5D & 0.05 & SDSS+SNLS & Trimmed & 0.100 & 9.998 & Fixed & -0.001 & 0.011 & -0.010 \\
DES-SV & $x_1, c$ & 5D & 0.05 & SDSS+SNLS & Trimmed & 0.100 & Fixed & 0.03 & -0.001 & 0.009 & -0.010 \\
DES-SV & $x_1, c$ & 5D & 0.00 & SDSS+SNLS & Trimmed & 0.100 & Fixed & Fixed & -0.001 & 0.010 & -0.008 \\
SVA1-GOLD & $x_1, c$ & 5D & 0.05 & SDSS+SNLS & Trimmed & 0.101 & Fixed & Fixed & -0.001 & 0.002 & -0.010 \\
\hline
\end{tabular}\\
}
\flushleft
\footnotetext{1}{$^{1}$ For each simulated event, the SN parameters are either linked to \mstellar, through the SDSS+SNLS sample or drawn from a parent population following the methodology of \citet{Scolnic2016}. See \S\ref{subsec:sim_x1c_input} for details.}\\
\footnotetext{2}{$^{2}$ For each simulated event, whether or not the value of \mstellar\ is chosen randomly from the HOSTLIB file or from a weighted distribution determined from the pseudo complete SDSS+SNLS sample. See \S\ref{subsec:sim_mstellar_input} for details.}\\
\footnotetext{3}{$^{3}$ For each HOSTLIB, whether or not events in the SDSS+SNLS sample with $x_1<-2$ and $x_1>2$ are included when determining the correlation between \mstellar\ and ($x_1,c$). See \S\ref{subsec:sim_mstellar_input} for details.}\\
\footnotetext{4}{$^{4}$ Fixed to $\mstep=10.0$ and $\gamma_\tau=0.01$ in the fiducial analysis.}\\
\footnotetext{5}{$^{5}$ $\Delta x = x_\textrm{fit} - x_\textrm{sim}$ where ($\alpha_\textrm{sim},\beta_\textrm{sim}\gamma_\textrm{sim}$)=(0.15,3.1,0.05).}\\
\end{table*}

\section{Summary and Conclusions} 
\label{sec:conclusions}

In this paper we have presented photometric measurements and derived physical parameters for the host galaxies of the \NDES\ SNe Ia discovered by the Dark Energy Survey (DES) Supernova Program (DES-SN) and used in the first DES-SN cosmological analysis (DES3YR). While host properties for the DES3YR analysis \citep{Brout2019a} were based on a relatively shallow catalogue (SVA1-GOLD), here we obtain host properties from deep stack photometry, based on all 5 years of DES-SN, fit to a library of SEDs to infer stellar masses and star-formation rates, we have shown the following: 

\begin{itemize}
\item The distribution of \mstellar\ and sSFR for the DES-SN sample is consistent with that derived from the SNLS survey, which spans a similar redshift range. The DES-SN sample has a higher fraction of low \mstellar\ galaxies than that determined by the intermediate redshift, SDSS and PS1 samples. The values derived for \mstellar\ are robust to the templates, IMF and photometric catalogue used. 
\item We observe a correlation between \mstellar\ and SN Ia light-curve width ($x_1$), as found previously for literature samples, but there is no evidence of a correlation with SN colour ($c$).
\item The correlation between \mstellar\ and Hubble residuals ($\Delta\mu$), parameterised as a `mass step' (\mstep) is observed at 2.4 and 2.1$\sigma$ for the DES3YR and DES-SN samples, respectively. The best-fit value of the strength of the mass step, $\gamma=0.040\pm0.019$\,mag is consistent with results derived for the Pantheon and PS1 analyses and robust to the methodology and underlying assumptions used to derive \mstellar. The value found here is larger than $\gamma=0.009\pm0.018$\,mag found by \citetalias{Brout2019a}. This difference is not due to host galaxy misassociation, but a combination of improved photometric measurements and an updated SED library. 
\item We find a dependence on the value of $\gamma$ based upon the methodology used to determine distances to each event. Within the BBC framework, we find that $\gamma$ is reduced by 0.026\,mag when using a 5D ($z,x_1,c,\alpha,\beta$) \mubias\ correction compared to a 1D  ($z$-only) correction. This conclusion is consistent across all other systematics considered. 
\item We find a strong correlation between \mstellar\ and the BBC $x_1$ component of the \mubias\ correction, suggesting that the BBC framework infers that some fraction of the mass step is not due to \mstellar, but is an uncorrected contribution due to $x_1$. 
\item To test this, and search for biases in the recovered value of $\gamma$, we simulate the DES-SN sample, introducing realistic correlations between \mstellar\ and ($x_1$,$c$). These input correlations are independent of the mass step. When \mstellar\ is independent of $x_1,c$ both 1D and 5D \mubias\ corrections successfully recover the input value of $\gamma$. When correlations between \mstellar\ and $x_1,c$ are introduced, we recover: 
\begin{align*}
    [\gamma_\textrm{1D}-\gamma_\textrm{5D}]_\textrm{sim } &= 0.012\pm0.002\,\textrm{mag} \\
    [\gamma_\textrm{1D}-\gamma_\textrm{5D}]_\textrm{data} &= 0.026\pm0.009\,\textrm{mag}.
\end{align*}
\item For our simulated samples including intrinsic correlations, we find that the value of $\gamma_\textrm{5D}$ is reduced relative to the simulated value by 0.0093\,mag. There is no evidence of a bias in $\alpha$ or $\beta$ for either a 1D or 5D \mubias\ correction. This indicates that the value of $\gamma$ found for DES-SN using a 5D \mubias\ correction is systematically underestimated by $\sim$0.01\,mag.
\end{itemize}

While significant attention has been given to the methodology used to determine \mstellar\ for SN Ia hosts, it is clear that the methodology used to determine distances to SNe~Ia plays an important role in the inferred mass step. Given the strong dependence between \mstellar\ and $x_1$ for SNe~Ia, the use of a 5D \mubias\ correction dependent on $z, x_1, c, \alpha$ and $\beta$ can result in a systematic under-estimation of the relationship between SN~Ia luminosity and host galaxy properties. This may potentially result in biases when estimating the cosmological parameters, subdominant to the current statistical and systematic error budget from SNe~Ia, but likely be important for future experiments, such as the Large Synoptic Survey Telescope (LSST). The underlying correlation between \mstellar\ and ($x_1$,$c$) also suggests the need for additional terms in the \mubias\ correction, with linked to \mstellar\ and $\gamma$ to fully encapsulate SN~Ia selection effects. 

Given the potential evolution in the distribution of \mstellar\ with redshift \citep{Rigault2013}, it is critical to consider the underlying relationship between SN Ia luminosity and local environment when estimating the distance to individual SN Ia. Upcoming samples of thousands of SNe Ia, both in the local Universe with IFU spectroscopy \citep{Galbany2016} and at high redshift from samples such as DES, LSST and the \textit{Wide Field InfraRed Survey Telescope} \citep[\textit{WFIRST};][]{Hounsell2018}, will allow us to constrain the to intrinsic correlation between host and SN, and probe its evolution with redshift. This is key to understanding the source of \mstep\ and ensuring the inferred cosmological parameters from SNe~Ia surveys are unbiased in the era of precision cosmology.

\section{Acknowledgements}

We acknowledge support from EU/FP7-ERC grant 615929. L.G. was funded by the European Union's Horizon 2020 research and innovation programme under the Marie Sk\l{}odowska-Curie grant agreement No. 839090. 

The UCSC team is supported in part by NASA grant NNG17PX03C, NSF grants AST-1518052 and AST-1815935, the Gordon \& Betty Moore Foundation, the Heising-Simons Foundation, and by fellowships from the Alfred P.\ Sloan Foundation and the David and Lucile Packard Foundation to R.J.F.

This work was completed in part with resources provided by the University of Chicago Research Computing Center. This research used resources of the National Energy Research Scientific Computing Center (NERSC), a U.S. Department of Energy Office of Science User Facility operated under Contract No. DE-AC02-05CH11231.

Funding for the DES Projects has been provided by the U.S. Department of Energy, the U.S. National Science Foundation, the Ministry of Science and Education of Spain, the Science and Technology Facilities Council of the United Kingdom, the Higher Education Funding Council for England, the National Center for Supercomputing Applications at the University of Illinois at Urbana-Champaign, the Kavli Institute of Cosmological Physics at the University of Chicago, the Center for Cosmology and Astro-Particle Physics at the Ohio State University, the Mitchell Institute for Fundamental Physics and Astronomy at Texas A\&M University, Financiadora de Estudos e Projetos, Funda{\c c}{\~a}o Carlos Chagas Filho de Amparo {\`a} Pesquisa do Estado do Rio de Janeiro, Conselho Nacional de Desenvolvimento Cient{\'i}fico e Tecnol{\'o}gico and the Minist{\'e}rio da Ci{\^e}ncia, Tecnologia e Inova{\c c}{\~a}o, the Deutsche Forschungsgemeinschaft and the Collaborating Institutions in the Dark Energy Survey. 

The Collaborating Institutions are Argonne National Laboratory, the University of California at Santa Cruz, the University of Cambridge, Centro de Investigaciones Energ{\'e}ticas, Medioambientales y Tecnol{\'o}gicas-Madrid, the University of Chicago, University College London, the DES-Brazil Consortium, the University of Edinburgh, the Eidgen{\"o}ssische Technische Hochschule (ETH) Z{\"u}rich, Fermi National Accelerator Laboratory, the University of Illinois at Urbana-Champaign, the Institut de Ci{\`e}ncies de l'Espai (IEEC/CSIC), the Institut de F{\'i}sica d'Altes Energies, Lawrence Berkeley National Laboratory, the Ludwig-Maximilians Universit{\"a}t M{\"u}nchen and the associated Excellence Cluster Universe, the University of Michigan, the National Optical Astronomy Observatory, the University of Nottingham, The Ohio State University, the University of Pennsylvania, the University of Portsmouth, SLAC National Accelerator Laboratory, Stanford University, the University of Sussex, Texas A\&M University, and the OzDES Membership Consortium.

Based in part on observations at Cerro Tololo Inter-American Observatory, National Optical Astronomy Observatory, which is operated by the Association of Universities for Research in Astronomy (AURA) under a cooperative agreement with the National Science Foundation.

The DES data management system is supported by the National Science Foundation under Grant Numbers AST-1138766 and AST-1536171. The DES participants from Spanish institutions are partially supported by MINECO under grants AYA2015-71825, ESP2015-66861, FPA2015-68048, SEV-2016-0588, SEV-2016-0597, and MDM-2015-0509, some of which include ERDF funds from the European Union. IFAE is partially funded by the CERCA program of the Generalitat de Catalunya. Research leading to these results has received funding from the European Research Council under the European Union's Seventh Framework Program (FP7/2007-2013) including ERC grant agreements 240672, 291329, and 306478. We acknowledge support from the Brazilian Instituto Nacional de Ci\^encia e Tecnologia (INCT) e-Universe (CNPq grant 465376/2014-2).

This manuscript has been authored by Fermi Research Alliance, LLC under Contract No. DE-AC02-07CH11359 with the U.S. Department of Energy, Office of Science, Office of High Energy Physics.

\appendix
\section{Host galaxy association}
\label{app:dlr}

The \lq Directional Light Radius\rq\ (DLR) methodology  \citep{Sullivan2006,Smith2012,Gupta2016,Sako2018} used to define the host galaxy of each DES SN Ia defines the distance from a SN event (at $x_\mathrm{SN}$, $y_\mathrm{SN}$) to the centre of a potential host galaxy (at $x_\mathrm{gal}$, $y_\mathrm{gal}$) according to
\begin{equation}
\label{eqn:d_dlr}
d_\text{DLR} = \frac{\text{separation from SN to galaxy}}{\text{DLR}}
\end{equation}
where DLR is the elliptical radius of a galaxy in the direction of the SN. This is based on the
elliptical shape determined by \textsc{SExtractor}, defined by semi-major ($r_A$) and semi-minor ($r_B$) axes together with a position angle ($\theta$). The DLR is then given by
\begin{equation}
\label{eqn:dlr}
\mathrm{DLR}^2={C_{xx}}x_r^2 + {C_{yy}}y_r^2 + {C_{xy}}x_ry_r,
\end{equation}
where $x_r=x_{\mathrm{SN}}-x_{\mathrm{gal}}$, $y_r=y_{\mathrm{SN}}-y_{\mathrm{gal}}$, $C_{xx}=\cos^2(\theta)/r_A^2+\sin^2(\theta)/r_B^2$, $C_{yy}=\sin^2(\theta)/r_A^2+\cos^2(\theta)/r_B^2$, and $C_{xy}=2\cos(\theta)\sin(\theta)(1/r_A^2-1/r_B^2)$. 

In short, the DLR method normalises the separation between a SN and a candidate host galaxy by the size of that galaxy in the direction of the SN, and then selects the galaxy with the smallest $d_\text{DLR}$ as the true host. Following \citet{Gupta2016} and \citet{Sako2018}, we only consider galaxies with $d_\text{DLR}<7$ to be candidates for the true host.

\section{Author Affiliations}
\label{app:affiliations}

$^{1}$ School of Physics and Astronomy, University of Southampton,  Southampton, SO17 1BJ, UK\\
$^{2}$ Department of Astronomy and Astrophysics, University of Chicago, Chicago, IL 60637, USA\\
$^{3}$ Kavli Institute for Cosmological Physics, University of Chicago, Chicago, IL 60637, USA\\
$^{4}$ Department of Physics, Duke University Durham, NC 27708, USA\\
$^{5}$ Department of Physics and Astronomy, University of Pennsylvania, Philadelphia, PA 19104, USA\\
$^{6}$ NASA Einstein Fellow\\
$^{7}$ Department of Physics, Bryn Mawr College, Bryn Mawr, PA 19010, USA\\
$^{8}$ School of Mathematics and Physics, University of Queensland,  Brisbane, QLD 4072, Australia\\
$^{9}$ Santa Cruz Institute for Particle Physics, Santa Cruz, CA 95064, USA\\
$^{10}$ Institute of Cosmology and Gravitation, University of Portsmouth, Portsmouth, PO1 3FX, UK\\
$^{11}$ PITT PACC, Department of Physics and Astronomy, University of Pittsburgh, Pittsburgh, PA 15260, USA\\
$^{12}$ Lawrence Berkeley National Laboratory, 1 Cyclotron Road, Berkeley, CA 94720, USA\\
$^{13}$ The Research School of Astronomy and Astrophysics, Australian National University, ACT 2601, Australia\\
$^{14}$ ARC Centre of Excellence for All-Sky Astrophysics (CAASTRO), Canberra, Australia\\
$^{15}$ Department of Physics and Astronomy, University of North Georgia, Dahlonega, GA 30597, USA\\
$^{16}$ Universit\'e Clermont Auvergne, CNRS/IN2P3, LPC, F-63000 Clermont-Ferrand, France\\
$^{17}$ Department of Astronomy, University of California, Berkeley, CA 94720-3411, USA\\
$^{18}$ Fermi National Accelerator Laboratory, P. O. Box 500, Batavia, IL 60510, USA\\
$^{19}$ INAF, Astrophysical Observatory of Turin, I-10025 Pino Torinese, Italy\\
$^{20}$ Sydney Institute for Astronomy, School of Physics, A28, The University of Sydney, NSW 2006, Australia\\
$^{21}$ Cerro Tololo Inter-American Observatory, National Optical Astronomy Observatory, Casilla 603, La Serena, Chile\\
$^{22}$ Departamento de F\'isica Matem\'atica, Instituto de F\'isica, Universidade de S\~ao Paulo, CP 66318, S\~ao Paulo, SP, 05314-970, Brazil\\
$^{23}$ Laborat\'orio Interinstitucional de e-Astronomia - LIneA, Rua Gal. Jos\'e Cristino 77, Rio de Janeiro, RJ - 20921-400, Brazil\\
$^{24}$ Instituto de Fisica Teorica UAM/CSIC, Universidad Autonoma de Madrid, 28049 Madrid, Spain\\
$^{25}$ CNRS, UMR 7095, Institut d'Astrophysique de Paris, F-75014, Paris, France\\
$^{26}$ Sorbonne Universit\'es, UPMC Univ Paris 06, UMR 7095, Institut d'Astrophysique de Paris, F-75014, Paris, France\\
$^{27}$ Department of Physics and Astronomy, Pevensey Building, University of Sussex, Brighton, BN1 9QH, UK\\
$^{28}$ Department of Physics \& Astronomy, University College London, Gower Street, London, WC1E 6BT, UK\\
$^{29}$ Kavli Institute for Particle Astrophysics \& Cosmology, P. O. Box 2450, Stanford University, Stanford, CA 94305, USA\\
$^{30}$ SLAC National Accelerator Laboratory, Menlo Park, CA 94025, USA\\
$^{31}$ Centro de Investigaciones Energ\'eticas, Medioambientales y Tecnol\'ogicas (CIEMAT), Madrid, Spain\\
$^{32}$ Department of Astronomy, University of Illinois at Urbana-Champaign, 1002 W. Green Street, Urbana, IL 61801, USA\\
$^{33}$ National Center for Supercomputing Applications, 1205 West Clark St., Urbana, IL 61801, USA\\
$^{34}$ INAF-Osservatorio Astronomico di Trieste, via G. B. Tiepolo 11, I-34143 Trieste, Italy\\
$^{35}$ Institute for Fundamental Physics of the Universe, Via Beirut 2, 34014 Trieste, Italy\\
$^{36}$ Observat\'orio Nacional, Rua Gal. Jos\'e Cristino 77, Rio de Janeiro, RJ - 20921-400, Brazil\\
$^{37}$ Department of Physics, IIT Hyderabad, Kandi, Telangana 502285, India\\
$^{38}$ Department of Astronomy/Steward Observatory, University of Arizona, 933 North Cherry Avenue, Tucson, AZ 85721-0065, USA\\
$^{39}$ Jet Propulsion Laboratory, California Institute of Technology, 4800 Oak Grove Dr., Pasadena, CA 91109, USA\\
$^{40}$ Institut d'Estudis Espacials de Catalunya (IEEC), 08034 Barcelona, Spain\\
$^{41}$ Institute of Space Sciences (ICE, CSIC),  Campus UAB, Carrer de Can Magrans, s/n,  08193 Barcelona, Spain\\
$^{42}$ Centre for Astrophysics \& Supercomputing, Swinburne University of Technology, Victoria 3122, Australia \\
$^{43}$ Department of Physics, Stanford University, 382 Via Pueblo Mall, Stanford, CA 94305, USA\\
$^{44}$ Department of Physics, ETH Zurich, Wolfgang-Pauli-Strasse 16, CH-8093 Zurich, Switzerland\\
$^{45}$ Center for Cosmology and Astro-Particle Physics, The Ohio State University, Columbus, OH 43210, USA\\
$^{46}$ Department of Physics, The Ohio State University, Columbus, OH 43210, USA\\
$^{47}$ Center for Astrophysics $\vert$ Harvard \& Smithsonian, 60 Garden Street, Cambridge, MA 02138, USA\\
$^{48}$ Australian Astronomical Optics, Macquarie University, North Ryde, NSW 2113, Australia\\
$^{49}$ Lowell Observatory, 1400 Mars Hill Rd, Flagstaff, AZ 86001, USA\\
$^{50}$ George P. and Cynthia Woods Mitchell Institute for Fundamental Physics and Astronomy, and Department of Physics and Astronomy, Texas A\&M University, College Station, TX 77843,  USA\\
$^{51}$ Department of Astronomy, The Ohio State University, Columbus, OH 43210, USA\\
$^{52}$ Department of Astrophysical Sciences, Princeton University, Peyton Hall, Princeton, NJ 08544, USA\\
$^{53}$ Instituci\'o Catalana de Recerca i Estudis Avan\c{c}ats, E-08010 Barcelona, Spain\\
$^{54}$ Institut de F\'{\i}sica d'Altes Energies (IFAE), The Barcelona Institute of Science and Technology, Campus UAB, 08193 Bellaterra (Barcelona) Spain\\
$^{55}$ Department of Physics, University of Michigan, Ann Arbor, MI 48109, USA\\
$^{56}$ Computer Science and Mathematics Division, Oak Ridge National Laboratory, Oak Ridge, TN 37831\\
$^{57}$ Max Planck Institute for Extraterrestrial Physics, Giessenbachstrasse, 85748 Garching, Germany\\
$^{58}$ Universit\"ats-Sternwarte, Fakult\"at f\"ur Physik, Ludwig-Maximilians Universit\"at M\"unchen, Scheinerstr. 1, 81679 M\"unchen, Germany\\

\bibliographystyle{mnras}
\bibliography{bib,inprep}

\begin{thebibliography}{}
\makeatletter
\relax
\def\mn@urlcharsother{\let\do\@makeother \do\$\do\&\do\#\do\^\do\_\do\%\do\~}
\def\mn@doi{\begingroup\mn@urlcharsother \@ifnextchar [ {\mn@doi@}
  {\mn@doi@[]}}
\def\mn@doi@[#1]#2{\def\@tempa{#1}\ifx\@tempa\@empty \href
  {http://dx.doi.org/#2} {doi:#2}\else \href {http://dx.doi.org/#2} {#1}\fi
  \endgroup}
\def\mn@eprint#1#2{\mn@eprint@#1:#2::\@nil}
\def\mn@eprint@arXiv#1{\href {http://arxiv.org/abs/#1} {{\tt arXiv:#1}}}
\def\mn@eprint@dblp#1{\href {http://dblp.uni-trier.de/rec/bibtex/#1.xml}
  {dblp:#1}}
\def\mn@eprint@#1:#2:#3:#4\@nil{\def\@tempa {#1}\def\@tempb {#2}\def\@tempc
  {#3}\ifx \@tempc \@empty \let \@tempc \@tempb \let \@tempb \@tempa \fi \ifx
  \@tempb \@empty \def\@tempb {arXiv}\fi \@ifundefined
  {mn@eprint@\@tempb}{\@tempb:\@tempc}{\expandafter \expandafter \csname
  mn@eprint@\@tempb\endcsname \expandafter{\@tempc}}}

\bibitem[\protect\citeauthoryear{{Bertin}}{{Bertin}}{2006}]{Bertin2006}
{Bertin} E.,  2006, in {Gabriel} C.,  {Arviset} C.,  {Ponz} D.,   {Enrique} S.,
   eds,  Astronomical Society of the Pacific Conference Series Vol. 351,
  Astronomical Data Analysis Software and Systems XV. p.~112

\bibitem[\protect\citeauthoryear{{Bertin}}{{Bertin}}{2011}]{Bertin2011}
{Bertin} E.,  2011, in {Evans} I.~N.,  {Accomazzi} A.,  {Mink} D.~J.,   {Rots}
  A.~H.,  eds,  Astronomical Society of the Pacific Conference Series Vol. 442,
  Astronomical Data Analysis Software and Systems XX. p.~435

\bibitem[\protect\citeauthoryear{{Bertin} \& {Arnouts}}{{Bertin} \&
  {Arnouts}}{1996}]{Bertin1996}
{Bertin} E.,  {Arnouts} S.,  1996, \mn@doi [\aaps] {10.1051/aas:1996164}, \href
  {http://adsabs.harvard.edu/abs/1996A%26AS..117..393B} {117, 393}

\bibitem[\protect\citeauthoryear{{Bertin}, {Mellier}, {Radovich}, {Missonnier},
  {Didelon}  \& {Morin}}{{Bertin} et~al.}{2002}]{Bertin2002}
{Bertin} E.,  {Mellier} Y.,  {Radovich} M.,  {Missonnier} G.,  {Didelon} P.,
  {Morin} B.,  2002, in {Bohlender} D.~A.,  {Durand} D.,   {Handley} T.~H.,
  eds,  Astronomical Society of the Pacific Conference Series Vol. 281,
  Astronomical Data Analysis Software and Systems XI. p.~228

\bibitem[\protect\citeauthoryear{{Betoule} et~al.,}{{Betoule}
  et~al.}{2014}]{Betoule2014}
{Betoule} M.,  et~al., 2014, \mn@doi [\aap] {10.1051/0004-6361/201423413},
  \href {http://adsabs.harvard.edu/abs/2014A%26A...568A..22B} {568, A22}

\bibitem[\protect\citeauthoryear{{Bonnett} et~al.,}{{Bonnett}
  et~al.}{2016}]{Bonnett2016}
{Bonnett} C.,  et~al., 2016, \mn@doi [\prd] {10.1103/PhysRevD.94.042005}, \href
  {http://adsabs.harvard.edu/abs/2016PhRvD..94d2005B} {94, 042005}

\bibitem[\protect\citeauthoryear{{Brout} et~al.,}{{Brout}
  et~al.}{2019a}]{Brout2019b}
{Brout} D.,  et~al., 2019a, \mn@doi [\apj] {10.3847/1538-4357/ab06c1}, \href
  {https://ui.adsabs.harvard.edu/abs/2019ApJ...874..106B} {874, 106}

\bibitem[\protect\citeauthoryear{{Brout} et~al.,}{{Brout}
  et~al.}{2019b}]{Brout2019a}
{Brout} D.,  et~al., 2019b, \mn@doi [\apj] {10.3847/1538-4357/ab08a0}, \href
  {https://ui.adsabs.harvard.edu/abs/2019ApJ...874..150B} {874, 150}

\bibitem[\protect\citeauthoryear{{Bruzual} \& {Charlot}}{{Bruzual} \&
  {Charlot}}{2003}]{Bruzual2003}
{Bruzual} G.,  {Charlot} S.,  2003, \mn@doi [\mnras]
  {10.1046/j.1365-8711.2003.06897.x}, \href
  {http://adsabs.harvard.edu/abs/2003MNRAS.344.1000B} {344, 1000}

\bibitem[\protect\citeauthoryear{{Burke} et~al.,}{{Burke}
  et~al.}{2018}]{Burke2018}
{Burke} D.~L.,  et~al., 2018, \mn@doi [\aj] {10.3847/1538-3881/aa9f22}, \href
  {http://adsabs.harvard.edu/abs/2018AJ....155...41B} {155, 41}

\bibitem[\protect\citeauthoryear{{Childress} et~al.,}{{Childress}
  et~al.}{2013a}]{Childress2013b}
{Childress} M.,  et~al., 2013a, \mn@doi [\apj] {10.1088/0004-637X/770/2/107},
  \href {http://adsabs.harvard.edu/abs/2013ApJ...770..107C} {770, 107}

\bibitem[\protect\citeauthoryear{{Childress} et~al.,}{{Childress}
  et~al.}{2013b}]{Childress2013}
{Childress} M.,  et~al., 2013b, \mn@doi [\apj] {10.1088/0004-637X/770/2/108},
  \href {http://adsabs.harvard.edu/abs/2013ApJ...770..108C} {770, 108}

\bibitem[\protect\citeauthoryear{{Childress}, {Wolf}  \& {Zahid}}{{Childress}
  et~al.}{2014}]{Childress2014}
{Childress} M.~J.,  {Wolf} C.,   {Zahid} H.~J.,  2014, \mn@doi [\mnras]
  {10.1093/mnras/stu1892}, \href
  {http://adsabs.harvard.edu/abs/2014MNRAS.445.1898C} {445, 1898}

\bibitem[\protect\citeauthoryear{{Chotard} et~al.,}{{Chotard}
  et~al.}{2011}]{Chotard2011}
{Chotard} N.,  et~al., 2011, \mn@doi [\aap] {10.1051/0004-6361/201116723},
  \href {http://adsabs.harvard.edu/abs/2011A%26A...529L...4C} {529, L4}

\bibitem[\protect\citeauthoryear{{Conley} et~al.,}{{Conley}
  et~al.}{2011}]{Conley2011}
{Conley} A.,  et~al., 2011, \mn@doi [\apjs] {10.1088/0067-0049/192/1/1}, \href
  {http://adsabs.harvard.edu/abs/2011ApJS..192....1C} {192, 1}

\bibitem[\protect\citeauthoryear{{D'Andrea} et~al.,}{{D'Andrea}
  et~al.}{2011}]{DAndrea2011}
{D'Andrea} C.~B.,  et~al., 2011, \mn@doi [\apj] {10.1088/0004-637X/743/2/172},
  \href {http://adsabs.harvard.edu/abs/2011ApJ...743..172D} {743, 172}

\bibitem[\protect\citeauthoryear{{D'Andrea} et~al.,}{{D'Andrea}
  et~al.}{2018}]{DAndrea2018}
{D'Andrea} C.~B.,  et~al., 2018, arXiv e-prints, \href
  {https://ui.adsabs.harvard.edu/\#abs/2018arXiv181109565D} {p.
  arXiv:1811.09565}

\bibitem[\protect\citeauthoryear{{DES Collaboration}}{{DES
  Collaboration}}{2019}]{DES2019}
{DES Collaboration} 2019, \mn@doi [\apjl] {10.3847/2041-8213/ab04fa}, \href
  {https://ui.adsabs.harvard.edu/abs/2019ApJ...872L..30A} {872, L30}

\bibitem[\protect\citeauthoryear{{Fioc} \& {Rocca-Volmerange}}{{Fioc} \&
  {Rocca-Volmerange}}{1997}]{Fioc1997}
{Fioc} M.,  {Rocca-Volmerange} B.,  1997, \aap, \href
  {http://adsabs.harvard.edu/abs/1997A%26A...326..950F} {326, 950}

\bibitem[\protect\citeauthoryear{{Flaugher} et~al.,}{{Flaugher}
  et~al.}{2015}]{Flaugher2015}
{Flaugher} B.,  et~al., 2015, \mn@doi [\aj] {10.1088/0004-6256/150/5/150},
  \href {http://adsabs.harvard.edu/abs/2015AJ....150..150F} {150, 150}

\bibitem[\protect\citeauthoryear{{Galbany} et~al.,}{{Galbany}
  et~al.}{2016}]{Galbany2016}
{Galbany} L.,  et~al., 2016, \mn@doi [\mnras] {10.1093/mnras/stv2620}, \href
  {http://adsabs.harvard.edu/abs/2016MNRAS.455.4087G} {455, 4087}

\bibitem[\protect\citeauthoryear{{Goldstein} et~al.,}{{Goldstein}
  et~al.}{2015}]{Goldstein2015}
{Goldstein} D.~A.,  et~al., 2015, \mn@doi [\aj] {10.1088/0004-6256/150/3/82},
  \href {http://adsabs.harvard.edu/abs/2015AJ....150...82G} {150, 82}

\bibitem[\protect\citeauthoryear{{Gupta} et~al.,}{{Gupta}
  et~al.}{2011}]{Gupta2011}
{Gupta} R.~R.,  et~al., 2011, \mn@doi [\apj] {10.1088/0004-637X/740/2/92},
  \href {http://adsabs.harvard.edu/abs/2011ApJ...740...92G} {740, 92}

\bibitem[\protect\citeauthoryear{{Gupta} et~al.,}{{Gupta}
  et~al.}{2016}]{Gupta2016}
{Gupta} R.~R.,  et~al., 2016, \mn@doi [\aj] {10.3847/0004-6256/152/6/154},
  \href {http://adsabs.harvard.edu/abs/2016AJ....152..154G} {152, 154}

\bibitem[\protect\citeauthoryear{{Guy} et~al.,}{{Guy} et~al.}{2007}]{Guy2007}
{Guy} J.,  et~al., 2007, \mn@doi [\aap] {10.1051/0004-6361:20066930}, \href
  {http://adsabs.harvard.edu/abs/2007A%26A...466...11G} {466, 11}

\bibitem[\protect\citeauthoryear{{Guy} et~al.,}{{Guy} et~al.}{2010}]{Guy2010}
{Guy} J.,  et~al., 2010, \mn@doi [\aap] {10.1051/0004-6361/201014468}, \href
  {http://adsabs.harvard.edu/abs/2010A%26A...523A...7G} {523, A7}

\bibitem[\protect\citeauthoryear{{Hamuy}, {Phillips}, {Maza}, {Suntzeff},
  {Schommer}  \& {Aviles}}{{Hamuy} et~al.}{1995}]{Hamuy1995}
{Hamuy} M.,  {Phillips} M.~M.,  {Maza} J.,  {Suntzeff} N.~B.,  {Schommer}
  R.~A.,   {Aviles} R.,  1995, \mn@doi [\aj] {10.1086/117251}, \href
  {http://adsabs.harvard.edu/abs/1995AJ....109....1H} {109, 1}

\bibitem[\protect\citeauthoryear{{Hamuy}, {Trager}, {Pinto}, {Phillips},
  {Schommer}, {Ivanov}  \& {Suntzeff}}{{Hamuy} et~al.}{2000}]{Hamuy2000}
{Hamuy} M.,  {Trager} S.~C.,  {Pinto} P.~A.,  {Phillips} M.~M.,  {Schommer}
  R.~A.,  {Ivanov} V.,   {Suntzeff} N.~B.,  2000, \mn@doi [\aj]
  {10.1086/301527}, \href {http://adsabs.harvard.edu/abs/2000AJ....120.1479H}
  {120, 1479}

\bibitem[\protect\citeauthoryear{{Hayden}, {Gupta}, {Garnavich}, {Mannucci},
  {Nichol}  \& {Sako}}{{Hayden} et~al.}{2013}]{Hayden2013}
{Hayden} B.~T.,  {Gupta} R.~R.,  {Garnavich} P.~M.,  {Mannucci} F.,  {Nichol}
  R.~C.,   {Sako} M.,  2013, \mn@doi [\apj] {10.1088/0004-637X/764/2/191},
  \href {http://adsabs.harvard.edu/abs/2013ApJ...764..191H} {764, 191}

\bibitem[\protect\citeauthoryear{{Hounsell} et~al.,}{{Hounsell}
  et~al.}{2018}]{Hounsell2018}
{Hounsell} R.,  et~al., 2018, \mn@doi [\apj] {10.3847/1538-4357/aac08b}, \href
  {http://adsabs.harvard.edu/abs/2018ApJ...867...23H} {867, 23}

\bibitem[\protect\citeauthoryear{{Johansson} et~al.,}{{Johansson}
  et~al.}{2013}]{Johansson2013}
{Johansson} J.,  et~al., 2013, \mn@doi [\mnras] {10.1093/mnras/stt1408}, \href
  {http://adsabs.harvard.edu/abs/2013MNRAS.435.1680J} {435, 1680}

\bibitem[\protect\citeauthoryear{{Jones}, {Riess}  \& {Scolnic}}{{Jones}
  et~al.}{2015}]{Jones2015}
{Jones} D.~O.,  {Riess} A.~G.,   {Scolnic} D.~M.,  2015, \mn@doi [\apj]
  {10.1088/0004-637X/812/1/31}, \href
  {http://adsabs.harvard.edu/abs/2015ApJ...812...31J} {812, 31}

\bibitem[\protect\citeauthoryear{{Jones} et~al.,}{{Jones}
  et~al.}{2018a}]{Jones2018}
{Jones} D.~O.,  et~al., 2018a, \mn@doi [\apj] {10.3847/1538-4357/aab6b1}, \href
  {http://adsabs.harvard.edu/abs/2018ApJ...857...51J} {857, 51}

\bibitem[\protect\citeauthoryear{{Jones} et~al.,}{{Jones}
  et~al.}{2018b}]{Jones2018b}
{Jones} D.~O.,  et~al., 2018b, \mn@doi [\apj] {10.3847/1538-4357/aae2b9}, \href
  {http://adsabs.harvard.edu/abs/2018ApJ...867..108J} {867, 108}

\bibitem[\protect\citeauthoryear{{Kelly}, {Hicken}, {Burke}, {Mandel}  \&
  {Kirshner}}{{Kelly} et~al.}{2010}]{Kelly2010}
{Kelly} P.~L.,  {Hicken} M.,  {Burke} D.~L.,  {Mandel} K.~S.,   {Kirshner}
  R.~P.,  2010, \mn@doi [\apj] {10.1088/0004-637X/715/2/743}, \href
  {http://adsabs.harvard.edu/abs/2010ApJ...715..743K} {715, 743}

\bibitem[\protect\citeauthoryear{{Kelsey} et~al.,}{{Kelsey}
  et~al.}{2020}]{Kelsey2020}
{Kelsey} L.,  et~al., 2020, in preparation

\bibitem[\protect\citeauthoryear{{Kessler} \& {Scolnic}}{{Kessler} \&
  {Scolnic}}{2017}]{Kessler2017}
{Kessler} R.,  {Scolnic} D.,  2017, \mn@doi [\apj]
  {10.3847/1538-4357/836/1/56}, \href
  {http://adsabs.harvard.edu/abs/2017ApJ...836...56K} {836, 56}

\bibitem[\protect\citeauthoryear{{Kessler} et~al.,}{{Kessler}
  et~al.}{2009}]{Kessler2009}
{Kessler} R.,  et~al., 2009, \mn@doi [\pasp] {10.1086/605984}, \href
  {http://adsabs.harvard.edu/abs/2009PASP..121.1028K} {121, 1028}

\bibitem[\protect\citeauthoryear{{Kessler} et~al.,}{{Kessler}
  et~al.}{2013}]{Kessler2013}
{Kessler} R.,  et~al., 2013, \mn@doi [\apj] {10.1088/0004-637X/764/1/48}, \href
  {http://adsabs.harvard.edu/abs/2013ApJ...764...48K} {764, 48}

\bibitem[\protect\citeauthoryear{{Kessler} et~al.,}{{Kessler}
  et~al.}{2015}]{Kessler2015}
{Kessler} R.,  et~al., 2015, \mn@doi [\aj] {10.1088/0004-6256/150/6/172}, \href
  {http://adsabs.harvard.edu/abs/2015AJ....150..172K} {150, 172}

\bibitem[\protect\citeauthoryear{{Kessler} et~al.,}{{Kessler}
  et~al.}{2019}]{Kessler2019}
{Kessler} R.,  et~al., 2019, \mn@doi [\mnras] {10.1093/mnras/stz463}, \href
  {http://adsabs.harvard.edu/abs/2019MNRAS.485.1171K} {485, 1171}

\bibitem[\protect\citeauthoryear{{Kim}, {Smith}, {Sullivan}  \& {Lee}}{{Kim}
  et~al.}{2018}]{Kim2018}
{Kim} Y.-L.,  {Smith} M.,  {Sullivan} M.,   {Lee} Y.-W.,  2018, \mn@doi [\apj]
  {10.3847/1538-4357/aaa127}, \href
  {http://adsabs.harvard.edu/abs/2018ApJ...854...24K} {854, 24}

\bibitem[\protect\citeauthoryear{{Kroupa}}{{Kroupa}}{2001}]{Kroupa2001}
{Kroupa} P.,  2001, \mn@doi [\mnras] {10.1046/j.1365-8711.2001.04022.x}, \href
  {http://adsabs.harvard.edu/abs/2001MNRAS.322..231K} {322, 231}

\bibitem[\protect\citeauthoryear{{LSST Dark Energy Science
  Collaboration}}{{LSST Dark Energy Science Collaboration}}{2012}]{LSST2012}
{LSST Dark Energy Science Collaboration} 2012, arXiv e-prints, \href
  {https://ui.adsabs.harvard.edu/abs/2012arXiv1211.0310L} {p. arXiv:1211.0310}

\bibitem[\protect\citeauthoryear{{Laigle} et~al.,}{{Laigle}
  et~al.}{2019}]{Laigle2019}
{Laigle} C.,  et~al., 2019, \mn@doi [\mnras] {10.1093/mnras/stz1054}, \href
  {https://ui.adsabs.harvard.edu/abs/2019MNRAS.486.5104L} {486, 5104}

\bibitem[\protect\citeauthoryear{{Lampeitl} et~al.,}{{Lampeitl}
  et~al.}{2010}]{Lampeitl2010}
{Lampeitl} H.,  et~al., 2010, \mn@doi [\apj] {10.1088/0004-637X/722/1/566},
  \href {http://adsabs.harvard.edu/abs/2010ApJ...722..566L} {722, 566}

\bibitem[\protect\citeauthoryear{{Lasker} et~al.,}{{Lasker}
  et~al.}{2019}]{Lasker2018}
{Lasker} J.,  et~al., 2019, \mn@doi [\mnras] {10.1093/mnras/stz619}, \href
  {https://ui.adsabs.harvard.edu/abs/2019MNRAS.485.5329L} {485, 5329}

\bibitem[\protect\citeauthoryear{{Le Borgne} \& {Rocca-Volmerange}}{{Le Borgne}
  \& {Rocca-Volmerange}}{2002}]{LeBorgne2002}
{Le Borgne} D.,  {Rocca-Volmerange} B.,  2002, \mn@doi [\aap]
  {10.1051/0004-6361:20020259}, \href
  {http://adsabs.harvard.edu/abs/2002A%26A...386..446L} {386, 446}

\bibitem[\protect\citeauthoryear{{Mannucci}, {Della Valle}, {Panagia},
  {Cappellaro}, {Cresci}, {Maiolino}, {Petrosian}  \& {Turatto}}{{Mannucci}
  et~al.}{2005}]{Mannucci2005}
{Mannucci} F.,  {Della Valle} M.,  {Panagia} N.,  {Cappellaro} E.,  {Cresci}
  G.,  {Maiolino} R.,  {Petrosian} A.,   {Turatto} M.,  2005, \mn@doi [\aap]
  {10.1051/0004-6361:20041411}, \href
  {http://adsabs.harvard.edu/abs/2005A%26A...433..807M} {433, 807}

\bibitem[\protect\citeauthoryear{{Mannucci}, {Della Valle}  \&
  {Panagia}}{{Mannucci} et~al.}{2006}]{Mannucci2006}
{Mannucci} F.,  {Della Valle} M.,   {Panagia} N.,  2006, \mn@doi [\mnras]
  {10.1111/j.1365-2966.2006.10501.x}, \href
  {http://adsabs.harvard.edu/abs/2006MNRAS.370..773M} {370, 773}

\bibitem[\protect\citeauthoryear{{Maraston}}{{Maraston}}{2005}]{Maraston2005}
{Maraston} C.,  2005, \mn@doi [\mnras] {10.1111/j.1365-2966.2005.09270.x},
  \href {http://adsabs.harvard.edu/abs/2005MNRAS.362..799M} {362, 799}

\bibitem[\protect\citeauthoryear{{March}, {Trotta}, {Berkes}, {Starkman}  \&
  {Vaudrevange}}{{March} et~al.}{2011}]{March2011}
{March} M.~C.,  {Trotta} R.,  {Berkes} P.,  {Starkman} G.~D.,   {Vaudrevange}
  P.~M.,  2011, \mn@doi [\mnras] {10.1111/j.1365-2966.2011.19584.x}, \href
  {https://ui.adsabs.harvard.edu/abs/2011MNRAS.418.2308M} {418, 2308}

\bibitem[\protect\citeauthoryear{{Mitchell}, {Lacey}, {Baugh}  \&
  {Cole}}{{Mitchell} et~al.}{2013}]{Mitchell2013}
{Mitchell} P.~D.,  {Lacey} C.~G.,  {Baugh} C.~M.,   {Cole} S.,  2013, \mn@doi
  [\mnras] {10.1093/mnras/stt1280}, \href
  {https://ui.adsabs.harvard.edu/abs/2013MNRAS.435...87M} {435, 87}

\bibitem[\protect\citeauthoryear{{Moreno-Raya}, {Galbany},
  {L{\'o}pez-S{\'a}nchez}, {Moll{\'a}}, {Gonz{\'a}lez-Gait{\'a}n},
  {V{\'\i}lchez}  \& {Carnero}}{{Moreno-Raya} et~al.}{2018}]{Moreno-Raya2018}
{Moreno-Raya} M.~E.,  {Galbany} L.,  {L{\'o}pez-S{\'a}nchez} {\'A}.~R.,
  {Moll{\'a}} M.,  {Gonz{\'a}lez-Gait{\'a}n} S.,  {V{\'\i}lchez} J.~M.,
  {Carnero} A.,  2018, \mn@doi [\mnras] {10.1093/mnras/sty185}, \href
  {https://ui.adsabs.harvard.edu/abs/2018MNRAS.476..307M} {476, 307}

\bibitem[\protect\citeauthoryear{{Morganson} et~al.,}{{Morganson}
  et~al.}{2018}]{Morganson2018}
{Morganson} E.,  et~al., 2018, \mn@doi [\pasp] {10.1088/1538-3873/aab4ef},
  \href {http://adsabs.harvard.edu/abs/2018PASP..130g4501M} {130, 074501}

\bibitem[\protect\citeauthoryear{{Oke} \& {Gunn}}{{Oke} \&
  {Gunn}}{1983}]{Oke83}
{Oke} J.~B.,  {Gunn} J.~E.,  1983, \mn@doi [\apj] {10.1086/160817}, \href
  {http://adsabs.harvard.edu/abs/1983ApJ...266..713O} {266, 713}

\bibitem[\protect\citeauthoryear{{Palmese} et~al.,}{{Palmese}
  et~al.}{2016}]{Palmese2016}
{Palmese} A.,  et~al., 2016, \mn@doi [\mnras] {10.1093/mnras/stw2062}, \href
  {https://ui.adsabs.harvard.edu/abs/2016MNRAS.463.1486P} {463, 1486}

\bibitem[\protect\citeauthoryear{{Palmese} et~al.,}{{Palmese}
  et~al.}{2019}]{Palmese2019}
{Palmese} A.,  et~al., 2019, arXiv e-prints, \href
  {https://ui.adsabs.harvard.edu/abs/2019arXiv190308813P} {p. arXiv:1903.08813}

\bibitem[\protect\citeauthoryear{{Pan} et~al.,}{{Pan} et~al.}{2014}]{Pan2014}
{Pan} Y.-C.,  et~al., 2014, \mn@doi [\mnras] {10.1093/mnras/stt2287}, \href
  {http://adsabs.harvard.edu/abs/2014MNRAS.438.1391P} {438, 1391}

\bibitem[\protect\citeauthoryear{{Perlmutter} et~al.,}{{Perlmutter}
  et~al.}{1999}]{Perlmutter1999}
{Perlmutter} S.,  et~al., 1999, \mn@doi [\apj] {10.1086/307221}, \href
  {http://adsabs.harvard.edu/abs/1999ApJ...517..565P} {517, 565}

\bibitem[\protect\citeauthoryear{{Perrett} et~al.,}{{Perrett}
  et~al.}{2010}]{Perrett2010}
{Perrett} K.,  et~al., 2010, \mn@doi [\aj] {10.1088/0004-6256/140/2/518}, \href
  {http://adsabs.harvard.edu/abs/2010AJ....140..518P} {140, 518}

\bibitem[\protect\citeauthoryear{{Phillips}}{{Phillips}}{1993}]{Phillips1993}
{Phillips} M.~M.,  1993, \mn@doi [\apjl] {10.1086/186970}, \href
  {http://adsabs.harvard.edu/abs/1993ApJ...413L.105P} {413, L105}

\bibitem[\protect\citeauthoryear{{Planck Collaboration} et~al.,}{{Planck
  Collaboration} et~al.}{2016}]{Planck2016}
{Planck Collaboration} et~al., 2016, \mn@doi [\aap]
  {10.1051/0004-6361/201525830}, \href
  {https://ui.adsabs.harvard.edu/abs/2016A&A...594A..13P} {594, A13}

\bibitem[\protect\citeauthoryear{{Riess}, {Press}  \& {Kirshner}}{{Riess}
  et~al.}{1996}]{Riess1996}
{Riess} A.~G.,  {Press} W.~H.,   {Kirshner} R.~P.,  1996, \mn@doi [\apj]
  {10.1086/178129}, \href {http://adsabs.harvard.edu/abs/1996ApJ...473...88R}
  {473, 88}

\bibitem[\protect\citeauthoryear{{Riess} et~al.,}{{Riess}
  et~al.}{1998}]{Riess1998}
{Riess} A.~G.,  et~al., 1998, \mn@doi [\aj] {10.1086/300499}, \href
  {http://adsabs.harvard.edu/abs/1998AJ....116.1009R} {116, 1009}

\bibitem[\protect\citeauthoryear{{Riess} et~al.,}{{Riess}
  et~al.}{2018}]{Riess2018}
{Riess} A.~G.,  et~al., 2018, \mn@doi [\apj] {10.3847/1538-4357/aaa5a9}, \href
  {http://adsabs.harvard.edu/abs/2018ApJ...853..126R} {853, 126}

\bibitem[\protect\citeauthoryear{{Rigault} et~al.,}{{Rigault}
  et~al.}{2013}]{Rigault2013}
{Rigault} M.,  et~al., 2013, \mn@doi [\aap] {10.1051/0004-6361/201322104},
  \href {http://adsabs.harvard.edu/abs/2013A%26A...560A..66R} {560, A66}

\bibitem[\protect\citeauthoryear{{Rigault} et~al.,}{{Rigault}
  et~al.}{2018}]{Rigault2018}
{Rigault} M.,  et~al., 2018, arXiv e-prints, \href
  {https://ui.adsabs.harvard.edu/abs/2018arXiv180603849R} {p. arXiv:1806.03849}

\bibitem[\protect\citeauthoryear{{Roman} et~al.,}{{Roman}
  et~al.}{2018}]{Roman2018}
{Roman} M.,  et~al., 2018, \mn@doi [\aap] {10.1051/0004-6361/201731425}, \href
  {http://adsabs.harvard.edu/abs/2018A%26A...615A..68R} {615, A68}

\bibitem[\protect\citeauthoryear{{Rykoff} et~al.,}{{Rykoff}
  et~al.}{2016}]{Rykoff2016}
{Rykoff} E.~S.,  et~al., 2016, \mn@doi [\apjs] {10.3847/0067-0049/224/1/1},
  \href {http://adsabs.harvard.edu/abs/2016ApJS..224....1R} {224, 1}

\bibitem[\protect\citeauthoryear{{Sako} et~al.,}{{Sako}
  et~al.}{2018}]{Sako2018}
{Sako} M.,  et~al., 2018, \mn@doi [\pasp] {10.1088/1538-3873/aab4e0}, \href
  {http://adsabs.harvard.edu/abs/2018PASP..130f4002S} {130, 064002}

\bibitem[\protect\citeauthoryear{{Salpeter}}{{Salpeter}}{1955}]{Salpeter1955}
{Salpeter} E.~E.,  1955, \mn@doi [\apj] {10.1086/145971}, \href
  {http://adsabs.harvard.edu/abs/1955ApJ...121..161S} {121, 161}

\bibitem[\protect\citeauthoryear{{Scannapieco} \& {Bildsten}}{{Scannapieco} \&
  {Bildsten}}{2005}]{Scannapieco2005}
{Scannapieco} E.,  {Bildsten} L.,  2005, \mn@doi [\apjl] {10.1086/452632},
  \href {http://adsabs.harvard.edu/abs/2005ApJ...629L..85S} {629, L85}

\bibitem[\protect\citeauthoryear{{Schlegel}, {Finkbeiner}  \&
  {Davis}}{{Schlegel} et~al.}{1998}]{Schlegel1998}
{Schlegel} D.~J.,  {Finkbeiner} D.~P.,   {Davis} M.,  1998, \mn@doi [\apj]
  {10.1086/305772}, \href {http://adsabs.harvard.edu/abs/1998ApJ...500..525S}
  {500, 525}

\bibitem[\protect\citeauthoryear{{Scolnic} \& {Kessler}}{{Scolnic} \&
  {Kessler}}{2016}]{Scolnic2016}
{Scolnic} D.,  {Kessler} R.,  2016, \mn@doi [\apjl]
  {10.3847/2041-8205/822/2/L35}, \href
  {http://adsabs.harvard.edu/abs/2016ApJ...822L..35S} {822, L35}

\bibitem[\protect\citeauthoryear{{Scolnic} et~al.,}{{Scolnic}
  et~al.}{2018}]{Scolnic2018}
{Scolnic} D.~M.,  et~al., 2018, \mn@doi [\apj] {10.3847/1538-4357/aab9bb},
  \href {http://adsabs.harvard.edu/abs/2018ApJ...859..101S} {859, 101}

\bibitem[\protect\citeauthoryear{{Smith} et~al.,}{{Smith}
  et~al.}{2012}]{Smith2012}
{Smith} M.,  et~al., 2012, \mn@doi [\apj] {10.1088/0004-637X/755/1/61}, \href
  {http://adsabs.harvard.edu/abs/2012ApJ...755...61S} {755, 61}

\bibitem[\protect\citeauthoryear{{Soumagnac} et~al.,}{{Soumagnac}
  et~al.}{2015}]{Soumagnac2015}
{Soumagnac} M.~T.,  et~al., 2015, \mn@doi [\mnras] {10.1093/mnras/stu1410},
  \href {http://adsabs.harvard.edu/abs/2015MNRAS.450..666S} {450, 666}

\bibitem[\protect\citeauthoryear{{Sullivan} et~al.,}{{Sullivan}
  et~al.}{2006}]{Sullivan2006}
{Sullivan} M.,  et~al., 2006, \mn@doi [\apj] {10.1086/506137}, \href
  {http://adsabs.harvard.edu/abs/2006ApJ...648..868S} {648, 868}

\bibitem[\protect\citeauthoryear{{Sullivan} et~al.,}{{Sullivan}
  et~al.}{2010}]{Sullivan2010}
{Sullivan} M.,  et~al., 2010, \mn@doi [\mnras]
  {10.1111/j.1365-2966.2010.16731.x}, \href
  {http://adsabs.harvard.edu/abs/2010MNRAS.406..782S} {406, 782}

\bibitem[\protect\citeauthoryear{{Sullivan} et~al.,}{{Sullivan}
  et~al.}{2011}]{Sullivan2011}
{Sullivan} M.,  et~al., 2011, \mn@doi [\apj] {10.1088/0004-637X/737/2/102},
  \href {http://adsabs.harvard.edu/abs/2011ApJ...737..102S} {737, 102}

\bibitem[\protect\citeauthoryear{{Suzuki} et~al.,}{{Suzuki}
  et~al.}{2012}]{Suzuki2012}
{Suzuki} N.,  et~al., 2012, \mn@doi [\apj] {10.1088/0004-637X/746/1/85}, \href
  {http://adsabs.harvard.edu/abs/2012ApJ...746...85S} {746, 85}

\bibitem[\protect\citeauthoryear{{Tripp}}{{Tripp}}{1998}]{Tripp1998}
{Tripp} R.,  1998, \aap, \href
  {http://adsabs.harvard.edu/abs/1998A%26A...331..815T} {331, 815}

\bibitem[\protect\citeauthoryear{{Wiseman} et~al.,}{{Wiseman}
  et~al.}{2020}]{Wiseman2020}
{Wiseman} P.,  et~al., 2020, arXiv e-prints, \href
  {https://ui.adsabs.harvard.edu/abs/2020arXiv200102640W} {p. arXiv:2001.02640}

\bibitem[\protect\citeauthoryear{{Wolf} et~al.,}{{Wolf}
  et~al.}{2016}]{Wolf2016}
{Wolf} R.~C.,  et~al., 2016, \mn@doi [\apj] {10.3847/0004-637X/821/2/115},
  \href {http://adsabs.harvard.edu/abs/2016ApJ...821..115W} {821, 115}

\makeatother
\end{thebibliography}

\begin{table*}
\section{Host galaxy magnitudes and derived properties}
\label{app:hostprop}
{\centering
\caption{Host galaxy photometric measurements and derived properties for the DES-SN sample.}
\label{tab:hostmags_masses}
\begin{tabular*}{\linewidth}{llcccccccc}
  \hline
  DES Name & SNID & Redshift$^{1}$ & g & r & i & z & log(\mstellar) & log(sSFR) & Catalogue \\
  \hline
DES13C3dgs &  1248677 &  $0.3502$ &  $21.80\pm0.01$ &  $21.04\pm0.01$ &  $20.81\pm0.01$ &  $20.59\pm0.01$ &  $9.57\pm0.01$ &  $-8.49$ &  W19 \\
DES13S1qv &  1250017 &  $0.1817$ &  $22.17\pm0.01$ &  $21.60\pm0.01$ &  $21.37\pm0.01$ &  $21.25\pm0.01$ &  $8.77\pm0.05$ &  $-8.73$ &  W19 \\
DES13C1hwx &  1253039 &  $0.4535$ &  $24.01\pm0.04$ &  $23.04\pm0.02$ &  $22.56\pm0.03$ &  $22.36\pm0.03$ &  $9.39\pm0.06$ &  $-9.91$ &  W19 \\
DES13E1goh &  1253101 &  $0.4596$ &  $25.48\pm0.11$ &  $24.32\pm0.05$ &  $24.22\pm0.07$ &  $23.81\pm0.06$ &  $8.57\pm0.08$ &  $-8.68$ &  W19 \\
DES13C1juw &  1253920 &  $0.1956$ &  $22.18\pm0.01$ &  $21.13\pm0.01$ &  $20.68\pm0.01$ &  $20.53\pm0.01$ &  $9.43\pm0.02$ &  $-19.43$ &  W19 \\
DES13X1kae &  1255502 &  $0.1482$ &  $19.22\pm0.01$ &  $18.34\pm0.01$ &  $17.84\pm0.01$ &  $17.09\pm0.01$ &  $10.70\pm0.01$ &  $-10.31$ &  W19 \\
DES13C1ryv &  1257366 &  $0.2114$ &  $18.74\pm0.01$ &  $17.82\pm0.01$ &  $17.50\pm0.01$ &  $17.31\pm0.01$ &  $10.72\pm0.04$ &  $-10.08$ &  W19 \\
DES13E1sae &  1257695 &  $0.1838$ &  $19.64\pm0.01$ &  $19.12\pm0.01$ &  $18.88\pm0.01$ &  $18.86\pm0.01$ &  $9.68\pm0.03$ &  $-8.66$ &  W19 \\
DES13E2tbn &  1258906 &  $0.3492$ &  $23.19\pm0.02$ &  $22.46\pm0.01$ &  $22.25\pm0.02$ &  $22.05\pm0.02$ &  $8.93\pm0.02$ &  $-8.25$ &  W19 \\
DES13S1sty &  1258940 &  $0.4259$ &  $21.22\pm0.01$ &  $19.80\pm0.01$ &  $19.33\pm0.01$ &  $19.01\pm0.01$ &  $10.78\pm0.02$ &  $-9.47$ &  W19 \\
DES13X3syi &  1259412 &  $0.3047$ &  $20.21\pm0.01$ &  $19.34\pm0.01$ &  $18.99\pm0.01$ &  $18.82\pm0.01$ &  $10.38\pm0.01$ &  $-9.61$ &  W19 \\
DES13X3woy &  1261579 &  $0.3222$ &  $20.23\pm0.01$ &  $18.75\pm0.01$ &  $18.16\pm0.01$ &  $17.98\pm0.01$ &  $11.16\pm0.02$ &  $-13.44$ &  W19 \\
DES13C3abht &  1262214 &  $0.69$ &  $26.45\pm0.26$ &  $25.08\pm0.11$ &  $24.51\pm0.08$ &  $24.17\pm0.09$ &  $9.01\pm0.15$ &  $-9.19$ &  W19 \\
DES13C3abhe &  1262715 &  $0.69$ &  $23.92\pm0.02$ &  $22.90\pm0.01$ &  $21.44\pm0.01$ &  $20.48\pm0.01$ &  $11.28\pm0.02$ &  $-11.73$ &  W19 \\
DES13S1acsq &  1263369 &  $0.3125$ &  $20.84\pm0.01$ &  $19.84\pm0.01$ &  $19.38\pm0.01$ &  $19.04\pm0.01$ &  $10.41\pm0.01$ &  $-9.38$ &  W19 \\
DES13S2acrg &  1263715 &  $0.2919$ &  $23.47\pm0.03$ &  $22.76\pm0.02$ &  $22.59\pm0.02$ &  $22.39\pm0.02$ &  $8.68\pm0.05$ &  $-8.78$ &  W19 \\
DES15E2bo &  1275946 &  $0.2321$ &  $25.16\pm0.10$ &  $24.38\pm0.07$ &  $24.19\pm0.08$ &  $24.08\pm0.11$ &  $7.93\pm0.08$ &  $-9.84$ &  W19 \\
DES15S2it &  1280217 &  $0.3590$ &  $24.80\pm0.08$ &  $24.15\pm0.06$ &  $23.85\pm0.06$ &  $23.62\pm0.08$ &  $8.32\pm0.09$ &  $-8.26$ &  W19 \\
DES15E2nk &  1281668 &  $0.3071$ &  $21.33\pm0.01$ &  $20.52\pm0.01$ &  $20.18\pm0.01$ &  $19.90\pm0.01$ &  $9.92\pm0.01$ &  $-9.15$ &  W19 \\
DES15S2og &  1281886 &  $0.3840$ &  $23.83\pm0.04$ &  $22.61\pm0.02$ &  $22.31\pm0.02$ &  $21.99\pm0.02$ &  $9.37\pm0.02$ &  $-9.24$ &  W19 \\
DES15X2asp &  1282736 &  $0.3689$ &  $21.68\pm0.01$ &  $20.40\pm0.01$ &  $19.93\pm0.01$ &  $19.66\pm0.01$ &  $10.48\pm0.03$ &  $-10.07$ &  W19 \\
DES15C1atm &  1283373 &  $0.2075$ &  $20.78\pm0.01$ &  $19.58\pm0.01$ &  $19.11\pm0.01$ &  $18.78\pm0.01$ &  $10.39\pm0.06$ &  $-10.34$ &  W19 \\
DES15X3atu &  1283878 &  $0.3135$ &  $20.21\pm0.01$ &  $18.85\pm0.01$ &  $18.37\pm0.01$ &  $17.77\pm0.01$ &  $11.21\pm0.01$ &  $-10.14$ &  W19 \\
DES15E2so &  1283936 &  $0.3690$ &  $22.66\pm0.01$ &  $21.78\pm0.01$ &  $21.42\pm0.01$ &  $21.18\pm0.01$ &  $9.57\pm0.03$ &  $-9.56$ &  W19 \\
DES15C3tz &  1285160 &  $0.7493$ &  $27.46\pm0.40$ &  $26.51\pm0.19$ &  $26.11\pm0.20$ &  $25.55\pm0.21$ &  $8.25\pm0.40$ &  $-8.58$ &  W19 \\
DES15E2uc &  1285317 &  $0.5649$ &  $24.40\pm0.05$ &  $23.79\pm0.04$ &  $23.60\pm0.05$ &  $23.69\pm0.07$ &  $8.49\pm0.06$ &  $-8.24$ &  W19 \\
DES15X3auw &  1286398 &  $0.1503$ &  $19.17\pm0.01$ &  $18.28\pm0.01$ &  $17.92\pm0.01$ &  $17.61\pm0.01$ &  $10.38\pm0.04$ &  $-9.81$ &  W19 \\
DES13X2agef &  1287626 &  $0.3040$ &  $24.74\pm0.08$ &  $24.03\pm0.05$ &  $23.85\pm0.05$ &  $23.81\pm0.11$ &  $8.19\pm0.07$ &  $-8.81$ &  W19 \\
DES15E2cwm &  1289288 &  $0.2897$ &  $20.48\pm0.01$ &  $19.41\pm0.01$ &  $18.95\pm0.01$ &  $18.65\pm0.01$ &  $10.58\pm0.01$ &  $-9.63$ &  W19 \\
DES15S2dyb &  1289555 &  $0.5588$ &  $23.83\pm0.04$ &  $23.09\pm0.03$ &  $22.76\pm0.03$ &  $22.72\pm0.04$ &  $9.01\pm0.04$ &  $-8.07$ &  W19 \\
DES15C3axd &  1289600 &  $0.4196$ &  $22.24\pm0.01$ &  $21.44\pm0.01$ &  $21.26\pm0.01$ &  $20.96\pm0.01$ &  $9.44\pm0.02$ &  $-8.39$ &  W19 \\
DES15C1aww &  1289656 &  $0.5394$ &  $25.62\pm0.16$ &  $24.51\pm0.08$ &  $24.19\pm0.08$ &  $24.30\pm0.15$ &  $8.51\pm0.10$ &  $-9.27$ &  W19 \\
DES15S2dye &  1289664 &  $0.2491$ &  $25.37\pm0.17$ &  $24.90\pm0.15$ &  $24.66\pm0.14$ &  $24.06\pm0.13$ &  $7.87\pm0.17$ &  $-8.73$ &  W19 \\
DES14C1eu &  1290779 &  $0.39$ &  $22.64\pm0.01$ &  $21.00\pm0.01$ &  $20.41\pm0.01$ &  $19.87\pm0.01$ &  $10.73\pm0.01$ &  $-10.42$ &  W19 \\
DES14C1es &  1290816 &  $0.2185$ &  $22.77\pm0.02$ &  $22.18\pm0.01$ &  $21.96\pm0.01$ &  $21.67\pm0.02$ &  $8.76\pm0.02$ &  $-8.67$ &  W19 \\
DES14X2dl &  1291080 &  $0.4438$ &  $25.97\pm0.20$ &  $24.44\pm0.07$ &  $24.13\pm0.07$ &  $23.93\pm0.09$ &  $8.67\pm0.09$ &  $-9.96$ &  W19 \\
DES14E2u &  1291090 &  $0.2920$ &  $21.03\pm0.01$ &  $20.45\pm0.01$ &  $20.28\pm0.01$ &  $20.10\pm0.01$ &  $9.38\pm0.05$ &  $-8.18$ &  W19 \\
DES15C2dyj &  1291794 &  $0.3951$ &  $21.79\pm0.01$ &  $20.42\pm0.01$ &  $19.92\pm0.01$ &  $19.61\pm0.01$ &  $10.57\pm0.04$ &  $-9.97$ &  W19 \\
DES14E1tb &  1291957 &  $0.3600$ &  $27.08\pm0.49$ &  $25.48\pm0.13$ &  $25.62\pm0.20$ &  $25.82\pm0.38$ &  $7.64\pm0.12$ &  $-9.88$ &  W19 \\
DES15C2dym &  1292145 &  $0.4925$ &  $24.30\pm0.05$ &  $23.26\pm0.02$ &  $22.99\pm0.03$ &  $22.62\pm0.03$ &  $9.10\pm0.06$ &  $-8.70$ &  W19 \\
DES15C3axo &  1292195 &  $0.8293$ &  $24.31\pm0.07$ &  $23.75\pm0.05$ &  $23.43\pm0.05$ &  $23.27\pm0.09$ &  $8.97\pm0.07$ &  $-7.91$ &  W19 \\
DES14S2qf &  1292332 &  $0.2735$ &  $21.77\pm0.01$ &  $20.42\pm0.01$ &  $19.93\pm0.01$ &  $19.61\pm0.01$ &  $10.16\pm0.05$ &  $-10.38$ &  W19 \\
DES14S2qb &  1292336 &  $0.2336$ &  $22.95\pm0.02$ &  $22.08\pm0.01$ &  $21.69\pm0.01$ &  $21.59\pm0.01$ &  $9.08\pm0.01$ &  $-9.28$ &  W19 \\
DES14X1qn &  1292560 &  $0.2248$ &  $22.87\pm0.02$ &  $22.44\pm0.03$ &  $22.20\pm0.02$ &  $21.87\pm0.03$ &  $8.57\pm0.03$ &  $-8.35$ &  W19 \\
DES14X2ags &  1293319 &  $0.2969$ &  $22.89\pm0.02$ &  $22.16\pm0.01$ &  $21.86\pm0.01$ &  $21.65\pm0.02$ &  $9.08\pm0.03$ &  $-8.85$ &  W19 \\
DES14E1anf &  1293758 &  $0.1472$ &  $17.93\pm0.01$ &  $17.13\pm0.01$ &  $16.72\pm0.01$ &  $16.60\pm0.01$ &  $10.71\pm0.02$ &  $-9.51$ &  W19 \\
DES14S1aot &  1294014 &  $0.3627$ &  $22.28\pm0.01$ &  $21.58\pm0.01$ &  $21.47\pm0.01$ &  $21.29\pm0.01$ &  $9.17\pm0.02$ &  $-8.20$ &  W19 \\
DES14X3aeb &  1294743 &  $0.3126$ &  $20.99\pm0.01$ &  $20.08\pm0.01$ &  $19.57\pm0.01$ &  $19.42\pm0.01$ &  $10.26\pm0.01$ &  $-9.79$ &  W19 \\
DES14X2aph &  1295027 &  $0.4254$ &  $22.41\pm0.01$ &  $21.28\pm0.01$ &  $20.92\pm0.01$ &  $20.64\pm0.01$ &  $10.02\pm0.01$ &  $-9.79$ &  W19 \\
DES14C1bdv &  1295256 &  $0.4425$ &  $25.79\pm0.15$ &  $25.05\pm0.10$ &  $24.62\pm0.10$ &  $24.11\pm0.15$ &  $8.20\pm0.29$ &  $-8.55$ &  W19 \\
DES14X3ajv &  1295305 &  $0.6117$ &  $25.69\pm0.14$ &  $24.82\pm0.06$ &  $24.45\pm0.07$ &  $24.32\pm0.09$ &  $8.55\pm0.10$ &  $-8.39$ &  W19 \\
DES14S1aoz &  1295921 &  $0.5239$ &  $24.18\pm0.09$ &  $23.69\pm0.07$ &  $23.31\pm0.07$ &  $23.15\pm0.08$ &  $8.55\pm0.09$ &  $-7.69$ &  W19 \\
DES14S2boa &  1296273 &  $0.3970$ &  $23.89\pm0.04$ &  $23.39\pm0.04$ &  $23.26\pm0.04$ &  $23.12\pm0.05$ &  $8.36\pm0.04$ &  $-8.08$ &  W19 \\
DES14S2bnq &  1296321 &  $0.1844$ &  $21.23\pm0.01$ &  $20.63\pm0.01$ &  $20.36\pm0.01$ &  $20.23\pm0.01$ &  $9.26\pm0.02$ &  $-8.92$ &  W19 \\
DES14X2bnz &  1296657 &  $0.1462$ &  $19.14\pm0.01$ &  $18.82\pm0.01$ &  $18.56\pm0.01$ &  $18.52\pm0.01$ &  $9.43\pm0.03$ &  $-8.14$ &  W19 \\
DES14X3amb &  1297026 &  $0.2585$ &  $20.71\pm0.01$ &  $19.43\pm0.01$ &  $18.77\pm0.01$ &  $18.49\pm0.01$ &  $10.82\pm0.01$ &  $-10.26$ &  W19 \\
DES14C3cwp &  1297465 &  $0.2778$ &  $21.09\pm0.01$ &  $20.29\pm0.01$ &  $19.96\pm0.01$ &  $19.65\pm0.01$ &  $9.95\pm0.01$ &  $-9.24$ &  W19 \\
DES14S2dbi &  1298281 &  $0.2351$ &  $23.45\pm0.04$ &  $22.21\pm0.01$ &  $21.84\pm0.01$ &  $21.59\pm0.01$ &  $9.18\pm0.04$ &  $-12.77$ &  W19 \\
DES14X2eei &  1298893 &  $0.1962$ &  $19.56\pm0.01$ &  $18.41\pm0.01$ &  $17.95\pm0.01$ &  $17.64\pm0.01$ &  $10.74\pm0.07$ &  $-10.30$ &  W19 \\
DES14C1fkl &  1299643 &  $0.3800$ &  $23.71\pm0.03$ &  $22.63\pm0.01$ &  $22.12\pm0.02$ &  $21.85\pm0.02$ &  $9.52\pm0.02$ &  $-9.73$ &  W19 \\
DES14C2fkd &  1299775 &  $0.1598$ &  $19.50\pm0.01$ &  $18.46\pm0.01$ &  $17.93\pm0.01$ &  $17.56\pm0.01$ &  $10.68\pm0.01$ &  $-10.30$ &  W19 \\
DES14C2fkf &  1299785 &  $0.3813$ &  $21.68\pm0.01$ &  $20.63\pm0.01$ &  $20.32\pm0.01$ &  $20.11\pm0.01$ &  $10.09\pm0.02$ &  $-9.73$ &  W19 \\
DES14X1fnt &  1300516 &  $0.3104$ &  $23.20\pm0.05$ &  $21.76\pm0.01$ &  $21.34\pm0.01$ &  $20.83\pm0.01$ &  $9.94\pm0.03$ &  $-10.02$ &  W19 \\
DES14C3foo &  1300912 &  $0.3376$ &  $21.90\pm0.01$ &  $21.09\pm0.01$ &  $20.89\pm0.01$ &  $20.69\pm0.01$ &  $9.53\pm0.02$ &  $-8.56$ &  W19 \\
  \hline
\end{tabular*}
}
\flushleft
\end{table*}
\begin{table*}
\caption*{Continued from above}
{\centering
\begin{tabular*}{\linewidth}{llcccccccc}
  \hline
  DES Name & SNID & Redshift$^{1}$ & g & r & i & z & log(\mstellar) & log(sSFR) & Catalogue \\
  \hline
DES14X3ftq &  1301933 &  $0.3299$ &  $22.16\pm0.01$ &  $21.28\pm0.01$ &  $20.92\pm0.01$ &  $20.66\pm0.01$ &  $9.69\pm0.01$ &  $-9.11$ &  W19 \\
DES14C2gwx &  1302058 &  $0.1984$ &  $20.72\pm0.01$ &  $19.95\pm0.01$ &  $19.57\pm0.01$ &  $19.36\pm0.01$ &  $9.81\pm0.01$ &  $-9.38$ &  W19 \\
DES15E1cwo &  1302141 &  $0.6088$ &  $25.37\pm0.17$ &  $24.89\pm0.13$ &  $24.46\pm0.12$ &  $24.35\pm0.16$ &  $8.35\pm0.21$ &  $-8.08$ &  W19 \\
DES14C3gqv &  1302187 &  $0.2195$ &  $19.10\pm0.01$ &  $18.09\pm0.01$ &  $17.59\pm0.01$ &  $17.25\pm0.01$ &  $10.95\pm0.01$ &  $-9.86$ &  W19 \\
DES15X3dyu &  1302523 &  $0.4239$ &  $22.01\pm0.01$ &  $20.90\pm0.01$ &  $20.44\pm0.01$ &  $20.11\pm0.01$ &  $10.28\pm0.01$ &  $-9.81$ &  W19 \\
DES14E2fyd &  1302648 &  $0.2319$ &  $21.17\pm0.01$ &  $19.98\pm0.01$ &  $19.53\pm0.01$ &  $19.24\pm0.01$ &  $10.21\pm0.07$ &  $-10.23$ &  W19 \\
DES14X2gxr &  1303004 &  $0.2949$ &  $23.90\pm0.04$ &  $23.36\pm0.03$ &  $22.72\pm0.02$ &  $22.41\pm0.03$ &  $8.98\pm0.04$ &  $-9.43$ &  W19 \\
DES14C1jkw &  1303279 &  $0.1724$ &  $19.26\pm0.01$ &  $18.41\pm0.01$ &  $17.95\pm0.01$ &  $17.26\pm0.01$ &  $10.70\pm0.02$ &  $-10.07$ &  W19 \\
DES14C2ikn &  1303496 &  $0.1798$ &  $19.86\pm0.01$ &  $18.93\pm0.01$ &  $18.40\pm0.01$ &  $18.17\pm0.01$ &  $10.39\pm0.01$ &  $-9.84$ &  W19 \\
DES15E2dzb &  1303883 &  $0.2591$ &  $20.69\pm0.01$ &  $19.47\pm0.01$ &  $18.99\pm0.01$ &  $18.62\pm0.01$ &  $10.65\pm0.03$ &  $-10.02$ &  W19 \\
DES15X2dzq &  1303952 &  $0.5626$ &  $21.43\pm0.01$ &  $19.96\pm0.01$ &  $19.20\pm0.01$ &  $18.84\pm0.01$ &  $11.26\pm0.05$ &  $-10.54$ &  W19 \\
DES15X2dzo &  1304127 &  $0.7286$ &  $26.99\pm0.35$ &  $26.34\pm0.26$ &  $25.56\pm0.17$ &  $24.88\pm0.14$ &  $9.02\pm0.34$ &  $-9.76$ &  W19 \\
DES14C3hud &  1304442 &  $0.2171$ &  $20.99\pm0.01$ &  $20.58\pm0.01$ &  $20.41\pm0.01$ &  $20.18\pm0.01$ &  $9.24\pm0.02$ &  $-8.34$ &  W19 \\
DES14C1ikl &  1304678 &  $0.2188$ &  $19.17\pm0.01$ &  $18.06\pm0.01$ &  $17.56\pm0.01$ &  $17.03\pm0.01$ &  $11.13\pm0.01$ &  $-10.20$ &  W19 \\
DES14X3jmx &  1305504 &  $0.4956$ &  $21.96\pm0.01$ &  $20.46\pm0.01$ &  $19.58\pm0.01$ &  $19.22\pm0.01$ &  $11.22\pm0.01$ &  $-10.48$ &  W19 \\
DES14X3kbb &  1305626 &  $0.3257$ &  $23.45\pm0.02$ &  $22.53\pm0.01$ &  $22.19\pm0.01$ &  $21.91\pm0.01$ &  $9.21\pm0.01$ &  $-9.20$ &  W19 \\
DES15S2eak &  1306029 &  $0.4289$ &  $22.28\pm0.01$ &  $21.52\pm0.01$ &  $21.15\pm0.01$ &  $20.92\pm0.01$ &  $9.53\pm0.02$ &  $-8.46$ &  W19 \\
DES14X3kvo &  1306073 &  $0.3286$ &  $21.96\pm0.01$ &  $20.54\pm0.01$ &  $19.94\pm0.01$ &  $19.76\pm0.01$ &  $10.43\pm0.03$ &  $-13.24$ &  W19 \\
DES14C2kct &  1306141 &  $0.3324$ &  $21.54\pm0.01$ &  $20.74\pm0.01$ &  $20.52\pm0.01$ &  $20.26\pm0.01$ &  $9.70\pm0.02$ &  $-8.77$ &  W19 \\
DES15C1eat &  1306360 &  $0.4494$ &  $21.48\pm0.01$ &  $20.00\pm0.01$ &  $19.56\pm0.01$ &  $19.14\pm0.01$ &  $10.81\pm0.03$ &  $-9.59$ &  W19 \\
DES15S1ebd &  1306390 &  $0.4070$ &  $22.93\pm0.02$ &  $21.38\pm0.01$ &  $20.80\pm0.01$ &  $20.41\pm0.01$ &  $10.34\pm0.04$ &  $-9.89$ &  W19 \\
DES15C1ebo &  1306537 &  $0.4795$ &  $24.70\pm0.06$ &  $23.42\pm0.03$ &  $22.76\pm0.02$ &  $22.37\pm0.02$ &  $9.75\pm0.04$ &  $-10.08$ &  W19 \\
DES15C1ebn &  1306626 &  $0.4095$ &  $23.64\pm0.03$ &  $22.90\pm0.02$ &  $22.75\pm0.02$ &  $22.62\pm0.03$ &  $8.73\pm0.03$ &  $-8.10$ &  W19 \\
DES15X1ebs &  1306785 &  $0.5787$ &  $23.57\pm0.09$ &  $21.54\pm0.01$ &  $20.53\pm0.01$ &  $19.98\pm0.01$ &  $11.07\pm0.05$ &  $-13.31$ &  W19 \\
DES15E1ebw &  1306980 &  $0.5489$ &  $22.60\pm0.02$ &  $21.48\pm0.01$ &  $21.01\pm0.01$ &  $20.68\pm0.01$ &  $10.28\pm0.02$ &  $-9.85$ &  W19 \\
DES15E1ece &  1306991 &  $0.4217$ &  $21.43\pm0.01$ &  $19.72\pm0.01$ &  $19.17\pm0.01$ &  $18.78\pm0.01$ &  $10.90\pm0.02$ &  $-10.21$ &  W19 \\
DES15S2eco &  1307277 &  $0.4107$ &  $22.39\pm0.02$ &  $21.07\pm0.01$ &  $20.59\pm0.01$ &  $20.23\pm0.01$ &  $10.32\pm0.01$ &  $-9.70$ &  W19 \\
DES15C3edd &  1307830 &  $0.3493$ &  $22.27\pm0.01$ &  $20.68\pm0.01$ &  $19.96\pm0.01$ &  $19.80\pm0.01$ &  $10.63\pm0.02$ &  $-13.00$ &  W19 \\
DES14C2kdr &  1308314 &  $0.4065$ &  $25.35\pm0.09$ &  $24.76\pm0.07$ &  $24.63\pm0.09$ &  $24.49\pm0.12$ &  $7.89\pm0.11$ &  $-8.11$ &  W19 \\
DES14S1kdq &  1308326 &  $0.3262$ &  $21.76\pm0.01$ &  $20.96\pm0.01$ &  $20.82\pm0.01$ &  $20.59\pm0.01$ &  $9.54\pm0.02$ &  $-8.88$ &  W19 \\
DES15X1eei &  1308568 &  $0.6431$ &  $24.21\pm0.05$ &  $23.87\pm0.05$ &  $23.51\pm0.03$ &  $23.45\pm0.05$ &  $8.64\pm0.04$ &  $-8.11$ &  W19 \\
DES15X3flq &  1308582 &  $0.3669$ &  $21.60\pm0.01$ &  $19.99\pm0.01$ &  $19.45\pm0.01$ &  $19.06\pm0.01$ &  $10.71\pm0.03$ &  $-10.21$ &  W19 \\
DES15C3efn &  1308884 &  $0.0772$ &  $16.55\pm0.01$ &  $16.11\pm0.01$ &  $15.75\pm0.01$ &  $15.72\pm0.01$ &  $10.34\pm0.02$ &  $-9.88$ &  W19 \\
DES15X2efk &  1308957 &  $0.6187$ &  $25.54\pm0.19$ &  $23.52\pm0.04$ &  $22.55\pm0.02$ &  $22.03\pm0.02$ &  $10.14\pm0.07$ &  $-12.97$ &  W19 \\
DES14C2mng &  1309288 &  $0.2684$ &  $19.55\pm0.01$ &  $18.46\pm0.01$ &  $18.00\pm0.01$ &  $17.76\pm0.01$ &  $10.88\pm0.02$ &  $-9.67$ &  W19 \\
DES14C3mpt &  1309492 &  $0.3335$ &  $22.99\pm0.01$ &  $22.13\pm0.01$ &  $21.89\pm0.01$ &  $21.53\pm0.01$ &  $9.30\pm0.02$ &  $-9.05$ &  W19 \\
DES14C3mpr &  1309749 &  $0.79$ &  $24.38\pm0.06$ &  $23.67\pm0.03$ &  $23.41\pm0.04$ &  $23.08\pm0.06$ &  $9.01\pm0.06$ &  $-7.86$ &  W19 \\
DES14X2mqz &  1310338 &  $0.44$ &  $25.17\pm0.08$ &  $24.90\pm0.08$ &  $25.21\pm0.14$ &  $24.90\pm0.15$ &  $7.40\pm0.10$ &  $-7.83$ &  W19 \\
DES14X2mgg &  1310395 &  $0.29$ &  $28.95\pm2.02$ &  $25.67\pm0.14$ &  $25.96\pm0.25$ &  $25.95\pm0.33$ &  $7.55\pm0.17$ &  $-12.06$ &  W19 \\
DES14S1lfk &  1312274 &  $0.4380$ &  $23.43\pm0.04$ &  $21.89\pm0.01$ &  $21.23\pm0.01$ &  $20.72\pm0.01$ &  $10.45\pm0.02$ &  $-10.20$ &  W19 \\
DES14X1oes &  1313594 &  $0.2883$ &  $21.99\pm0.02$ &  $21.29\pm0.02$ &  $20.52\pm0.01$ &  $20.01\pm0.01$ &  $10.09\pm0.02$ &  $-9.87$ &  W19 \\
DES15X3itc &  1314897 &  $0.3369$ &  $20.50\pm0.01$ &  $19.28\pm0.01$ &  $18.82\pm0.01$ &  $18.40\pm0.01$ &  $10.86\pm0.01$ &  $-9.68$ &  W19 \\
DES14S1qid &  1315192 &  $0.2254$ &  $20.12\pm0.01$ &  $19.17\pm0.01$ &  $18.74\pm0.01$ &  $18.44\pm0.01$ &  $10.40\pm0.01$ &  $-9.54$ &  W19 \\
DES14S2pkz &  1315259 &  $0.2280$ &  $18.23\pm0.01$ &  $17.15\pm0.01$ &  $16.65\pm0.01$ &  $16.40\pm0.01$ &  $11.33\pm0.01$ &  $-9.78$ &  W19 \\
DES14S2pon &  1315296 &  $0.4126$ &  $22.07\pm0.01$ &  $21.19\pm0.01$ &  $21.02\pm0.01$ &  $20.77\pm0.01$ &  $9.56\pm0.02$ &  $-8.43$ &  W19 \\
DES14C1qty &  1316385 &  $0.1478$ &  $19.89\pm0.01$ &  $19.08\pm0.01$ &  $18.71\pm0.01$ &  $18.16\pm0.01$ &  $10.12\pm0.01$ &  $-9.84$ &  W19 \\
DES14S1rah &  1316431 &  $0.1981$ &  $26.17\pm0.23$ &  $25.77\pm0.19$ &  $25.76\pm0.23$ &  $26.06\pm0.46$ &  $6.52\pm0.32$ &  $-8.00$ &  W19 \\
DES14S1rag &  1316437 &  $0.4820$ &  $22.86\pm0.02$ &  $22.00\pm0.01$ &  $21.82\pm0.01$ &  $21.67\pm0.02$ &  $9.27\pm0.02$ &  $-8.35$ &  W19 \\
DES14C3rap &  1316465 &  $0.3291$ &  $22.97\pm0.01$ &  $22.22\pm0.01$ &  $21.96\pm0.01$ &  $21.77\pm0.01$ &  $9.07\pm0.02$ &  $-8.56$ &  W19 \\
DES15X1ith &  1317164 &  $0.1547$ &  $19.41\pm0.01$ &  $18.05\pm0.01$ &  $17.37\pm0.01$ &  $17.05\pm0.01$ &  $11.08\pm0.02$ &  $-21.08$ &  W19 \\
DES14X2raq &  1317277 &  $0.2371$ &  $21.58\pm0.01$ &  $20.32\pm0.01$ &  $19.86\pm0.01$ &  $19.55\pm0.01$ &  $10.04\pm0.06$ &  $-10.38$ &  W19 \\
DES14X2rao &  1317286 &  $0.28$ &  $25.57\pm0.14$ &  $24.99\pm0.11$ &  $25.32\pm0.20$ &  $24.82\pm0.18$ &  $7.46\pm0.13$ &  $-8.69$ &  W19 \\
DES14E1rpk &  1317454 &  $0.57$ &  --- &  --- &  --- &  --- &  --- &  --- &  --- \\
DES15X3kqv &  1317666 &  $0.1414$ &  $19.63\pm0.01$ &  $18.86\pm0.01$ &  $18.49\pm0.01$ &  $18.28\pm0.01$ &  $9.97\pm0.01$ &  $-9.48$ &  W19 \\
DES15S2kqw &  1319366 &  $0.2361$ &  $21.06\pm0.01$ &  $20.46\pm0.01$ &  $20.18\pm0.01$ &  $20.03\pm0.01$ &  $9.49\pm0.02$ &  $-8.54$ &  W19 \\
DES14C2rsj &  1319821 &  $0.3088$ &  $20.22\pm0.01$ &  $19.50\pm0.01$ &  $19.21\pm0.01$ &  $18.92\pm0.01$ &  $10.21\pm0.01$ &  $-8.93$ &  W19 \\
DES14E2slo &  1320166 &  $0.45$ &  --- &  --- &  --- &  --- &  --- &  --- &  --- \\
DES14X3tdv &  1322229 &  $0.5261$ &  $23.40\pm0.02$ &  $22.50\pm0.01$ &  $22.12\pm0.01$ &  $21.87\pm0.01$ &  $9.61\pm0.02$ &  $-9.62$ &  W19 \\
DES14C3tvk &  1322979 &  $0.5294$ &  $27.37\pm0.35$ &  $25.66\pm0.09$ &  $25.37\pm0.09$ &  $25.78\pm0.27$ &  $8.05\pm0.08$ &  $-9.87$ &  W19 \\
DES14C2vnf &  1324542 &  $0.5415$ &  $24.81\pm0.09$ &  $24.06\pm0.06$ &  $23.65\pm0.06$ &  $23.40\pm0.07$ &  $8.96\pm0.15$ &  $-9.54$ &  W19 \\
DES14C3uje &  1325358 &  $0.7793$ &  $25.68\pm0.10$ &  $25.61\pm0.11$ &  $25.04\pm0.09$ &  $24.59\pm0.12$ &  $8.30\pm0.15$ &  $-8.68$ &  W19 \\
DES15E2kvn &  1327978 &  $0.2073$ &  $19.21\pm0.01$ &  $18.02\pm0.01$ &  $17.56\pm0.01$ &  $17.20\pm0.01$ &  $11.03\pm0.04$ &  $-10.18$ &  W19 \\
DES15E1kst &  1328066 &  $0.4489$ &  $23.54\pm0.03$ &  $21.93\pm0.01$ &  $21.39\pm0.01$ &  $21.09\pm0.01$ &  $9.94\pm0.04$ &  $-10.98$ &  W19 \\
DES15E1kvp &  1328105 &  $0.4402$ &  $20.21\pm0.01$ &  $19.01\pm0.01$ &  $18.60\pm0.01$ &  $18.24\pm0.01$ &  $11.07\pm0.01$ &  $-9.73$ &  W19 \\
  \hline
\end{tabular*}
}
\flushleft
\end{table*}
\begin{table*}
\caption*{Continued from above}
{\centering
\begin{tabular*}{\linewidth}{llcccccccc}
  \hline
  DES Name & SNID & Redshift$^{1}$ & g & r & i & z & log(\mstellar) & log(sSFR) & Catalogue \\
  \hline
DES15C3kuw &  1329166 &  $0.7316$ &  $24.83\pm0.05$ &  $24.02\pm0.03$ &  $23.52\pm0.02$ &  $23.24\pm0.03$ &  $9.20\pm0.04$ &  $-8.53$ &  W19 \\
DES15C3kue &  1329196 &  $0.7693$ &  $24.19\pm0.04$ &  $23.57\pm0.03$ &  $23.07\pm0.02$ &  $22.66\pm0.03$ &  $9.32\pm0.04$ &  $-8.30$ &  W19 \\
DES15X3kxu &  1329312 &  $0.3444$ &  $20.96\pm0.01$ &  $19.64\pm0.01$ &  $19.15\pm0.01$ &  $18.74\pm0.01$ &  $10.82\pm0.01$ &  $-9.83$ &  W19 \\
DES15X2kvt &  1329615 &  $0.4039$ &  $21.66\pm0.01$ &  $20.07\pm0.01$ &  $19.37\pm0.01$ &  $19.11\pm0.01$ &  $10.96\pm0.03$ &  $-20.96$ &  W19 \\
DES15E1kwg &  1330031 &  $0.1041$ &  $20.66\pm0.01$ &  $20.09\pm0.01$ &  $19.80\pm0.01$ &  $19.72\pm0.01$ &  $9.03\pm0.03$ &  $-10.08$ &  W19 \\
DES15C2kyh &  1330426 &  $0.2595$ &  $25.39\pm0.17$ &  $24.53\pm0.09$ &  $24.67\pm0.15$ &  $24.26\pm0.18$ &  $7.83\pm0.10$ &  $-10.15$ &  W19 \\
DES15X3lab &  1330642 &  $0.6339$ &  $24.22\pm0.04$ &  $23.52\pm0.02$ &  $23.03\pm0.03$ &  $22.83\pm0.03$ &  $9.30\pm0.06$ &  $-9.41$ &  W19 \\
DES15S2lam &  1330903 &  $0.5638$ &  $23.87\pm0.05$ &  $22.12\pm0.01$ &  $21.34\pm0.01$ &  $20.95\pm0.01$ &  $10.32\pm0.04$ &  $-10.78$ &  W19 \\
DES15E1lew &  1331123 &  $0.2291$ &  $19.46\pm0.01$ &  $18.19\pm0.01$ &  $17.71\pm0.01$ &  $17.34\pm0.01$ &  $11.09\pm0.05$ &  $-10.42$ &  W19 \\
DES15X3lqs &  1331993 &  $0.7196$ &  $24.56\pm0.13$ &  $23.41\pm0.04$ &  $21.95\pm0.02$ &  $16.49\pm0.01$ &  $11.20\pm0.01$ &  $-14.30$ &  W19 \\
DES15S2lmu &  1332059 &  $0.58$ &  --- &  --- &  --- &  --- &  --- &  --- &  --- \\
DES15X2lnb &  1332413 &  $0.6087$ &  $25.42\pm0.15$ &  $24.68\pm0.10$ &  $24.47\pm0.19$ &  $24.49\pm0.15$ &  $8.32\pm0.14$ &  $-8.40$ &  W19 \\
DES15S2lot &  1333246 &  $0.4379$ &  $21.31\pm0.01$ &  $19.98\pm0.01$ &  $19.51\pm0.01$ &  $19.17\pm0.01$ &  $10.76\pm0.02$ &  $-9.70$ &  W19 \\
DES15C3lvt &  1333438 &  $0.3995$ &  $21.55\pm0.02$ &  $20.16\pm0.01$ &  $19.75\pm0.01$ &  $19.42\pm0.01$ &  $10.53\pm0.02$ &  $-9.56$ &  SVA1 \\
DES15S2mau &  1334084 &  $0.1335$ &  $19.25\pm0.01$ &  $18.61\pm0.01$ &  $18.24\pm0.01$ &  $18.05\pm0.01$ &  $9.94\pm0.01$ &  $-9.28$ &  W19 \\
DES15S2max &  1334087 &  $0.2663$ &  $20.04\pm0.01$ &  $18.65\pm0.01$ &  $18.15\pm0.01$ &  $17.85\pm0.01$ &  $11.00\pm0.10$ &  $-13.07$ &  W19 \\
DES15E1mar &  1334302 &  $0.4549$ &  $22.50\pm0.01$ &  $21.06\pm0.01$ &  $20.45\pm0.01$ &  $20.01\pm0.01$ &  $10.63\pm0.01$ &  $-9.87$ &  W19 \\
DES15S1lyi &  1334423 &  $0.3577$ &  $21.94\pm0.01$ &  $21.24\pm0.01$ &  $21.20\pm0.01$ &  $20.93\pm0.01$ &  $9.27\pm0.02$ &  $-8.43$ &  W19 \\
DES15X1mbc &  1334448 &  $0.5167$ &  $23.58\pm0.06$ &  $21.46\pm0.01$ &  $20.48\pm0.01$ &  $20.03\pm0.01$ &  $11.09\pm0.05$ &  $-13.87$ &  W19 \\
DES15X1mav &  1334470 &  $0.5188$ &  $27.22\pm0.78$ &  $26.01\pm0.28$ &  $25.92\pm0.23$ &  $25.47\pm0.24$ &  $8.00\pm0.39$ &  $-9.13$ &  W19 \\
DES15X3lya &  1334597 &  $0.2890$ &  $24.18\pm0.05$ &  $22.99\pm0.02$ &  $22.58\pm0.02$ &  $22.37\pm0.02$ &  $9.04\pm0.07$ &  $-10.51$ &  W19 \\
DES15X2lxw &  1334620 &  $0.1961$ &  $20.03\pm0.01$ &  $19.21\pm0.01$ &  $18.81\pm0.01$ &  $18.64\pm0.01$ &  $10.11\pm0.01$ &  $-9.38$ &  W19 \\
DES15X2lxv &  1334644 &  $0.2881$ &  $19.87\pm0.01$ &  $18.40\pm0.01$ &  $17.78\pm0.01$ &  $17.61\pm0.01$ &  $11.34\pm0.01$ &  $-16.68$ &  W19 \\
DES15X2mei &  1334645 &  $0.2312$ &  $20.44\pm0.01$ &  $19.35\pm0.01$ &  $18.88\pm0.01$ &  $18.69\pm0.01$ &  $10.37\pm0.03$ &  $-10.17$ &  W19 \\
DES15C3lyd &  1334707 &  $0.6488$ &  $23.34\pm0.01$ &  $22.94\pm0.01$ &  $22.67\pm0.01$ &  $22.47\pm0.02$ &  $9.02\pm0.01$ &  $-8.13$ &  W19 \\
DES15C3lzl &  1334879 &  $0.6393$ &  $28.09\pm0.39$ &  $27.47\pm0.26$ &  $26.58\pm0.15$ &  $27.74\pm0.88$ &  $7.06\pm0.37$ &  $-7.48$ &  W19 \\
DES15E2mhj &  1335472 &  $0.4989$ &  $23.03\pm0.02$ &  $21.74\pm0.01$ &  $21.27\pm0.01$ &  $20.88\pm0.01$ &  $10.21\pm0.02$ &  $-9.81$ &  W19 \\
DES15X2mey &  1335564 &  $0.6077$ &  $24.20\pm0.06$ &  $23.20\pm0.03$ &  $22.77\pm0.03$ &  $22.51\pm0.03$ &  $9.41\pm0.05$ &  $-8.36$ &  W19 \\
DES15S1mjm &  1335694 &  $0.2591$ &  $18.89\pm0.01$ &  $18.36\pm0.01$ &  $18.04\pm0.01$ &  $17.99\pm0.01$ &  $10.24\pm0.02$ &  $-8.14$ &  W19 \\
DES15C3mgv &  1335717 &  $0.3048$ &  $21.74\pm0.01$ &  $20.97\pm0.01$ &  $20.71\pm0.01$ &  $20.45\pm0.01$ &  $9.62\pm0.01$ &  $-8.96$ &  W19 \\
DES15C3mga &  1335718 &  $0.6993$ &  $28.72\pm1.30$ &  $26.55\pm0.20$ &  $25.70\pm0.12$ &  $25.02\pm0.13$ &  $9.26\pm0.25$ &  $-10.61$ &  W19 \\
DES15C1mhp &  1335868 &  $0.62$ &  --- &  --- &  --- &  --- &  --- &  --- &  --- \\
DES15E2mhy &  1336008 &  $0.4380$ &  $23.17\pm0.02$ &  $22.12\pm0.01$ &  $21.78\pm0.01$ &  $21.39\pm0.01$ &  $9.74\pm0.01$ &  $-9.73$ &  W19 \\
DES15E2mhv &  1336009 &  $0.3416$ &  $23.08\pm0.02$ &  $22.15\pm0.01$ &  $21.93\pm0.01$ &  $21.59\pm0.01$ &  $9.31\pm0.02$ &  $-9.06$ &  W19 \\
DES15S2mpg &  1336453 &  $0.1848$ &  $20.69\pm0.01$ &  $20.27\pm0.01$ &  $20.11\pm0.01$ &  $20.03\pm0.01$ &  $9.16\pm0.03$ &  $-8.35$ &  W19 \\
DES15S2mpl &  1336480 &  $0.2560$ &  $20.69\pm0.01$ &  $19.96\pm0.01$ &  $19.60\pm0.01$ &  $19.44\pm0.01$ &  $9.96\pm0.01$ &  $-9.10$ &  W19 \\
DES15X2mpm &  1336687 &  $0.2337$ &  $22.56\pm0.02$ &  $22.12\pm0.02$ &  $22.03\pm0.02$ &  $21.93\pm0.02$ &  $8.57\pm0.03$ &  $-8.49$ &  W19 \\
DES15E2msq &  1337117 &  $0.5539$ &  $25.33\pm0.11$ &  $24.33\pm0.06$ &  $24.20\pm0.07$ &  $23.84\pm0.08$ &  $8.64\pm0.08$ &  $-8.38$ &  W19 \\
DES15E1mvj &  1337221 &  $0.6688$ &  $25.64\pm0.16$ &  $24.97\pm0.11$ &  $24.29\pm0.08$ &  $24.21\pm0.13$ &  $8.83\pm0.16$ &  $-9.47$ &  W19 \\
DES15E1mvi &  1337228 &  $0.5788$ &  $23.67\pm0.04$ &  $22.62\pm0.02$ &  $22.22\pm0.02$ &  $21.90\pm0.02$ &  $9.73\pm0.05$ &  $-9.62$ &  W19 \\
DES15X1mvl &  1337272 &  $0.5027$ &  $24.76\pm0.10$ &  $24.03\pm0.05$ &  $23.84\pm0.06$ &  $23.54\pm0.07$ &  $8.46\pm0.10$ &  $-8.02$ &  W19 \\
DES15X1mvs &  1337325 &  $0.6387$ &  $25.27\pm0.31$ &  $23.57\pm0.08$ &  $22.45\pm0.03$ &  $22.00\pm0.02$ &  $10.36\pm0.10$ &  $-12.71$ &  W19 \\
DES15S1mvv &  1337649 &  $0.2491$ &  $23.62\pm0.04$ &  $22.82\pm0.02$ &  $22.61\pm0.02$ &  $22.44\pm0.03$ &  $8.70\pm0.03$ &  $-9.03$ &  W19 \\
DES15X3mwb &  1337655 &  $0.85$ &  --- &  --- &  --- &  --- &  --- &  --- &  --- \\
DES15C1mvy &  1337687 &  $0.3195$ &  $24.21\pm0.03$ &  $23.61\pm0.03$ &  $23.54\pm0.04$ &  $23.37\pm0.05$ &  $8.23\pm0.05$ &  $-8.35$ &  W19 \\
DES15C1mvx &  1337703 &  $0.5334$ &  $22.95\pm0.02$ &  $21.19\pm0.01$ &  $20.42\pm0.01$ &  $20.02\pm0.01$ &  $10.64\pm0.03$ &  $-11.55$ &  W19 \\
DES15X2mzv &  1337838 &  $0.3119$ &  $25.59\pm0.21$ &  $24.75\pm0.11$ &  $24.63\pm0.14$ &  $24.39\pm0.18$ &  $7.95\pm0.19$ &  $-8.86$ &  W19 \\
DES15E1nei &  1338128 &  $0.3117$ &  $22.68\pm0.01$ &  $22.01\pm0.01$ &  $21.86\pm0.01$ &  $21.64\pm0.01$ &  $8.95\pm0.03$ &  $-8.21$ &  W19 \\
DES15E1neh &  1338170 &  $0.3890$ &  $22.55\pm0.01$ &  $21.09\pm0.01$ &  $20.66\pm0.01$ &  $20.36\pm0.01$ &  $10.13\pm0.02$ &  $-9.69$ &  W19 \\
DES15X1mwg &  1338233 &  $0.5980$ &  $22.18\pm0.02$ &  $21.07\pm0.01$ &  $20.00\pm0.01$ &  $19.57\pm0.01$ &  $11.21\pm0.01$ &  $-10.59$ &  W19 \\
DES15X1mzz &  1338266 &  $0.6476$ &  $24.04\pm0.06$ &  $23.54\pm0.06$ &  $23.09\pm0.04$ &  $22.86\pm0.04$ &  $9.08\pm0.05$ &  $-8.13$ &  W19 \\
DES15X1ney &  1338278 &  $0.5639$ &  $23.24\pm0.05$ &  $22.08\pm0.02$ &  $21.55\pm0.01$ &  $21.31\pm0.01$ &  $10.05\pm0.03$ &  $-9.65$ &  W19 \\
DES15S2myz &  1338387 &  $0.5838$ &  $25.78\pm0.20$ &  $24.79\pm0.12$ &  $24.74\pm0.13$ &  $24.23\pm0.12$ &  $8.51\pm0.14$ &  $-8.40$ &  W19 \\
DES15S2mwz &  1338430 &  $0.5089$ &  $22.14\pm0.02$ &  $20.59\pm0.01$ &  $19.92\pm0.01$ &  $19.51\pm0.01$ &  $10.94\pm0.03$ &  $-9.84$ &  W19 \\
DES15S2mxe &  1338471 &  $0.5309$ &  $22.83\pm0.02$ &  $21.56\pm0.01$ &  $21.12\pm0.01$ &  $20.78\pm0.01$ &  $10.23\pm0.02$ &  $-9.62$ &  W19 \\
DES15X3naa &  1338675 &  $0.3306$ &  $22.29\pm0.01$ &  $21.61\pm0.01$ &  $21.36\pm0.01$ &  $21.10\pm0.01$ &  $9.27\pm0.02$ &  $-8.47$ &  W19 \\
DES15X2nkl &  1339002 &  $0.3025$ &  $21.07\pm0.01$ &  $19.80\pm0.01$ &  $19.25\pm0.01$ &  $18.98\pm0.01$ &  $10.65\pm0.04$ &  $-10.27$ &  W19 \\
DES15X2nkz &  1339149 &  $0.4678$ &  $22.39\pm0.02$ &  $21.42\pm0.01$ &  $20.99\pm0.01$ &  $20.86\pm0.01$ &  $9.77\pm0.03$ &  $-8.66$ &  W19 \\
DES15C2njv &  1339392 &  $0.1804$ &  $19.98\pm0.01$ &  $18.93\pm0.01$ &  $18.49\pm0.01$ &  $18.26\pm0.01$ &  $10.29\pm0.03$ &  $-10.72$ &  W19 \\
DES15C2nfs &  1339450 &  $0.5505$ &  $21.12\pm0.01$ &  $19.78\pm0.01$ &  $19.13\pm0.01$ &  $18.81\pm0.01$ &  $11.14\pm0.05$ &  $-10.42$ &  W19 \\
DES15C1nhv &  1340454 &  $0.4210$ &  $22.45\pm0.02$ &  $21.37\pm0.01$ &  $20.98\pm0.01$ &  $20.65\pm0.01$ &  $10.03\pm0.01$ &  $-9.78$ &  W19 \\
DES15E2nlz &  1341370 &  $0.4090$ &  $24.76\pm0.07$ &  $23.65\pm0.04$ &  $23.51\pm0.04$ &  $23.30\pm0.06$ &  $8.70\pm0.05$ &  $-9.08$ &  W19 \\
DES15X1nxy &  1341894 &  $0.3121$ &  $21.46\pm0.01$ &  $20.47\pm0.01$ &  $20.05\pm0.01$ &  $19.79\pm0.01$ &  $10.11\pm0.01$ &  $-9.59$ &  W19 \\
DES15S2ocv &  1342255 &  $0.2141$ &  $21.09\pm0.01$ &  $20.29\pm0.01$ &  $19.92\pm0.01$ &  $19.71\pm0.01$ &  $9.75\pm0.01$ &  $-9.31$ &  W19 \\
DES15C3nym &  1343208 &  $0.4993$ &  $22.65\pm0.01$ &  $21.13\pm0.01$ &  $20.52\pm0.01$ &  $20.17\pm0.01$ &  $10.54\pm0.04$ &  $-10.42$ &  W19 \\
  \hline
\end{tabular*}
}
\flushleft
\end{table*}
\begin{table*}
\caption*{Continued from above}
{\centering
\begin{tabular*}{\linewidth}{llcccccccc}
  \hline
  DES Name & SNID & Redshift$^{1}$ & g & r & i & z & log(\mstellar) & log(sSFR) & Catalogue \\
  \hline
DES15C2odp &  1343337 &  $0.3389$ &  $21.73\pm0.01$ &  $20.99\pm0.01$ &  $20.72\pm0.01$ &  $20.43\pm0.01$ &  $9.64\pm0.02$ &  $-8.79$ &  W19 \\
DES15X1odo &  1343401 &  $0.3829$ &  $21.43\pm0.01$ &  $20.08\pm0.01$ &  $19.29\pm0.01$ &  $18.97\pm0.01$ &  $11.00\pm0.02$ &  $-10.13$ &  W19 \\
DES15E1ods &  1343533 &  $0.3680$ &  $21.39\pm0.01$ &  $19.98\pm0.01$ &  $19.48\pm0.01$ &  $19.14\pm0.01$ &  $10.69\pm0.03$ &  $-9.84$ &  W19 \\
DES15C3odz &  1343759 &  $0.5080$ &  $22.77\pm0.01$ &  $21.96\pm0.01$ &  $21.75\pm0.01$ &  $21.51\pm0.01$ &  $9.36\pm0.02$ &  $-8.05$ &  W19 \\
DES15S1oeh &  1343871 &  $0.6379$ &  $24.80\pm0.09$ &  $24.10\pm0.06$ &  $23.58\pm0.05$ &  $23.53\pm0.07$ &  $8.86\pm0.09$ &  $-8.17$ &  W19 \\
DES15X2ogh &  1344692 &  $0.3795$ &  $20.81\pm0.01$ &  $19.13\pm0.01$ &  $18.54\pm0.01$ &  $18.26\pm0.01$ &  $11.04\pm0.12$ &  $-12.47$ &  W19 \\
DES15X1ojh &  1345553 &  $0.3189$ &  $22.57\pm0.03$ &  $21.30\pm0.01$ &  $20.73\pm0.01$ &  $20.44\pm0.01$ &  $10.14\pm0.03$ &  $-10.23$ &  W19 \\
DES15X1oox &  1345582 &  $0.4588$ &  $24.15\pm0.13$ &  $23.04\pm0.05$ &  $22.20\pm0.03$ &  $21.79\pm0.02$ &  $10.28\pm0.06$ &  $-9.32$ &  W19 \\
DES15X1oqk &  1345594 &  $0.4224$ &  $21.11\pm0.01$ &  $19.59\pm0.01$ &  $19.00\pm0.01$ &  $18.63\pm0.01$ &  $11.14\pm0.04$ &  $-9.95$ &  W19 \\
DES15C1olp &  1346137 &  $0.3635$ &  $21.29\pm0.01$ &  $20.49\pm0.01$ &  $20.25\pm0.01$ &  $19.91\pm0.01$ &  $9.89\pm0.02$ &  $-8.76$ &  W19 \\
DES15C3omh &  1346387 &  $0.3445$ &  $24.60\pm0.13$ &  $23.33\pm0.05$ &  $22.63\pm0.03$ &  $23.01\pm0.09$ &  $9.04\pm0.08$ &  $-12.48$ &  W19 \\
DES15C2oxn &  1346956 &  $0.3335$ &  $27.02\pm0.21$ &  $26.51\pm0.16$ &  $26.54\pm0.23$ &  $26.29\pm0.32$ &  $6.90\pm0.25$ &  $-8.19$ &  W19 \\
DES15C2oxo &  1346966 &  $0.3355$ &  $21.66\pm0.01$ &  $20.96\pm0.01$ &  $20.80\pm0.01$ &  $20.54\pm0.01$ &  $9.45\pm0.02$ &  $-8.31$ &  W19 \\
  \hline
\end{tabular*}
}
\flushleft
\footnote{a}{Redshift quoted to 4 decimal places (d.p.) when determined from galaxy emission / absorption features or 2 d.p. when determined from SN template matches.}
\end{table*}

\label{lastpage}
\end{document}